\documentclass[a4paper,11 pt]{article}
\pdfoutput=1 
\usepackage{jheppub} 
\usepackage{orcidlink}

\usepackage[T1]{fontenc}
\usepackage{feynman}
\usepackage{amsmath}
\usepackage{caption}
\usepackage{subcaption,longtable,stmaryrd}
\usepackage{bigints}
\usepackage{multicol}
\usepackage{blindtext}
\usepackage{tikz}
\usepackage{enumitem}
\usepackage[customcolors,shade]{hf-tikz}
\usepackage{graphicx}
\usepackage[bitstream-charter]{mathdesign}

\usepackage[a4paper,
      bindingoffset=0.6in,
      left=1.4in,
         right=-0.1 in,
         top=1.9in,
      bottom=0.5in,
         footskip=.4in]{geometry}
          \DeclareSymbolFont{usualmathcal}{OMS}{cmsy}{m}{n}
\DeclareSymbolFontAlphabet{\mathcal}{usualmathcal}
\DeclareSymbolFont{rsfs}{U}{rsfs}{m}{n}
\DeclareSymbolFontAlphabet{\mathscrsfs}{rsfs}
\DeclareSymbolFont{rmlargesymbols}{OMX}{mdbch}{m}{n}
\DeclareMathSymbol{\rmintop}{\mathop}{rmlargesymbols}{82}
\DeclareMathSymbol{\rmointop}{\mathop}{rmlargesymbols}{72}
\usepackage[customcolors,shade]{hf-tikz}
\newcommand{\rmint}{\rmintop\nolimits}
\definecolor{darkbrown}{rgb}{0.73, 0.26, 0.187}
\definecolor{alizarin}{rgb}{0.73, 0.26, 0.187}
\title{Observables from classical black hole scattering in Scalar-Tensor theory of gravity from worldline quantum field theory}

\author[a]{Arpan Bhattacharyya
,}

\author[b,c]{Debodirna Ghosh 
}
\author[a]{Saptaswa Ghosh
,} 
\author[a]{and Sounak Pal 
}
\affiliation[a]{\it Indian Institute of Technology Gandhinagar,\\ Gujarat-382055, India}
\affiliation[b]{\it Institute of Mathematical Sciences,\\ IV Cross Road, C.I.T. Campus, Taramani, Chennai 600113, India}
\affiliation[c]{Homi Bhabha National Institute, \\Training School Complex, Anushakti Nagar, Mumbai 400094, India}
\emailAdd{abhattacharyya@iitgn.ac.in}
\emailAdd{debodirna@imsc.res.in}
\emailAdd{saptaswaghosh@iitgn.ac.in}
\emailAdd{palsounak@iitgn.ac.in}
\abstract{In this paper, we compute the two observables, impulse and waveform, in a black hole scattering event for the Scalar-Tensor theory of gravity with a generic scalar potential using the techniques of Worldline Quantum Field Theory. We mainly investigate the corrections to the above mentioned observables due to the extra scalar degree of freedom. For the computation of impulse, we consider the most general scenario by making the scalar field massive and then show that each computed diagram has a smooth massless limit. We compute the waveform for scalar and graviton up to 2PM, taking the scalar as massless. Furthermore, we discuss when the scalar has mass, how the radiation integrals get more involved than the massless case. We also arrive at some analytical results using the stationary phase approximation. 

}
\begin{document}
\maketitle
\flushbottom
\newpage
\section{Introduction} \label{intro}
The detection of gravitational waves through a network of ground-based laser-interferometers provide a new way of ‘‘listening’’ to the Universe in
the high-frequency band \cite{LIGOScientific:2014oec,LIGOScientific:2016aoc,LIGOScientific:2016sjg,LIGOScientific:2016vlm,LIGOScientific:2017bnn,LIGOScientific:2019hgc}. A future space-based ones promises to open the low-frequency band, and also pulsar timing arrays are designed to explore gravitational waves at nano-hertz frequencies. In addition to
providing astrophysical information, these
observations will provide us a scope to test Einstein’s theory of general relativity in the strong-field, dynamical regime. While most tests so far constrain theory-agnostic deviations from general relativity (GR), current efforts aim at calculating waveform templates in specific modified gravity theories. The LIGO/VIRGO observations of black hole/Neutron star inspirals/mergers require high precision analytical computations of classical potential and radiation coming out from the binary system \cite{Purrer:2019jcp}. The high-precision computations are equivalent to the computations of scattering cross-section in elementary particle physics using the tools of Quantum Field Theory (QFT). Inspired by the QFT formalism, the perturbative quantum gravity formalism has been proven to be very efficient in investigating the classical gravitational interaction of black holes or Neutron stars.\par 
In recent years, after the direct detection of gravitational waves (GW), renewed interest has been spurred to test the legitimacy of General relativity (GR) directly at all possible length scales. One of the main reasons for this test is that although General Relativity (GR) has had remarkable success in describing classical gravity, including the phenomenal discovery of Gravitational waves (GW), the theory is incomplete. It is widely expected that there should be a consistent quantum theory of gravity, which is UV complete. Also, GR has not been able to explain the late-time acceleration of our universe: the phenomenon of dark energy. The late-time acceleration of the universe occurs at Hubble scales, a length scale much larger than the size of a gravitational wave emitter. Despite these shortcomings, GR and the Standard Model of particle physics have been two bedrocks of theoretical physics. Hence, to incorporate the phenomenon of dark energy, Einstein's GR needs to be adjusted so that, within reasonable limits, the adjusted theory yields the GR theory. There are two ways one can modify GR: adding higher curvature terms and adding extra gravitational degrees of freedom (DOF) \cite{stelle1,Alexeev:1996vs,Lehebel:2018zga,Volkov:2016ehx,Kunz:2006ca}. The simplest way to add extra DOF is by adding a scalar field along with the massless spin-2 graviton field \cite{Damour:1992we,Horbatsch:2015bua,Schon:2021pcv,Rainer:1996gw,DeFelice:2011bh}. The addition of a scalar field is not ad-hoc but has several phenomenological motivations: The scalar field is considered one of the potential candidates of dark matter and can also explain the accelerating nature of our universe \cite{DeFelice:2011bh,Gsponer:2021obj}.
 Most importantly, adding an extra degree of freedom can have several effects on GW observables. The new degree of freedom can be observed as extra polarisation while studying the radiation emission during a scattering process. Extra polarisations induce rapid energy loss and can be measured in the propagation of GWs. In this paper, our main motivation is to find the scalar polarisation of the emission of radiation. Along with the observational motivations, the appearance scalar field also has a theoretical motivation. After the discovery of General relativity, several attempts for a  grand unification were made by Weyl and Kaluza \cite{weyl}. Kaluza's proposal, which was later known as the Kaluza-Klein (KK) theory, attempted to unify the Einstein-Maxwell theory by considering a five-dimensional spacetime with the fifth dimension compactified and behaving as a 4-vector. It plays the role of an electromagnetic potential. This idea led to the path towards formulating the Scalar-Tensor theory proposed by Jordan, where the scalar is coupled non-minimally to the tensor field. With the motivation to study the nature of the contribution coming from the scalar degree of freedom as an extra polarisation in radiation emission, one can explore novel amplitude techniques to find various GW observables. \par
{\color{black} Several classical approaches to investigate the binary inspirals and merger problem of black holes and neutron stars \cite{Blanchet:2013haa, Schafer:2018kuf,cite-key, Pati:2000vt} have surfaced in recent years. } 
The gravitational classical potential and gravitational wave radiation up to 3PN has been investigated using the traditional methods in \cite{Tagoshi:2000zg,Faye:2006gx,Blanchet:2006gy,Blanchet:2004ek,Damour:2000ni,Itoh:2003fy,Boetzel:2019nfw,Mishra:2013rna,Kumar:2023bdf}. Recent computations up to 4.5PN has been done in \cite{Fujita:2010xj,Faye:2014fra,Blanchet:2023sbv}. Later, the investigation has been continued to the finite size effects of the binary system, including spins and tidal deformations \cite{PhysRevD.12.329,Kidder:1992fr,Cho:2022syn,Steinhoff:2007mb}. Apart from General Relativity, those analysis are extensively done in modified gravity theories \cite{Zhang:2017srh,Bernard:2022noq,AbhishekChowdhuri:2022ora, Zhang:2018prg,Saffer:2018jmx,Lin:2018ken,Li:2022grj,Shiralilou:2021mfl}. In parallel with the traditional computations, a field-theoretic approach to the computation of classical potential and radiation in GR has emerged in recent years \cite{Goldberger:2004jt,Goldberger:2009qd,Kol:2007bc,Goldberger:2007hy,Porto:2016pyg,Foffa:2013qca,cite-key2,Levi:2018nxp}. Furthermore, the EFT tools has been extensively used in modified gravity theories \cite{Bhattacharyya:2023kbh,Diedrichs:2023foj,Huang:2018pbu,Bernard:2023eul}.\par 
In addition to the inspiral problem, gravitational waves can be generated by scattering events. The study of classical gravitational scattering in post-Newtonian and post-Minkowskian regimes have been studied in \cite{hyp,hyp1,hyp2,hyp3,hyp4,hyp5,hyp6,hyp7,hyp8,hyp9,hyp10,hyp11,hyp12,hyp13,hyp16,hyp14,Damour:2016gwp,Bini:2017wfr,Bini:2017xzy,Damour:2017zjx,Damour:2019lcq,Bini:2020flp,Bini:2020uiq,Damour:2020tta,Bini:2020rzn,Bini:2021gat,Bini:2022enm,Damour:2022ybd,Bini:2022wrq,Rettegno:2023ghr,Bini:2023fiz,Ceresole:2023wxg}. Inspired by the scattering amplitude computation in QFT and invited by Damour in \cite{Damour:2017zjx}, the scattering of black holes/Neutron stars has gained interest in recent years by the amplitude community. A worldline Effective Field theory based on Post-Minkowskian (PM) expansion (where one has resumed PN expansion for every order in $G_N$) in Newton's constant has been developed in \cite{Kalin:2020mvi} for conservative binary dynamics. Later, the investigations continued with finite-size effects, including spins and the tidal effects \cite{Bini:2020flp,Cheung:2020sdj,Kalin:2020lmz,Haddad:2020que}. The higher PM computations can be found in \cite{Kalin:2020fhe,Dlapa:2021npj,Dlapa:2021vgp,Kalin:2022hph,Jinno:2022sbr,Dlapa:2022lmu,Dlapa:2023hsl,Riva:2021vnj}. In very recent past, the gravitational two-body problem has been investigated using more direct QFT techniques especially focused on on-shell methods of scattering amplitude \cite{PhysRevD.7.2317,Holstein:2004dn,Neill:2013wsa,Bjerrum-Bohr:2013bxa,Luna:2017dtq,Bjerrum-Bohr:2018xdl,Kosower:2018adc,Cristofoli:2021vyo,DeAngelis:2023lvf,Brandhuber:2023hhl,Aoude:2023dui,Brandhuber:2023hhy,Georgoudis:2023eke,Herderschee:2023fxh} \footnote{This references are by no means exhausted. Interested readers are referred to the citations and references of \cite{Buonanno:2022pgc} for more details.}. Those intricate methods have been extensively used in computing the classical gravitational potential in 2PM and 3PM \cite{Bjerrum-Bohr:2018xdl,Cheung:2018wkq,Cristofoli:2019neg,Cheung:2020gyp,Bern:2019nnu,Bern:2019crd} \footnote{In the context of soft theorem, two-body gravitational waveform upto 3PM order has been discussed in \cite{ashoke,ashoke1,ashoke2,ashoke3,ashoke4,ashoke5,ashoke6}.}. The worldline EFT and scattering amplitude approaches give the same results, and the question of efficiency depends on the taste. The prime question arises: Why do these two quite different approaches give the same results? The gap between the two approaches has been bridged by Plefka et al. in \cite{Mogull:2020sak}. It has been shown that the Feynmann-Schwinger representation of a graviton-dressed scalar propagator serves as the key to the scattering amplitude of two massive scalars. They provide a precise link between the scattering amplitude and the operator expectation value in the so-called Worldline Quantum Field Theory (WQFT). The formalism is similar to the formalism of Worldline Effective Field Theory (WEFT), but the main difference is that in WQFT, the worldline degrees of freedom (specifically, the worldline fluctuations) are quantized. Using this method, the results from scattering amplitudes and PM-EFT have been investigated in 
 the non-spinning case first \cite{Jakobsen:2021smu} and then generalized into spinning cases \cite{Jakobsen:2021lvp,Jakobsen:2021zvh,Jakobsen:2023ndj,Jakobsen:2023hig}. Further, the implementation has been done in higher PM order, including the tidal effects also \cite{Jakobsen:2023pvx}. The WQFT formalism has also been used to investigate the phenomena of gravitational lensing \cite{Bastianelli:2021nbs}. Eventually, the classical double copy relation has been studied using WQFT in \cite{Shi:2021qsb, Diaz-Jaramillo:2021wtl,Comberiati:2022cpm}.\par
\textit{All in all, till now, most of the scattering events have been investigated in GR. We take the step to compute the observables in a non-GR theory, primarily focused on the massive Scalar-Tensor theory of gravity, where we have a non-trivial cubic and quartic scalar potential. Our bigger goal is to inspect, apart from the mass of the scalar, whether we can put constraints on the cubic ($\lambda_3$) and quartic coupling ($\lambda_4$) from gravitational wave observations.} To do this, we take the primary step forward and compute the two main observables in a scattering event: impulse and waveform. The impulse is directly connected to the scattering angle, from which one can get information about the bound state of the system through the connection with the EOB prescription \cite{Buonanno:1998gg}. Rather, the more interesting part comes from the waveform computation, which is the key to computing the gravitational wave phase. \textit{In the game of waveform computation from WQFT, the main technical challenge is to compute the integrals. At this point, we make explicit computations of massive waveform integrals and propose methods for handling such integrals analytically, which are, to the best of our knowledge, new results.}\par
The paper is organized as follows: In Section~(\ref{sec2}), we briefly review the WQFT formalism and summarize the main results of the formalism in \cite{Mogull:2020sak}. In Section~(\ref{sec3}), we mention the changes in the results of \cite{Mogull:2020sak} due to the presence of an extra massive scalar degree of freedom. Then, we derive all the worldline Feynman rules from the scalar and graviton. In Section~(\ref{sec1}), we use the Feynman rules derived earlier to compute the corrections to the impulse due to the scalar field. We compute all the Feynman diagrams involving the self-interaction of the scalar field and scalar-graviton interaction. In Section~(\ref{sec5}), we start computing the radiation integrands and then move to the computation of the time-domain waveform, assuming the scalar field is massless. In Section~(\ref{sec6}), we first discuss how the massive radiation integral gets complicated due to the presence of more involved phase factors. Next, we list all the massive radiation integrands. We propose a method to compute those integrals using the stationary phase method, which is a very good approximation in the limit $|x|\to \infty$. We computed all the massive integrals in the stationary phase approximation and discussed the underlying subtleties. In Appendix~(\ref{appnew}), we give a short derivation of the worldline effective action in the presence of extra scalar degrees of freedom. In Appendix~(\ref{sec7}), we discuss the connection between impulse and the scattering amplitude through the Eikonal phase. In Appendix~(\ref{app1}), (\ref{App2}), (\ref{appnew1}),  we give some more details about the integrals used in the main text. In Appendix~(\ref{app22}), we provide another method of approximating the worldline radiation integral using the large velocity approximation.\\\\\\
\textbf{\textit{Notations and Conventions}}:\\
\textbullet $\,\,$ Metric sign convention: ($+,-,-,-$).\\
\textbullet $\,\,$ All computations are done in the unit where, $(c,\hbar)=1$.\\
\textbullet $\,\,$ The impact parameter $b$ is purely spacelike, $b^2=b^\mu b_\mu=-|b|^2$.\\
\textbullet $\,\,$ Black hole velocity parametrization: $v_1=(\gamma,\gamma\beta,0,0)$,  $v_2=(1,0,0,0)$.\\
\textbullet $\,\,$ Planck mass: $m_p=\frac{1}{\sqrt{8\pi G_N}}$, where $G_{N}$ is Newton's constant.\\ 
\textbullet $\,\,$ \textcolor{black}{Scaled delta function: $\hat\delta^{(D)}(\cdots)\equiv (2\pi)^{D}\delta^{(D)}(\cdots)$.}\\
\textbullet$\,\,\,\,$\textcolor{black}{Incomplete beta function: $B_{z}(a,b):=\rmint_0^z dt\,t^{a-1}(1-t)^{b-1}$.}\\ 
\textbullet$ \,$ \textcolor{black}{$J_n$ and $K_n$ s are Bessel functions of the first kind and modified Bessel functions of the second kind, respectively.}\\
\textbullet$ \,$ \textcolor{black}{\textbf{MeigerG} function is defined as,\\
\begin{align}
    G_{p q}^{m n} \left(z,r\Big|
\begin{array}{c}
 a_1\text{...}.a_p \\
 b_1\text{...} b_q \\
\end{array}\right)
=\frac{r}{2 \pi  i}\rmint_{-i\infty}^{i\infty} \frac{   \Gamma  \left(b_1+r s\right) \Gamma\left(-a_n-r s+1\right) \Gamma  \left(-a_1-r s+1\right) \Gamma\left(b_m+r s\right)}{  \Gamma  \left(a_p+r s\right) \left(-b_q-r s+1\right) \Gamma  \left(a_{n+1}+r s\right) \Gamma  \left(-b_{m+1}-r s+1\right)}z^{-s}\,ds\,.\nonumber
\end{align}}
\section{A quick tour to worldline quantum field theory}\label{sec2}
In this section, we briefly summarize the results of the Worldline Quantum Field Theory (WQFT) approach to the scattering problem. In \cite{Mogull:2020sak}, precise correspondence has been derived between the scalar-graviton S-matrix element and the one-point functions of the worldline operators. In this string-inspired approach, one maps black hole scattering in worldline theory to field scattering in the field theory approach. Using the gravitationally dressed Green's functions, one can write S-matrix elements as the expectation value of operators present in the worldline theory.

Now, for spinless black holes in GR, the action can be written in the EFT framework as\footnote{To get the linearized action one should expand $\sqrt{-g}$ as,
\begin{align}
    \begin{split}
        \sqrt{-g}=1+\frac{1}{2m_p} h-\frac{1}{4m_p^2} h^{\mu\nu}h_{\mu\nu}+\frac{1}{8m_p^2}h^2,\,\, h=\text{Tr}( h_{\mu\nu}).\nonumber
    \end{split}
\end{align}},
$$S=S_{EH}+S_{gf}+\sum_i S_{pm}^i\,,$$
where $S_{EH}$ is the usual Einstein-Hilbert action with $S_{gf}$ is the gauge-fixing action, which in the weak field approximation is given by,

$$S_{gf}=\rmint d^D x\Big(\partial_\nu h^{\mu\nu}-\frac{1}{2}\partial^\mu h^{\nu}_{\nu}\Big)^2\,.$$

This imposes the de-Donder gauge condition. Now, for an extended object, the worldline action consists of the Wilson coefficients $c_v$ and $c_R$ as follows,
\begin{align}S_{pm}=-m\rmint d\tau +c_R\rmint d\tau R(x)+c_v\rmint d\tau R_{\mu\nu}(x)\dot{x}^\mu \dot{x}^\nu+\cdots\,.\end{align}
In our paper, we will drop the second and third terms above to remove the complicacy due to the finite-size effects. In principle, there will be an infinite tower of terms consisting of higher-order derivatives of the metric represented as dots.

Now, assuming a fixed background, we can write Green's function of the field theory as a two-point correlator : 
\begin{align}
G_i(x,x')=\mathcal{Z}_i^{-1}\rmint \mathcal{D}[\phi_i]\phi_i(x)\phi_i^{\dagger}(x')e^{i S_i},
\end{align}
where, $S_i=\rmint d^D x \sqrt{-g}(g^{\mu\nu}\partial_\mu\phi_i^{\dagger}\partial_\nu\phi_i-m^2\phi_i^{\dagger}\phi_i-\zeta R\phi_i^{\dagger}\phi_i)$.\par
 To this end, we have to integrate out certain field degrees of freedom with respect to a relevant scale of the problem. This will give rise to the one-loop effective action, which can be represented by a worldline path integral \cite{Strassler:1992zr,Feal:2022iyn,Ahmadiniaz:2022yam}. Then, from this path integral, we can identify the point particle action in the background of $h_{\mu\nu}$.
Now for our interest, we evaluate the S-matrix element of two scalars with or without a final state graviton in certain \textit{classical limit} as,
\begin{align}
&\langle \Omega|T\{h_{\mu\nu}\phi_1(x_1)\phi_1^{\dagger}(x_1')\phi_2(x_2)\phi_2^{\dagger}(x_2')\}|\Omega\rangle\\&
=\mathcal{Z}_i^{-1}\rmint \mathcal{D}[h_{\mu\nu},\phi_i]h_{\mu\nu}(x)\prod_{i=1}^2\phi_i(x_i)\phi_i^{\dagger}(x_i')e^{i S'}\,.
\end{align}
Fourier transforming this relation and using LSZ reduction in the $\hbar \rightarrow 0 $ limit we get \cite{Mogull:2020sak},
\begin{align}&\langle \phi_1\phi_2(+h)|S|\phi_1\phi_2\rangle\\&
=\mathcal{Z}^{-1}\rmint d^D[x_i,x_i',x]e^{ip\cdot x-ip_i'\cdot x_i'-ik\cdot x}\times \rmint [\mathcal{D}h_{\alpha\beta}]\,\epsilon^{\mu\nu}(k)h_{\mu\nu}(x)\prod_{i=1}^2 G_i(x_i,x_i')e^{i S_{EH}+S_{gf}}\bigg{|}_{\textrm{Amputated,connected}}\,.
\end{align}
The study of the S-matrices consists of applying the LSZ reduction. Cutting the propagators on external legs, we convert correlators into S-matrices and send the external legs to infinity, where they interact weakly. We put the scalar legs on-shell, and the integration over the finite time domain extends $\tau \in (-\infty, \infty)$.
The key relation connecting the QFT form factor to the WQFT correlator is given by \cite{Mogull:2020sak},
\hfsetfillcolor{white!10}
\hfsetbordercolor{black}
\begin{align}
\begin{split}
\tikzmarkin[disable rounded corners=false]{g1}(0.1,-0.6)(-0.1,0.8) \frac{\Xi(b,v;\{\epsilon^{l},k_{l}\})}{\Xi_0}=\delta\Bigg(\sum_{l=1}^N k_l\cdot v\Bigg) \exp\Big({\sum_{l=1}^N k_l\cdot b} \Big)\mathcal{F}(p,p'|\{\epsilon^{l},k_{l}\})
\tikzmarkend{g1}
\end{split}
\end{align}

where,
\begin{align}
    \Xi(b,v;\{\epsilon^{l},k_{l}\})=\rmint \mathcal{D}[x]\rmint \mathcal{D}[a,b,c]\exp[-i\rmint d\sigma \Big[\frac{1}{4}g_{\mu\nu}(\dot{x}^\mu\dot{x}^\nu+a^\mu a^\nu+b^\mu c^\nu)\Big].
\end{align}
Having the gravitationally dressed propagator in momentum space, we put the external scalar legs on-shell to perform the LSZ reduction. Effectively, it can be represented as,
\\
\\
\begin{align}
\begin{split}
\hspace{-2cm}\mathcal{F}(p,p'|\{\epsilon^{l},k_{l}\})=\langle p'|\prod_{i=1}^N\epsilon_i\cdot h(k_i)|p\rangle= \begin{minipage}[h]{0.05\linewidth}
	\vspace{-1 cm}\scalebox{0.2}{\includegraphics{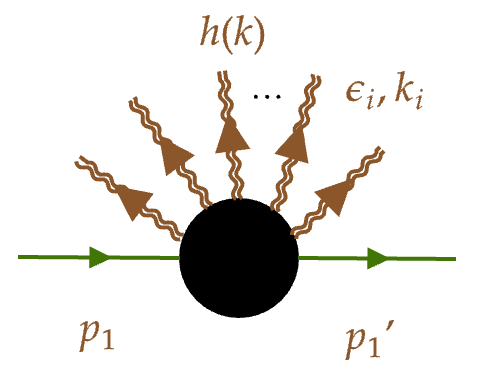}}\end{minipage}\,.
\end{split}\end{align}
For details, we refer the reader to \cite{Mogull:2020sak}.
In the following section, we will discuss how to derive the Feynman rules for the scattering events of two black holes (including extra scalar modes), which are considered to be point-like objects and, hence, spinless.

\section{Derivation of the Feynman rules}\label{sec3}
One of the key ingredients to compute the observables in a scattering problem is the partition function. 
We compute the partition function for a binary system in the Scalar-Tensor theory of gravity with a generic scalar potential. The gravitational action is given by,
\begin{align}
    \begin{split}
        S_{g}=\rmint d^4x \sqrt{- g}\Big(-\frac{m_p^2}{2} R+\frac{1}{2} {g}^{\alpha\beta}\partial_{\alpha}{\varphi}\partial_{\beta}{\varphi}-\frac{1}{2}m^2{\varphi}^2-\frac{\lambda_3}{\textcolor{black}{3!}} m_p\, \varphi^3-\frac{\lambda_4}{\textcolor{black}{4!}}{\varphi}^4\Big)\in S_{EH}+S_{scalar}.\label{3.1m}
    \end{split}
\end{align}
In this theory, the mass of the point particle, which is the avatar of the scalar field $\phi_i$, is not constant but rather depends on the extra scalar degree of freedom $\varphi$ \cite{eardley}. Hence, the matter action has the following form,
\begin{align}
    \begin{split}
        S_m=\sum_{i=1}^2\rmint d^4 x \sqrt{-g}(g^{\mu\nu}\partial_\mu\phi_i^{\dagger}\partial_\nu\phi_i-m_i(\varphi)^2\phi_i^{\dagger}\phi_i-\zeta R\phi_i^{\dagger}\phi_i)\,.
    \end{split}
\end{align}
Therefore, the WQFT partition function takes the form \footnote{A short derivation of the worldline effective action in presence of extra scalar degrees of freedom appearing in the partition function is given in Appendix~(\ref{appnew}).} ,
\begin{align}
    \begin{split}
        {Z}_{\textrm{WQFT}}=\mathcal{N}\times \rmint \mathcal{D}[h_{\mu\nu},\,{\varphi}]\rmint \mathcal{D}[x_i,a_i,b_i,c_i]e^{iS_{g}}\exp\Big\{-i\sum_{i=1}^{2}\rmint d\tau_{i}\frac{m_i(\varphi)}{2}g_{\mu\nu}(\dot x_{i}^{\mu}\dot x_{i}^{\nu}+a_{i}^{\mu}a_{i}^{\nu}+b_{i}^{\mu}c_{i}^{\nu})\Big\}\label{3.2}
    \end{split}
\end{align}
where, $a_i,b_i$ s are Lee-Yang Ghost and can be ignored in the classical limit. Therefore, the $n$-body partition function (non-spinning) in WQFT can be written as,
\begin{align}
    \begin{split}
        Z_{\textrm{WQFT}}^{(n)}=\mathcal{N}\rmint \mathcal{D}h_{\mu\nu}\mathcal{D}\varphi\,e^{iS_{EH}+iS_{scalar}}\Big(\prod_{i=1}^{n}\mathcal{D}x_{i}\,e^{iS_{pm}^i}\Big)\,.\label{3.3}
    \end{split}
\end{align}
It has been noticed that the partition function is related to the Eikonal phase  $\chi$ by the following exponentiation \cite{Amati:1990xe, Mogull:2020sak}.
\begin{align}
    \begin{split}
Z_{\textrm{WQFT}}:=e^{i\chi}\,.
    \end{split}
\end{align}
Now $\chi$ can be calculated by computing the connected Feynman diagrams. In order to compute the path integral in \eqref{3.3} one could start by demanding that a gravitational field is a fluctuation over Minkowski space, i.e., one can decompose the metric $g_{\mu\nu}$ as,
\begin{align}
    \begin{split}
g_{\mu\nu}=\eta_{\mu\nu}+\frac{h_{\mu\nu}}{m_p}.
    \end{split}
\end{align}
and, eventually, the worldline degree of freedom can be expanded as,
\begin{align}
    \begin{split}
        x^{\mu}(\tau)=b^{\mu}+v^{\mu}\tau+z^{\mu}(\tau)
    \end{split}
\end{align}
where, $z^{\mu}(\tau)$ is the worldline fluctuation about the straightline geodesic. Once we have the well-defined partition function, one could compute the classical observables as a correlation function in WQFT.
\begin{align}
    \begin{split}
        \mathcal{O}(b_i,v_i):=\langle\hat{\mathcal{O}}(\hat{x},\hat{h}_{\mu\nu},\hat\varphi)\rangle_{\textrm{WQFT}}=\frac{1}{Z^{(n)}_{\textrm{WQFT}}}\rmint \mathcal{D}h_{\mu\nu}\mathcal{D}\varphi\,e^{iS_{EH}+iS_{scalar}}\Big(\prod_{i=1}^{n}\mathcal{D}x_{i}e^{iS_{pm}^i}\Big)\mathcal{O}(x_i,h,\varphi)
    \end{split}
\end{align}
where the worldline action can be written in Polyakov form.
The point particle action has the following form,
\begin{align}
    \begin{split}
        S_{pm}=-\sum_{i=1}^{n}\rmint_{-\infty}^{\infty}d\tau_i\frac{m_i( \varphi)}{2}(g_{\mu\nu}\dot x^{\mu}_i\dot{x}^{\nu}_i+1).\label{3.4m}
    \end{split}
\end{align}
We intend to compute the two main observables, impulse and waveform, schematically defined as,
\begin{align}
    \begin{split}
      \langle \hat h_{\mu\nu}(k),\hat\varphi(k),\hat z^{\mu}(\omega)\rangle=Z_{\textrm{WQFT}}^{-1}\rmint\mathcal{D}h_{\mu\nu}\mathcal{D}\varphi\,e^{iS_{EH}+iS_{scalar}}\Big(\prod_{i=1}^{2}\mathcal{D}x_{i}e^{iS_{pm}^i}\Big)\{h_{\mu\nu}(k),\varphi(k),z^{\mu}(\omega)\}\,.
    \end{split}
\end{align}
Before going to the computations of the partition function, we first derive all the Feynman rules. In Scalar-Tensor theory, the violation of the equivalence principle automatically implies that the mass of the binaries depends on the extra gravitational polarisation ${\varphi}$. As there is a self-interaction term so one can expand $m_{a}(\varphi)$ around one of the chosen vacuums ${\varphi}_{0}$, which for our case is zero, as,
\begin{align}
    \begin{split}
        m_a({\varphi})=m_a(0)\Big \{1+s_a\frac{\varphi}{m_p}+g_a\frac{\varphi^2}{m_p^2}+\mathcal{O}(\varphi^3)\Big\}.
    \end{split}
\end{align}
In order to derive the Feynman rules it would be better to decompose the fields  ($\chi(x)\in(h_{\mu\nu},\varphi)$) in their Fourier mode assuming the black hole worldline has following the fluctuation due to the radiation reaction: $x_{(i)}^{\mu}=b_{(i)}^{\mu}+v_{(i)}^{\mu}\tau+z_{(i)}^{\mu}(\tau)$. where $z_{i}^{\mu}(\tau)$ is the worldline fluctuations about the straight line geodesic. Now, we expand the point particle action at different orders in $z$ to get the worldline vertices. In our theory we have gravitational field ($ h_{\mu\nu}$) and scalar field (${\varphi}$). One can decompose the fields in Fourier modes when it couples with the worldline, as,
\begin{align}
    \begin{split}
        &  \chi(x_i)=\rmint_{k}e^{ik\cdot(b_i+v_i\tau+z_i)} \chi(-k)\,.
    \end{split}
\end{align}
Now, expanding the exponential at different orders in $z$ and is given by,
\begin{align}
    \begin{split}
         \chi (x_i)=\rmint_{k}e^{ik\cdot(b_i+v_i \tau)}\prod_{n=0}^{\infty}\frac{i^{n}}{n!}(k\cdot z)^{n}  \chi(-k)\,.\label{3.8m}
    \end{split}
\end{align}
Again, we decompose the worldline fluctuations as,
\begin{align}
    \begin{split}
        z^{\mu}(\tau)=\rmint_{\omega}e^{i\omega \tau}z^{\mu}(-\omega)\,.
    \end{split}\end{align}
From the quadratic part of the Einstein-Hilbert action, one can identify the graviton propagator as,
\begin{align}
   \langle  h_{\mu\nu}(x_1) h_{\rho\sigma}(x_2)\rangle\equiv \begin{minipage}[h]{0.12\linewidth}
	\vspace{4pt}
	\scalebox{1.8}{\includegraphics[width=\linewidth]{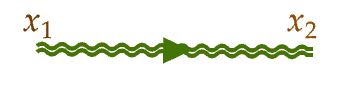}}
\end{minipage}\hspace{1.2 cm}=iP_{\mu\nu;\rho\sigma}\rmint_{k}\frac{e^{-ik\cdot(x_1-x_2)}}{k^2+i\epsilon}
\end{align}
where,
\begin{align}
    \begin{split} \label{tensor}
        P_{\mu\nu,\rho\sigma}=\textcolor{black}{\frac{1}{2}}\Big(\eta_{\mu\rho}\eta_{\nu\sigma}+\eta_{\mu\sigma}\eta_{\nu\rho}-\eta_{\mu\nu}\eta_{\rho\sigma}\Big).
    \end{split}
\end{align}
and the propagator for the scalar degrees of freedom has the following form,
\begin{align}
    \begin{split}
        \langle \varphi(x_1)\varphi(x_2)\rangle\equiv \begin{minipage}[h]{0.12\linewidth}
	\vspace{4pt}
	\scalebox{1.7}{\includegraphics[width=\linewidth]{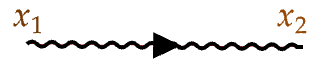}}
   \end{minipage}\hspace{1.3 cm}= i\rmint_{k}\frac{e^{-ik\cdot(x_1-x_2)}}{k^2-m^2+i\epsilon} \,.
    \end{split}
\end{align}
Apart from the field degrees of freedom, we have another dynamical degree of freedom: worldline fluctuations $z^{\mu}(\omega)$. One can identify the propagator by inspecting the quadratic part of the worldline action (\ref{3.4m}).
\begin{align}
    \begin{split}
        S_{\textrm{pm}}=-m(\varphi_{0})\rmint d\tau \Big[1+\eta_{\mu\nu}v^{\mu}\dot z^{\nu}+\frac{1}{2}\eta_{\mu\nu}\dot z^{\mu}\dot z^{\nu}\Big]\label{3.12m}
    \end{split}
\end{align}
The propagator of $z^{\mu}$ can be identified from the quadratic part of (\ref{3.12m}).
\begin{align}
    \begin{split}
      \Big  \langle z^{\mu}(\tau_1)z^{\nu}(\tau_2)\Big\rangle\equiv \begin{minipage}[h]{0.12\linewidth}
	\vspace{4pt}
	\scalebox{1.5}{\includegraphics[width=\linewidth]{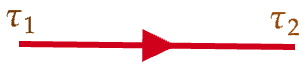}}
   \end{minipage}\hspace{1 cm}=-i\frac{\eta^{\mu\nu}}{m}\rmint_{\omega}\frac{e^{-i \omega (\tau_1-\tau_2)}}{(\omega\pm i\epsilon)^2}\,.
    \end{split}
\end{align}
Now, we are in a position to derive the worldline Feynman rules. The key ingredient is the worldline action $S_{\textrm{pm}}$ in (\ref{3.4m}).
\begin{align}
    \begin{split}
        S_{\textrm{pm}}\Big|_{\textrm{int.}}\in \rmint d\tau \Big[\Big(-\frac{m_a}{2}-\frac{m_a s_a}{2m_p} \varphi-\frac{m_a g_a}{2m_p^2} \varphi^2\Big)(\eta_{\mu\nu}+\frac{1}{m_p}h_{\mu\nu})(v^{\mu}+\dot z^{\mu})(v^{\nu}+\dot z^{\nu})\Big]\,.
    \end{split}
\end{align}
We can rewrite (\ref{3.8m}) by inserting the Fourier mode of the worldline fluctuation in the following way.
\begin{align}
    \begin{split}
         \chi=\sum_{m=0}^{\infty}\frac{i^m}{m!}\rmint_{k,\omega_1,...,\omega_m}e^{ik\cdot b}e^{i(k\cdot v+\sum_{i=1}^{m} \omega_{i})\tau}\prod_{i=1}^{m}[k\cdot z(-\omega_i)] \chi(-k).\label{3.15m}
    \end{split}
\end{align}
The expansion in \eqref{3.15m} helps to derive the Feynman rules upto $\mathcal{O}(z^{j})$. Now, we list down the Feynman rules coming from the scalar-worldline interaction:
\vspace{-0.4 cm}
\subsection*{\underline{Vertex for scalar field:}}
    \textbullet $\,\,$ $\mathcal{O}{(z^0)}:- i\frac{m_a s_a}{2m_p} \,e^{ik\cdot b_a}\,\hat\delta(k\cdot v)\rightarrow  \begin{minipage}[h]{0.12\linewidth}
	\vspace{4pt}
	\scalebox{1.5}{\includegraphics[width=\linewidth]{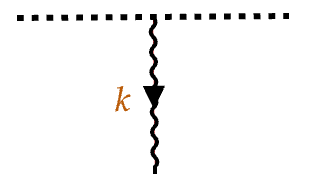}}
\end{minipage}$\\
 \textbullet $\,\,$ $\mathcal{O}(z):\frac{m_a s_a}{2m_p}\,e^{ik\cdot b}\,\hat{\delta}(k\cdot v+\omega)(2\omega v_{\rho}+k_{\rho})\rightarrow \begin{minipage}[h]{0.12\linewidth}
	\vspace{4pt}
	\scalebox{1.5}{\includegraphics[width=\linewidth]{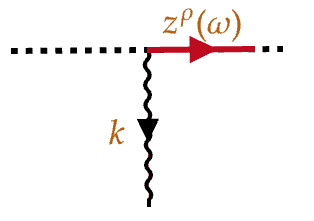}}
\end{minipage}$\\
 \textbullet $\,\,$ $ \mathcal{O}(z): -\frac{m_a s_a}{2m_p^2}e^{i(k_1+k_2)\cdot b_a}\hat\delta(k_1\cdot v_1+k_2\cdot v_1+\omega)\{(k_{1\rho}+k_{2\rho})v^{\mu}v^{\nu}+2\omega v^{(\mu}\delta_{\rho}^{\nu)}\}\rightarrow
\begin{minipage}[h]{0.12\linewidth}
	\vspace{4pt}
	\scalebox{1.8}{\includegraphics[width=\linewidth]{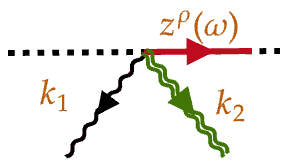}}
\end{minipage}$\\
 \textbullet $\,\,$ $ \mathcal{O}(z): -\frac{m_a g_a}{2m_p^2}e^{i(k_1+k_2)\cdot b_a}\hat\delta(k_1\cdot v_1+k_2\cdot v_1+\omega)\{(k_{1\rho}+k_{2\rho})v^{\mu}v^{\nu}+2\omega v_\rho\}\rightarrow
\begin{minipage}[h]{0.12\linewidth}
	\vspace{4pt}
	\scalebox{1.8}{\includegraphics[width=\linewidth]{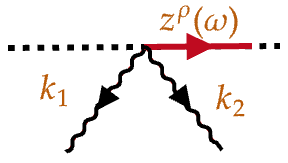}}
\end{minipage}$\\
   \textbullet $\,\,$ $\mathcal{O}{(z^2)}:i\frac{m_a s_a}{2m_p} \,e^{ik\cdot b_a}\,\hat\delta(k\cdot v+\omega_1+\omega_2)  (\frac{1}{2} k_{\rho_1} k_{\rho_2} +\omega_1 k_{\rho_2}v_{\rho_1}+\omega_2 k_{\rho_1}v_{\rho_2}+\omega_1\omega_2 \eta_{\rho_1\rho_2})\rightarrow  \begin{minipage}[h]{0.12\linewidth}
	\vspace{4pt}
	\scalebox{1.6}{\includegraphics[width=\linewidth]{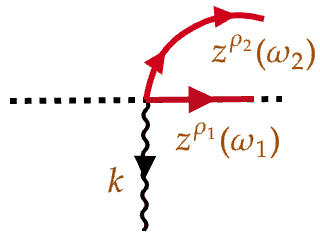}}
\end{minipage}$
\subsection*{\underline{Vertex for gravitational field:}}
     \textbullet $\,\,$ $\mathcal{O}{(z^0)}:- i\frac{m_a }{2 m_p} \,e^{ik\cdot b_a}\,\hat\delta(k\cdot v)v_a^\mu v_a^\nu \rightarrow  \begin{minipage}[h]{0.12\linewidth}
	\vspace{4pt}
	\scalebox{1.5}{\includegraphics[width=\linewidth]{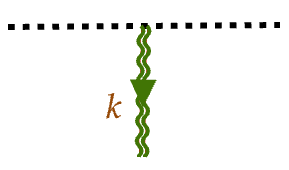}}
\end{minipage}$\\
 \textbullet $\,\,$ $\mathcal{O}(z^1):\frac{m_a }{2 m_p}\,e^{ik\cdot b}\,\hat{\delta}(k\cdot v+\omega)(2\omega v^{(\mu}\delta^{\nu)}_\rho+v^\mu v^\nu k_{\rho})\rightarrow \begin{minipage}[h]{0.12\linewidth}
	\vspace{4pt}
	\scalebox{1.5}{\includegraphics[width=\linewidth]{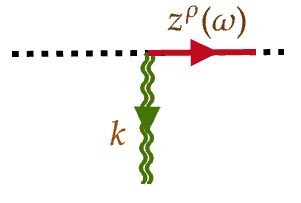}}
\end{minipage}$\\
 \textbullet $\,\,$ $\mathcal{O}{(z^2)}:i\frac{m_a }{m_p} \,e^{ik\cdot b_a}\,\hat\delta(k\cdot v+\omega_1+\omega_2) (\frac{1}{2} k_{\rho_1} k_{\rho_2} v^\mu v^\nu +\omega_1 k_{\rho_2}v^{(\mu}\delta^{\nu)}_{\rho_1}+\omega_2 k_{\rho_1}v^{(\mu}\delta^{\nu)}_{\rho_2}+\omega_1\omega_2 \delta^{(\mu}_{\rho_1}\delta^{\nu)}_{\rho_2})\rightarrow  \begin{minipage}[h]{0.12\linewidth}
	\vspace{4pt}
	\scalebox{1.5}{\includegraphics[width=\linewidth]{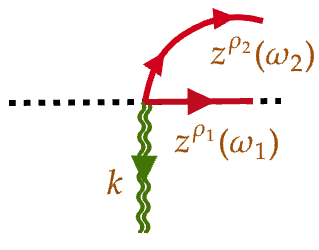}}
\end{minipage}$
\section{Computation of impulse}\label{sec1} 
In this section, we will compute the impulse in a purely conservative setting. We mainly focus on the corrections to the impulse coming from the purely scalar degree of freedom or the interaction between scalar and graviton.
The impulse is defined as,
\begin{align}
    \begin{split}
        \Delta p^{\mu}_{i}=m_i \rmint_{-\infty}^{\infty}d\tau_{i}\Big\langle\frac{d^2 z_{i}^{\mu}}{d\tau_{i}^2}\Big\rangle_{\text{WQFT}}=-m_i \omega^2\langle z_{i}^{\mu}(\omega)\rangle|_{\text{WQFT}}\Big |_{\omega=0}.\label{4.1w}
    \end{split}
\end{align}
Now we will compute (\ref{4.1w}) order by order in Newton's constant $G_N$ upto 2PM order. 
\subsection{1PM contribution to impulse}
\textbullet $\,\,$ The simplest diagram at  1 PM order consists only of a scalar field that has the following form, 
\begin{align}
    \begin{split}
        [\Delta p_1^{\mu}]_{(a)}=\begin{minipage}[h]{0.12\linewidth}
	\vspace{4pt}
	\scalebox{1.4}{\includegraphics[width=\linewidth]{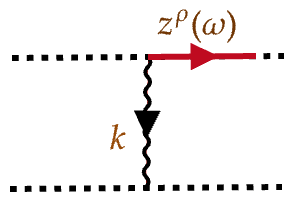}}
\end{minipage}
\hspace{1 cm}&=-i\frac{m_1s_1m_2s_2}{4m_p^2}\rmint_{k_1,k_2}e^{ik_1\cdot b_1+ik_2\cdot b_2}\frac{\hat\delta(k_2\cdot v_2)\hat\delta(\omega-k_1\cdot v_1)}{k_1^2-m^2}\hat\delta^{(4)}(k_1+k_2)\\&  \hspace{9 cm}\times(k_1^{\mu}+2\omega v_1^\mu)\Big|_{\omega=0}\,,\\ &
=\frac{\partial}{\partial b_{1\mu}}\underbrace{\Big[-\frac{m_1s_1m_2s_2}{4m_p^2}\rmint_{k}e^{ik\cdot(b_1-b_2)}\frac{\delta(k\cdot v_1)\hat\delta(k\cdot v_2)}{k^2-m^2}\Big]}_{\chi}\,.
    \end{split}
\end{align}

To do this integral, we need to choose the coordinates suitably. We study the dynamics from the frame of the second black hole. Hence in the mentioned parametrization of velocities
satisfying $v_1\cdot v_2=\gamma$.
Hence \footnote{Note that $(2\pi)$ is associated with each one-dimensional delta function and $(2\pi)^4$ factor with a four-dimensional delta function. Now, for each of the momentum integrals, there is a $\frac{1}{(2\pi)^4}$ factor associated with the integration measure. One has to take into account all of these factors. Furthermore, there will be a $(2\pi)$ factor whenever we get a ${\bf K_0}$ or ${\bf J_0}$  and $(2\pi)^2$ factor whenever we get a $\boldsymbol{\arctan}$ after doing an integral. One needs to consider these to get the correct factor of $\pi$ at the end. In all the subsequent integrals, we have taken this into account. },
\begin{align}
    \begin{split}
    \chi&=-\frac{m_1m_2s_1s_2}{4m_p^2}\rmint_{k}\frac{\hat{\delta}{(\gamma k^{(0)}-\gamma\beta k^{(1)})}\hat\delta(k^{0})}{k^2-m^2}e^{ik\cdot b}\,,\\ &
    =\frac{\pi ^2 m_1m_2s_1s_2}{m_p^2\sqrt{\gamma^2-1}}
   \rmint_{\tilde k}\frac{e^{-i\tilde k \cdot b}}{\tilde k^2+m^2}=\frac{m_1m_2s_1s_2}{8\pi m_p^2\sqrt{\gamma^2-1}}K_{0}(m|b|)\,.
    \end{split}
\end{align}
Therefore,
\hfsetfillcolor{gray!10}
\hfsetbordercolor{black!150}
\begin{align}
\begin{split} \label{e1:barwq2}
\tikzmarkin[disable rounded corners=false]{e}(0.1,-0.55)(-0.1,0.8)   [\Delta p_1^{\mu}]_{(a)}&=-\frac{m_1m_2s_1s_2}{8\pi m_p^2\sqrt{\gamma^2-1}} \frac{b^\mu}{|b|}m K_1(m|b|)\,.
\tikzmarkend{e}
\end{split}
\end{align} 
In the massless limit the impulse took the form,
\begin{align}
    \begin{split}
          [\Delta p_1^{\mu}]_{(a)}\Big|_{m\to 0}=-\frac{m_1m_2s_1s_2}{8\pi m_p^2\sqrt{\gamma^2-1}} \frac{b^\mu}{|b|^2}.
    \end{split}
\end{align}
\vspace{0.5cm}
\textbullet $\,\,$ One more diagram comes from the self-interaction vertex, which contributes at 1 PM order.
\begin{align}
    \begin{split}
        \mathcal{O}(z,\lambda_2):-\frac{\lambda_3}{\textcolor{black}{3!}} m_p\rmint d^4x\,\varphi(x)^3.
    \end{split}
\end{align}
Then\footnote{\textcolor{black}{Again note that the combinatorial factor associated with this diagram is 3!. We have multiplied it by that. For the subsequent diagrams, we will also multiply by the suitable combinatorial factors from the beginning.} },
\begin{align}
    \begin{split}
        [\Delta p_1^{\mu}]_{(b)}=\begin{minipage}[h]{0.12\linewidth}
	\scalebox{1.5}{\includegraphics[width=\linewidth]{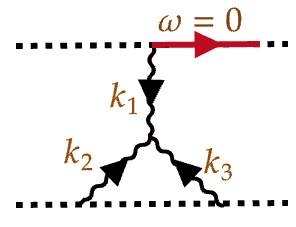}}
\end{minipage}&=-i\lambda_3\frac{m_1s_1}{2}\times \Big(\frac{m_2s_2}{2m_p}\Big)^2\rmint_{k_i}\hat\delta^{(4)}\Big(\sum_ik_i\Big)\frac{k_1^\mu\hat\delta(k_1\cdot v_1)\hat\delta(k_2\cdot v_2)\hat\delta(k_3\cdot v_2)}{\prod_{i=1}^3(k_i^2-m^2)}e^{ik_1\cdot b_1}e^{i(k_2+k_3)\cdot b_2}\,,\\ &
=-i\lambda_3\frac{m_1s_1}{2}\times \Big(\frac{m_2s_2}{2m_p}\Big)^2\rmint_{k_{1,2}}\frac{k_1^\mu\hat\delta(k_1\cdot v_1)\hat\delta(k_2\cdot v_2)\hat\delta(k_1\cdot v_2)}{(k_1^2-m^2)(k_2^2-m^2)[(k_1+k_2)^2-m^2]}e^{ik_1\cdot (b_1-b_2)}\,.
    \end{split}
\end{align}
Now using the proper velocity parametrization and the integral results derived in \cite{Bhattacharyya:2023kbh}, we get,
\begin{align}
    \begin{split}
         [\Delta p_1^{\mu}]_{(b)}&=-i\lambda_3\frac{m_1s_1}{2(2\pi)^3}\times \Big(\frac{m_2s_2}{2m_p}\Big)^2\frac{1}{\gamma\beta}\rmint d^2 l_1\frac{l_1^\mu e^{il_1\cdot b}}{(\vec l_1^2+m^2)|\vec l_1|}\arctan\Big(\frac{|\vec l_1|}{2m}\Big)\,,\\ &
        =\lambda_3\frac{m_1s_1}{2(2\pi)^2}\times \Big(\frac{m_2s_2}{2m_p}\Big)^2\frac{1}{\gamma\beta}\frac{\partial}{\partial b_{1\mu}}\underbrace{\rmint_{0}^{\infty}dl_1\,\frac{J_{0}(|b|l_1)}{ l_1^2+m^2}\arctan\Big(\frac{l_1}{2m}\Big)}_{I_3(m,b)}\,,\\ &
        =\lambda_3\frac{m_1s_1}{2(2\pi)^2}\times \Big(\frac{m_2s_2}{2m_p}\Big)^2\frac{1}{\gamma\sqrt{\gamma^2-1}}\frac{b^\mu}{|b|}\frac{\partial I_1(m,|b|)}{\partial |b|}\,.
    \end{split}
\end{align}  
Finally, we get, 
\hfsetfillcolor{gray!10}
\hfsetbordercolor{black}
\begin{equation}\label{e:barwq2}
\tikzmarkin[disable rounded corners=false]{c}(0.3,-0.6)(-0.1,0.8)  [\Delta p_1^{\mu}]_{(b)}=\lambda_3\frac{m_1s_1}{8\pi^2}\times \Big(\frac{m_2s_2}{2m_p}\Big)^2\frac{1}{\gamma\sqrt{\gamma^2-1}}\frac{b^\mu}{|b|}\frac{\partial I_1(m,|b|)}{\partial |b|}\tikzmarkend{c}\,.
\end{equation}\\
\textcolor{black}{To the best of our knowledge, this integral does not have any closed-form expression. One can perform a numerical analysis to solve the integral. Moreover, the integral has a smooth massless limit.} In the massless limit the contribution gives,
\begin{align}
    \begin{split}
         [\Delta p_1^{\mu}]_{(b)}\Big|_{m\to 0}=-\lambda_3 \frac{\pi m_1s_1}{8\pi}\times \Big(\frac{m_2s_2}{4m_p}\Big)^2\frac{1}{\gamma\sqrt{\gamma^2-1}}\frac{b^\mu}{|b|}\,.
    \end{split}
\end{align}
Then, the total impulse (due to the scalar field) at 1PM order is the sum of (\ref{e1:barwq2}) and (\ref{e:barwq2})  as well as the terms that come from interchanging the worldline one and two\,.
\begin{equation}
     \Delta p_1^{\mu}\Big|^{\textrm{1PM}, \textrm{Total}}_{\textrm{scalar}}=   [\Delta p_1^{\mu}]_{(a)}+ [\Delta p_1^{\mu}]_{(b)}+1\leftrightarrow 2\,.
\end{equation}
Finally, collecting all individual expressions we get, 
\hfsetfillcolor{yellow!6}
\hfsetbordercolor{brown}
\textcolor{black}{
\begin{align}
\begin{split}
\tikzmarkin[disable rounded corners=false]{ca}(0.8,-1)(-0.3,0.8)
    \Delta p_1^{\mu}\Big|^{\textrm{1PM}, \textrm{Total}}_{\textrm{scalar}}= &-\frac{m_1m_2s_1s_2}{8\pi m_p^2\sqrt{\gamma^2-1}} \frac{b^\mu}{|b|}m K_1(m|b|)+\lambda_3\frac{m_1s_1}{8\pi^2}\times \Big(\frac{m_2s_2}{2m_p}\Big)^2\frac{1}{\gamma\sqrt{\gamma^2-1}}\frac{b^\mu}{|b|}\frac{\partial I_1(m,|b|)}{\partial |b|}\,+1\leftrightarrow 2,\tikzmarkend{ca}
\end{split}
\end{align}
where, $$I_1(m,|b|)=\rmint_{0}^{\infty}dx\,\frac{J_{0}(|b|x)}{ x^2+m^2}\arctan\Big(\frac{x}{2m}\Big)\,.$$}\\ 
In the next subsection, we extend the impulse computation to 2PM order. \\
\subsection{2PM contribution to the impulse}
In this subsection, we compute the 2PM contribution to impulse, mainly focusing on the scalar field contribution.\\\\
\textbullet $\,\,$  At $\mathcal{O}(z)$ simplest diagram comes from the graviton-scalar interaction vertex:
\begin{align}
    \begin{split}
        \mathcal{O}(z): -\frac{1}{2}m^2\rmint d^4x \frac{h}{2m_p}\varphi^2\,.
    \end{split}
\end{align}
The corresponding contribution to the impulse is given by,
\begin{align}
    \begin{split}
        [\Delta p_1^{\mu}]_{(c)}& = \begin{minipage}[h]{0.12\linewidth}
	\scalebox{1.5}{\includegraphics[width=\linewidth]{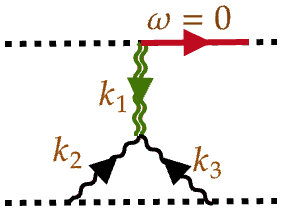}}
\end{minipage}\hspace{0.7 cm}\\ &
=-im^2\frac{m_1 m_2^2s_2^2}{8m_p^4}\rmint_{k_i}\hat\delta^{(4)}\Big(\sum_{i=1}^{3}k_i\Big)\frac{\hat\delta(k_1\cdot v_1)\hat\delta(k_2\cdot v_2)\hat\delta(k_3\cdot v_2)}{k_1^2(k_2^2-m^2)(k_3^2-m^2)}e^{ik_1\cdot b_1}e^{ik_2\cdot b_2}e^{ik_3\cdot b_2}k_1^\mu \underbrace{P_{\rho\rho,\sigma\delta}\,v_1^\sigma v_1^\delta}_{-1}\,,\\ &
=im^2\frac{m_1 m_2^2s_2^2}{8m_p^4}\rmint_{k_{1,2}}\frac{k_1^\mu\hat\delta(k_1\cdot v_1)\hat\delta(k_2\cdot v_2)\hat\delta(k_1\cdot v_2)}{k_1^2(k_2^2-m^2)[(k_1+k_2)^2-m^2]}e^{ik_1\cdot (b_1-b_2)}\,,\\ &
=m^2\frac{m_1m_2^2s_2^2}{32\pi^2 m_p^4}\frac{1}{\gamma\sqrt{\gamma^2-1}}\frac{b^\mu}{|b|}\partial_{|b|}\underbrace{\rmint _{0}^{\infty}dl\,\frac{J_{0}(bl)}{l^2}\arctan\Big(\frac{l}{2m}\Big)}_{I_2(m,|b|)}.\label{4.23m}
    \end{split}
\end{align}
Therefore, the contribution to the impulse from this diagram has the following form.
\hfsetfillcolor{gray!10}
\hfsetbordercolor{black}
\begin{equation}\label{e:barwq244}
\tikzmarkin[disable rounded corners=false]{+}(0.3,-0.6)(-0.1,0.8)  [\Delta p_1^{\mu}]_{(c)}=m^2\frac{m_1m_2^2s_2^2}{32 \pi^2 m_p^4}\frac{1}{\gamma\sqrt{\gamma^2-1}}\frac{b^\mu}{|b|}\partial_{|b|}{I_2(m,|b|)}\tikzmarkend{+}\,.
\end{equation}\\
Like (\ref{e:barwq2}), it also does not have any closed-form expression. Furthermore, it is evident from \eqref{4.23m} that it becomes identically zero in the massless limit as it is proportional to $m^2$.\\\\
\textbullet $\,\,$ Another contributing diagram may appear from the scalar-graviton interaction vertex where one scalar and one graviton line connect with one worldline, and the other scalar line connects with another worldline.
\begin{align}
    \begin{split}
         [\Delta p_1^{\mu}]_{(d)}& =
        \begin{minipage}[h]{0.12\linewidth}
	\scalebox{1.5}{\includegraphics[width=\linewidth]{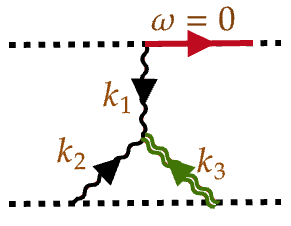}}
\end{minipage}\\ &
        =im^2 \frac{m_1 s_1 m_2^2 s_2}{8 m_p^4}\rmint_{k_{1,2}}\frac{k_1^\mu\hat\delta(k_1\cdot v_1)\hat\delta(k_1\cdot v_2)\hat\delta(k_2\cdot v_2)}{(k_1^2-m^2)(k_2^2-m^2)(k_1+k_2)^2}e^{ik_1\cdot b}\\ &
        =m^2 \frac{m_1 s_1 m_2^2 s_2}{32 \pi^2 m_p^4\gamma\sqrt{\gamma^2-1}}\frac{b^\mu}{|b|}\partial_{|b|} \underbrace{\rmint_{k_1}dk \frac{ J_{0}(|b|k_1)}{k_1^2+m^2}\arctan\Big(\frac{k_1}{m}\Big)}_{\Bar{I}_2(m,|b|)}
    \end{split}
\end{align}
Hence, the contribution to the impulse from this diagram has the following form:
\hfsetfillcolor{gray!10}
\hfsetbordercolor{black}
\begin{equation}\label{e:barwq248}
\tikzmarkin[disable rounded corners=false]{++}(0.3,-0.6)(-0.1,0.8)  [\Delta p_1^{\mu}]_{(d)}=m^2 \frac{m_1 s_1 m_2^2 s_2}{32 \pi^2 m_p^4\gamma\sqrt{\gamma^2-1}}\frac{b^\mu}{|b|}\partial_{|b|} \bar{I}_2(m,|b|)
\tikzmarkend{++}\,.
\end{equation}
\\
\textbullet $\,\,$  Now, we concentrate on the scalar interaction diagrams. We compute the contribution to the impulse at 2PM order coming from the scalar self-interaction vertex:
\begin{align}
    \begin{split}
        \mathcal{O}(z):-\frac{\lambda_4}{\textcolor{black}{4!}}\rmint d^4x\,\varphi^4(x)\,.
    \end{split}
\end{align}
The corresponding contribution to the impulse is shown below,
\begin{align}
    \begin{split}
          [\Delta p_1^{\mu}]_{(e)}&=\begin{minipage}[h]{0.12\linewidth}
	\scalebox{1.5}{\includegraphics[width=\linewidth]{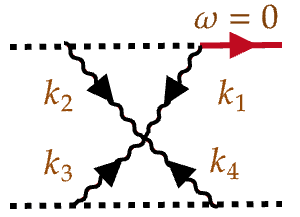}}
\end{minipage}\\ & =-i\lambda_4\Big(\frac{m_1s_1m_2s_2}{4m_p^2}\Big)^2\rmint_{\{k_i\}}\hat\delta^{(4)}\Big(\sum_{i=1}^{4}k_{i}\Big)\hat\delta(k_1\cdot v_1)\hat\delta(k_2\cdot v_1)\hat\delta(k_3\cdot v_2)\hat\delta(k_4\cdot v_2)\prod_{j=1}^{4}\frac{1}{k_j^2-m^2}\\ & 
\hspace{0.8cm}\times k_1^{\mu} \,e^{i(k_1+k_2)\cdot b_2}e^{i(k_3+k_4)\cdot b_2}\,,\\ &
=-i\lambda_4\Big(\frac{m_1s_1m_2s_2}{4m_p^2}\Big)^2\rmint_{k_i\ne k_4}\hat\delta(k_1\cdot v_1)\hat\delta(k_2\cdot v_1)\hat\delta(k_3\cdot v_2)\hat\delta(k_1\cdot v_2+k_2\cdot v_2)\\ &\hspace{0.8cm}\times\frac{k_1^\mu e^{i(k_1+k_2)\cdot (b_1-b_2)}}{\prod_{j=1}^{3}(k_j^2-m^2)[(k_1+k_2+k_3)^2-m^2]}\,,\\ &
\xrightarrow[]{k_1+k_2\rightarrow q}-i\lambda_4\Big(\frac{m_1s_1m_2s_2}{4m_p^2}\Big)^2\rmint_{q,k_1,k_3}\frac{k_1^\mu\hat\delta(k_1\cdot v_1)\hat\delta(q\cdot v_1)\hat\delta(k_3\cdot v_2)\hat\delta(q\cdot v_2)}{(k_1^2-m^2)(k_3^2-m^2)[(q-k_1)^2-m^2][(q+k_3)^2-m^2]}e^{iq\cdot b}\,.\\ &
   \label{4.26m} \end{split}
\end{align}
In \eqref{4.26m} one can see that the delta function constraints set $q$ and $k_3$ two and three-dimensional vectors, respectively and $k_1$ a three-dimensional vector with scaled variable: $
        \bar k_1^{(1)}=\frac{k_1^{(1)}}{\gamma}.
   $ 
Therefore, the integral in  (\ref{4.26m}) can be re-written as,
\begin{align}
\begin{split}
   &   [\Delta p_1^{\mu}]_{(e)}\\ &=-i\lambda_4\Big(\frac{m_1s_1m_2s_2}{4m_p^2}\Big)^2 \int e^{iq\cdot b}\,\frac{q^\mu}{q^2}\,\hat \delta(q\cdot v_1)\hat\delta(q\cdot v_2)\int_{k_3,k_1}\frac{\hat\delta(k_1\cdot v_1)\hat\delta(k_3\cdot v_2)\left((k_1^2-m^2)+q^2-((k_1-q)^2-m^2)\right)}{(k_1^2-m^2)(k_3^2-m^2)[(q-k_1)^2-m^2][(q+k_3)^2-m^2]}\,, \\ &
   =-i\lambda_4\Big(\frac{m_1s_1m_2s_2}{4m_p^2}\Big)^2\int e^{iq\cdot b} \frac{q^\mu}{q^2}\,\hat \delta(q\cdot v_1)\hat\delta(q\cdot v_2)\Bigg(\int_{k_1}\frac{\hat\delta(k_1\cdot v_1)}{(k_1-q)^2-m^2}\int_{k_3}\frac{\hat\delta(k_3\cdot v_2)}{(k_3^2-m^2)((k_3+q)^2-m^2)}\\ &+q^2\int_ {k_1}\frac{\hat\delta(k_1\cdot v_1)}{(k_1^2-m^2)((k_1-q)^2-m^2)}\int _{k_3}\frac{\hat\delta(k_3\cdot v_2)}{(k_3^2-m^2)((k_3+q)^2-m^2)}-\int_{k_1}\frac{\hat\delta(k_1\cdot v_1)}{k_1^2-m^2}\int_{k_3}\frac{\hat\delta(k_3\cdot v_2)}{(k_3^2-m^2)((k_3+q)^2-m^2)}\Bigg)\,,\\ &
   =-i\lambda_4\Big(\frac{m_1s_1m_2s_2}{16\pi m_p^2}\Big)^2\int e^{iq\cdot b} \frac{q^\mu}{-q^2}\,\hat \delta(q\cdot v_1)\hat\delta(q\cdot v_2)\arctan^2\left(\frac{\sqrt{-q^2}}{2m}\right)\,,\\ &
  =- \lambda_4\Big(\frac{m_1s_1m_2s_2}{16\pi m_p^2}\Big)^2\frac{b^\mu}{\sqrt{\gamma^2-1}|b|} \int_0^\infty dq\, J_{1}(q\,b)\arctan^2\left(\frac{q^2}{2m}\right)\,.
   \label{4.19f}
    \end{split}
\end{align}
\hfsetfillcolor{gray!10}
\hfsetbordercolor{black}
\begin{equation}\label{e11}
\tikzmarkin[disable rounded corners=false]{++}(0.3,-0.6)(-0.1,0.8)    [\Delta p_1^{\mu}]_{(e)}=- \lambda_4\Big(\frac{m_1s_1m_2s_2}{16\pi m_p^2}\Big)^2\frac{b^\mu}{\sqrt{\gamma^2-1}|b|} \int_0^\infty dq\, J_{1}(q\,b)\arctan^2\left(\frac{q^2}{2m}\right)
\tikzmarkend{++}\,.
\end{equation}
\\
The massless counterpart takes the following form,
\begin{align}
    \begin{split}
        [\Delta p_1^{\mu}]_{(e)}\Big|_{m\rightarrow 0}=- \lambda_4\Big(\frac{m_1s_1m_2s_2}{32m_p^2}\Big)^2\frac{b^\mu}{\sqrt{\gamma^2-1}|b|^2}\,.
    \end{split}
\end{align}
\\
\textbullet $\,\,$ Another diagram involving $\lambda_4 \varphi^4$ vertex will contribute to the impulse, where we will have three scalar lines connected with one worldline, and the other scalar line connects with another worldline.
\begin{align}
    \begin{split}
         [\Delta p_1^{\mu}]_{(f)}&=\begin{minipage}[h]{0.12\linewidth}
	\scalebox{1.5}{\includegraphics[width=\linewidth]{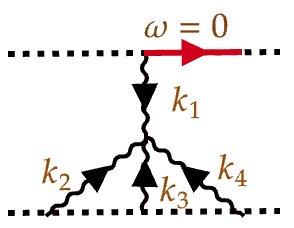}}
\end{minipage}\\ &=-i\lambda_4 \Big(\frac{m_2 s_2}{2m_p}\Big)^3\Big(\frac{m_1 s_1}{2m_p}\Big)\rmint_{\{k_i\}}\hat\delta^{(4)}\Big(\sum_{i=1}^{4}k_{i}\Big)\hat\delta(k_1\cdot v_1)\hat\delta(k_2\cdot v_2)\hat\delta(k_3\cdot v_2)\hat\delta(k_4\cdot v_2)\prod_{j=1}^{4}\frac{1}{k_j^2-m^2}\\ & 
\hspace{5 cm}\times k_1^{\mu} \,e^{ik_1\cdot b_1}e^{i(k_2+k_3+k_4)\cdot b_2}\,,\\ &
=-i\lambda_4 \Big(\frac{m_2 s_2}{2m_p}\Big)^3\Big(\frac{m_1 s_1}{2m_p}\Big)\rmint e^{ik_1\cdot b}k_1^\mu \,\frac{\hat\delta(k_1\cdot v_1)\hat\delta(k_1\cdot v_2)}{k_1^2-m^2}\rmint \frac{\hat\delta(q\cdot v_2)\hat\delta(k_3\cdot v_2)}{[(q-k_1)^2-m^2](k_3^2-m^2)[(q+k_3)^2-m^2]}\,,\\ &
=\frac{\lambda_4}{(2\pi)^2} \Big(\frac{m_2 s_2}{2m_p}\Big)^3\Big(\frac{m_1 s_1}{2 m_p}\Big)\frac{1}{\gamma\sqrt{\gamma^2-1}}\frac{b^\mu}{|b|}\partial_{|b|}\rmint  d^2 k_1\,\frac{e^{-i\vec k_1\cdot \vec b}}{\vec k_1^2+m^2}\,\hat{I}(|k_1|,m)\,.
\label{4.19c}
    \end{split}
\end{align}
We can further simplify \eqref{4.19c} using Schwinger parametrization.
\begin{align}
    \begin{split}
       \hat{I}(|k_1|,m)=\frac{1}{(2\pi)^4}\rmint d^3 q\frac{1}{|\vec q\,|} \rmint_{0}^{\infty}d\alpha\, e^{-\alpha (\vec q-\vec k_1)^2}e^{-\alpha m^2}\arctan\Big(\frac{|\vec q\,|}{2m}\Big)\,.
    \end{split}
\end{align}
Now in \eqref{4.19c} do the angular part of, $q$ integral first,
\begin{align}
    \begin{split}
         [\Delta p_1^{\mu}]_{(f)} &\sim \frac{1}{(2\pi)^3}\rmint d\alpha\rmint d^3q \frac{e^{-\alpha (q^2+m^2)}}{|q|}\arctan\Big(\frac{|\vec q\,|}{2m}\Big)\rmint_{0}^\infty dk_1\frac{k_1}{ k_1^2+m^2}J_{0}(k_1|b|)e^{-\alpha k_1^2+2\alpha \vec q\cdot \vec k_1}\,,\\ &
        =\frac{1}{(2\pi)^2}\rmint d\alpha \,\frac{e^{-\alpha m^2}}{\alpha}\rmint dk_1\,\frac{1}{ k_1^2+m^2}J_{0}(k_1|b|)e^{-\alpha k_1^2}\rmint dq\,\sinh(2q k_1 \alpha)\arctan\Big(\frac{q}{2m}\Big)e^{-\alpha q^2}\,,\\ &
        =\frac{1}{(2\pi)^2}\rmint_0^\infty dq \,dk_1\,\frac{J_{0}(k_1 b)}{k_1^2+m^2}\arctan\Big(\frac{q}{2m}\Big)\textrm{arctanh}\Big(\frac{2qk_1}{m^2+q^2+k_1^2}\Big)\,.\label{4.21cc}
    \end{split}
\end{align}
The $q$ integral can be done by taking the \textbf{logarithmic} representation of the \textbf{ArcTan} and \textbf{ArcTanh} in the following way,
\begin{align}
    \begin{split}
       & \rmint dq \arctan\Big(\frac{q}{2m}\Big)\,\textrm{arctanh}\Big(\frac{2q k_1}{q^2+k_1^2+m^2}\Big)\,,\\ &
       =\frac{i}{4}\rmint dq\, \log\Bigg[\frac{1-\frac{iq}{2m}}{1+\frac{iq}{2m}}\Bigg]\log\Big(1+\frac{4q k_1}{m^2+(q-k_1)^2}\Big)\,,\\ &
       =\frac{\pi}{2}\Big[2k_1-2m\arctan\Big(\frac{k_1}{m}\Big)-6m\arctan\Big(\frac{k_1}{3m}\Big)-k_1\log\Big(1+\frac{k_1^2}{m^2}\Big)-k_1\log\Big(1+\frac{k_1^2}{9m^2}\Big)-2k_1 \log(3m^2)\Big]\,.\label{4.25j}
    \end{split}
\end{align}
Therefore the impulse becomes,
\begin{align}
    \begin{split}
         [\Delta p_1^{\mu}]_{(f)}&\sim \frac{1}{32\pi^3}\rmint dk_1\frac{J_{0}(k_1b)}{k_1^2+m^2}\Big[2k_1-2m\arctan\Big(\frac{k_1}{m}\Big)-6m\arctan\Big(\frac{k_1}{3m}\Big)\\ &
        \hspace{5 cm}-k_1\log\Big(1+\frac{k_1^2}{m^2}\Big)-k_1\log\Big(1+\frac{k_1^2}{9m^2}\Big)-2k_1 \log(3m^2)\Big]\,,\\ &
        =\frac{1}{32\pi^3}\Big[2(1-\log3m^2)K_{0}(|b|m)-\rmint dk_1\frac{J_{0}(k_1 b)}{k_1^2+m^2}\Theta(k_1,m)\Big]
    \end{split}
\end{align}
where,
\begin{align}
    \begin{split}
        \Theta(k_1,m)=-2m\arctan\Big(\frac{k_1}{m}\Big)-6m\arctan\Big(\frac{k_1}{3m}\Big)-k_1 \log\Big[\Big(1+\frac{k_1^2}{m^2}\Big)\Big(1+\frac{k_1^2}{9m^2}\Big)\Big]\,.
    \end{split}
\end{align}
After inserting the prefactors, we get\\\\
%
\hfsetfillcolor{gray!10}
\hfsetbordercolor{black}
\begin{equation}\label{e:barwq246}
\tikzmarkin[disable rounded corners=false]{=}(18,-0.9)(-0.1,0.8)  [\Delta p_1^{\mu}]_{(f)}=\frac{\lambda_4}{32\pi^3} \Big(\frac{m_2 s_2}{2m_p}\Big)^3\Big(\frac{m_1 s_1}{2m_p }\Big)\frac{1}{\gamma\sqrt{\gamma^2-1}}\frac{b^\mu}{|b|}\partial_{|b|}\Big(2(1-\log(3m^2/\mu^2))K_{0}(|b|m)-\rmint_0^{\infty} dk_1\frac{J_{0}(k_1 |b|)}{k_1^2+m^2}\Theta(k_1,m)\Big)\tikzmarkend{=}\,
\end{equation}\\
where $\mu$ is the UV cutoff.\textcolor{black}{ As can be seen, the integral in \eqref{4.21cc} does not have any closed-form, but one can derive the massless limit}\footnote{As one can see, the impulse has a logarithmic behaviour w.r.t $|b|$. Hence, one should divide it by a UV cutoff $b_0$ to make it dimensionless. However, it does not contribute to the finite part of the impulse. The same thing can be said whenever some logarithmic terms appear. },
\begin{align}
    \begin{split}
      [\Delta p_1^{\mu}]_{(f)}\Big|_{m\rightarrow 0}   &=\frac{\lambda_4 }{32\pi^3}\Big(\frac{m_2 s_2}{2m_p}\Big)^3\Big( \frac{m_1 s_1}{2 m_p}\Big)\frac{1}{\gamma\sqrt{\gamma^2-1}}\frac{b^\mu}{|b|}\Big[-\frac{2}{|b|}+4\partial_{|b|}\rmint_0^\infty dk_1\frac{ \,J_{0}(k_1 |b|)}{k_1}\log(k_1)\Big]\,,\\ &
     =\frac{\lambda_4}{8\pi^3} \Big(\frac{m_2 s_2}{2m_p}\Big)^3\Big( \frac{m_1 s_1}{2 m_p}\Big)\frac{1}{\gamma\sqrt{\gamma^2-1}}\frac{b^\mu}{|b|^2}\Big[\log (|b|/b_0)+\gamma_E-\frac{1}{2}-\log 2\Big]\,.
    \end{split}
\end{align}\\\\
\textbullet $\,\,$ Now we will deal with the derivative interactions. The simplest scalar-gravitation derivative interaction, which contributes to the impulse at 2PM order:
\begin{align}
    \begin{split}
        \mathcal{O}(z):\rmint d^4x \frac{h^{\alpha\beta}}{m_p}\,\partial_{\alpha}\varphi\partial_{\beta}\varphi\,.
    \end{split}
\end{align}
\vspace{-0.53 cm}
The vertex factor has the following form,
\begin{align}
    \begin{split}
        \mathcal{V}_{\alpha\beta}(k_1,k_2,k_3)=  -\rmint_{\{k_{i}\}}  \hat\delta\Big(\sum_{i=1}^{3}k_{i}\Big)   (k_2)_{\alpha} (k_3)_{\beta}\,.
    \end{split}
\end{align}
The contribution to the impulse is given by\footnote{Time-symmetric Feynman propagators imply a elastic scattering of the black holes. In principle, one can also begin with the retarded graviton propagators to take care of radiative effects, but we will use the first one.},
\begin{align}
    \begin{split}
         [\Delta p_1^{\mu}]_{(g)} &=\begin{minipage}[h]{0.12\linewidth}
	\scalebox{1.5}{\includegraphics[width=\linewidth]{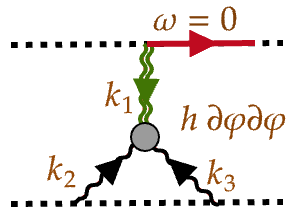}}
\end{minipage}\hspace{0.7 cm}\\ &
=i\Big(\frac{m_2s_2}{2m_p}\Big)^2\times \frac{m_1}{ m_p^2}\rmint_{k_i}\hat\delta^{(4)}\Big(\sum_{i}k_i\Big)k_2^\alpha k_3^\beta\,\frac{k_1^\mu P_{\sigma\delta;\alpha\beta }v_1^\sigma v_1^\delta\,\hat\delta(k_1\cdot v_1)\hat\delta(k_2\cdot v_2)\hat\delta(k_3\cdot v_2)}{k_1^2\prod_{i=2,3}(k_i^2-m^2)}e^{ik_1\cdot b_1}e^{i(k_2+k_3)\cdot b_2}\,,\\ &
=\Big(\frac{m_2s_2}{2m_p}\Big)^2\times \frac{m_1}{2m_p^2}\\ &
\hspace{2 cm}\times\frac{\partial}{\partial b_{1\mu}} \underbrace{\rmint_{k_{1,2}}\Big[2(k_2\cdot v_1)(k_1+k_2)\cdot v_1-\,k_2\cdot (k_1+k_2)\Big]\frac{\hat\delta(k_1\cdot v_1)\hat\delta(k_2\cdot v_2)\hat\delta(k_1\cdot v_2)}{k_1^2(k_2^2-m^2)[(k_1+k_2)^2-m^2]}e^{ik_1\cdot b}}_{I_{4}(m,|b|)}\,.\label{4.37} 
    \end{split}
\end{align}
The computation in \eqref{4.37} is an involved one and should be done carefully. We show below the details of the integration for individual terms.
\begin{align}
    \begin{split}
        I_4(m,|b|)\Big|_{(1)}&\equiv 2\rmint_{k_{1,2}}(k_2\cdot v_1)^2\frac{\hat\delta(k_1\cdot v_1) \hat\delta(k_2\cdot v_2)\hat\delta(k_1\cdot v_2)}{k_1^2(k_2^2-m^2)[(k_1+k_2)^2-m^2]}e^{i k_1\cdot b}\,,\\ &
         =(v_1)_{\mu}(v_1)_{\nu}\rmint_{k_{1,2}}k_2^{\mu}\,k_2^{\nu}\frac{\hat\delta(\gamma k_1^{(0)}-\gamma\beta k_1^{(1)})\hat\delta( k_2^0)\hat\delta(k_1^0)}{k_1^2(k_2^2-m^2)[(k_1+k_2)^2-m^2]}e^{i k_1 \cdot b}\,,\\ &
    =\textcolor{black}{\frac{1}{(2\pi)^5}}\frac{v_{1i}v_{1j}    }{\gamma\beta}\rmint d^2 k_1\frac{e^{i k_1 \cdot b}}{\vec k_1^2}
    \underbrace{\rmint d^3 k \frac{k_i k_j}{(\vec k^2+m^2)[(\vec k+\vec k_1)^2+m^2]}}_{\mathcal{C}_{ij}}\,.
    \end{split}
\end{align}
Using the Passarino-Veltman reduction one can deduce the form of $\mathcal{C}_{ij}$,
\begin{align}
    \begin{split}
\mathcal{C}_{ij}=&\underbrace{\Big[\textcolor{black}{(2\pi)^2}\frac{3}{8 |\vec k_1|}\arctan\Big(\frac{|\vec k_1|}{2m}\Big)+\textcolor{black}{(2\pi)^2}\frac{m^2}{2 |\vec k_1|^3}\arctan\Big(\frac{|\vec k_1|}{2m}\Big)-\textcolor{black}{\frac{\pi^2}{2}}\frac{m}{|\vec k_1|^2}\Big]}_{\Xi_1(|k_1|,m)}k_1^i k_1^j\\ &
       \underbrace{- \Big[|\vec k_1|\textcolor{black}{\frac{(2\pi)^2}{8}}\arctan\Big(\frac{|\vec k_1|}{2m}\Big)+\textcolor{black}{\frac{(2\pi)^2}{2}}\frac{m^2}{ |\vec k|}\arctan\Big(\frac{|\vec k_1|}{2m}\Big)+m\textcolor{black}{\frac{\pi^2}{2}}\Big]}_{\Xi_2 (|k_1|,m)}\delta^{ij}\,,\\ &
={\Xi}_1(|k_1|,m)k_1^i k_1^j+{\Xi}_{2}(|k_1|,m)\delta^{ij}\,.  \label{new11}
    \end{split}
\end{align}
Therefore,
\begin{align}
    \begin{split} \label{eq:4.35}
        I_{4}(m,|b|)\Big|_{(1)}=\frac{2(2\pi)}{\gamma\sqrt{\gamma^2-1}}\Bigg[-v_{1i}v_{1j}\partial_{b_i}\partial_{b_j}\rmint_{0}^{\infty}dl\,\frac{J_{0}(bl)}{l}\,{\Xi}_1\Big(l,m\Big)+(\gamma^2-1)\rmint_{0}^\infty dl\,\frac{J_{0}(bl)}{l}{\Xi}_2\Big(l,m\Big)\Bigg]\,,
    \end{split}
\end{align}
where ${\Xi}_1(l,m)$ and ${\Xi}_2(l,m)$ are defined in (\ref{new11}). 
Other parts of the integrals $I_4(m,b)$ can be computed as follows,
\begin{align}
    \begin{split}
        I_{4}(m,|b|)\Big|_{(2)}\equiv 2\rmint_{k_{1,2}}(k_2\cdot v_1)(k_1\cdot v_1)\hat\delta(k_1\cdot v_1)\cdots \rightarrow 0.
    \end{split}
\end{align}
and,
\begin{align}
    \begin{split}
        I_{4}(m,|b|)\Big|_{(3)}&\equiv- \rmint_{k_{1,2}}k_1\cdot k_2\,\frac{\hat\delta(k_1\cdot v_1) \hat\delta(k_2\cdot v_2)\hat\delta(k_1\cdot v_2)}{k_1^2(k_2^2-m^2)[(k_1+k_2)^2-m^2]}e^{i k_1\cdot b}\,,\\ &
        =\textcolor{black}{\frac{1}{(2\pi)^5}}\frac{1}{\gamma\sqrt{\gamma^2-1}}\rmint d^2 k_1 \,\frac{k_1^i e^{ik_1\cdot b}}{k_1^2}\rmint d^3 k_2 \frac{k_2^i}{(\vec k_2^2+m^2)[(\vec k_1+\vec k_2)^2+m^2]}\,,\\ &
        =\textcolor{black}{\frac{1}{(2\pi)^2}}\frac{1}{2\gamma\sqrt{\gamma^2-1}}\rmint_{0}^{\infty} dl \,J_{0}(|b|l)\arctan\Big(\frac{l}{2m}\Big) \label{eq:4.37}
    \end{split}
\end{align}
 and,
 \begin{align}
     \begin{split}
         I_{4}(m,|b|)\Big|_{(4)}&\equiv -\rmint_{k_{1,2}}k_2^2\,\frac{\hat\delta(k_1\cdot v_1) \hat\delta(k_2\cdot v_2)\hat\delta(k_1\cdot v_2)}{k_1^2(k_2^2-m^2)[(k_1+k_2)^2-m^2]}e^{i k_1\cdot b}\,,\\ &
         =-\textcolor{black}{\frac{1}{(2\pi)^5}}\frac{1}{\gamma\sqrt{\gamma^2-1}}\rmint d^2 k_1\frac{e^{-i\vec k_1\cdot \vec b }}{-\vec k_1^2}\rmint d^3 k_2 \frac{-\vec k_2^2}{(\vec k_2^2+m^2)[(\vec k_1+\vec k_2)^2+m^2]}\,,\\ &
         =-\textcolor{black}{\frac{1}{(2\pi)^4}}\frac{1}{\gamma\sqrt{\gamma^2-1}}\rmint_0^{\infty} dl\, \frac{J_{0}(bl)}{l}\Big(l^2\,{\Xi}_1(l,m)+3\,{\Xi}_2(l,m)\Big)\,, \label{eq:4.38}
     \end{split}
 \end{align}
 where ${\Xi}_1(l,m)$ and ${\Xi}_2(l,m)$ are defined in (\ref{new11}). 
\textcolor{black}{After (\ref{eq:4.35}), (\ref{eq:4.37}) and (\ref{eq:4.38}) we get the full answer for $I_{4}(m,|b|)$ mentioned in (\ref{4.37}). Then we get, the }\\
 \hfsetfillcolor{gray!10}
\hfsetbordercolor{black}
\begin{equation}\label{e:barwqpp}
\tikzmarkin[disable rounded corners=false]{l}(0.15,-0.6)(-0.1,0.8)  [\Delta p_1^{\mu}]_{(g)}=\Big(\frac{m_2s_2}{2m_p}\Big)^2\times \frac{m_1}{2m_p^2}\frac{b^\mu}{|b|}\,\partial_{|b|} I_{4}(m,|b|)\,.\tikzmarkend{l}
\end{equation}
Similiarly, the massless limit can be taken by using the fact that $\arctan(\infty)=\frac{\pi}{2}$ and taking the massless limit of $\Xi(l,m)$.\\
\textbullet $\,\,$ Another interesting diagram will contribute to the impulse from the previously mentioned derivative interaction, where one scalar line connects with one worldline, and the other scalar line and the graviton line connect with the other worldline.
\begin{align}
    \begin{split}
         [\Delta p_1^{\mu}]_{(h)}&=\begin{minipage}[h]{0.12\linewidth}
	\scalebox{1.5}{\includegraphics[width=\linewidth]{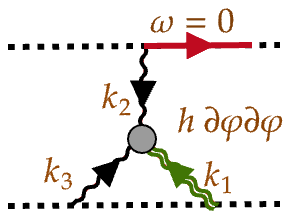}}
\end{minipage}\\ &=i\frac{m_1 s_1 m_2^2 s_2}{4m_p^4}\rmint_{k_i}\hat\delta^{(4)}(\sum_i k_i)k_2^\alpha k_3^\beta \frac{k_2^\mu P_{\alpha\beta;\sigma\delta}v_2^\sigma v_2^\delta\hat\delta(k_2\cdot v_1)\hat\delta(k_1\cdot v_2)\hat\delta(k_3\cdot v_2)}{k_1^2(k_2^2-m^2)(k_3^2-m^2)}e^{ik_2\cdot b_1}e^{i(k_1+k_3)\cdot b_2}\,,\\  &
        =-i\frac{m_1 s_1 m_2^2 s_2}{8m_p^4}\rmint_{k_3,k_2}e^{ik_2\cdot b}k_2^\mu\,\hat\delta(k_2\cdot v_1)\hat\delta(k_2\cdot v_2)\frac{k_2\cdot k_3\,\hat\delta(k_3\cdot v_2)}{(k_2^2-m^2)(k_2+k_3)^2(k_3^2-m^2)}\,,\\ &
        =\frac{m_1 s_1 m_2^2 s_2}{128 \pi^2 m_p^4}\frac{1}{\gamma\sqrt{\gamma^2-1}}\frac{b^\mu}{|b|}\partial_{|b|}\underbrace{\rmint_0^\infty dk \frac{k^2}{k^2+m^2}J_{0}(k|b|)\arctan\Big(\frac{k}{m}\Big)}_{\bar{I}_4(m,|b|)}\,.
    \end{split}
\end{align}
Hence, the contribution from the diagram reads,
\hfsetfillcolor{gray!10}
\hfsetbordercolor{black}
\begin{equation}\label{e:barwq24}
\tikzmarkin[disable rounded corners=false]{+++}(0.3,-0.6)(-0.1,0.8)  [\Delta p_1^{\mu}]_{(h)}=\frac{m_1 s_1 m_2^2 s_2}{128 \pi^2 m_p^4}\frac{1}{\gamma\sqrt{\gamma^2-1}}\frac{b^\mu}{|b|}\partial_{|b|}\bar{I}_4(m,|b|)
\tikzmarkend{+++}\,.
\end{equation}
\textbullet $\,\,$ Another interesting 3-point worldline vertex, which comes from the derivative interaction,  contributing to the impulse at 2PM order, 
\begin{align}
    \begin{split} \label{vert1}
        \mathcal{O}(z): -\frac{m_1s_1}{2m_p^2}\rmint_{k_1,k_2,\omega}e^{i(k_1+k_2)\cdot b_1}\hat\delta(k_1\cdot v_1+k_2\cdot v_1+\omega)\{(k_{1\rho}+k_{2\rho})v^{\mu}v^{\nu}+2\omega v^{(\mu}\delta_{\rho}^{\nu)}\}\varphi(-k_1) h_{\mu\nu}(-k_2)z^{\rho}(-\omega)\,.
    \end{split}
\end{align}
The corresponding contribution to the impulse is given by, 
\begin{align}
    \begin{split}
         [\Delta p_1^{\mu}]_{(i)}=\begin{minipage}[h]{0.12\linewidth}
	\scalebox{1.4}{\includegraphics[width=\linewidth]{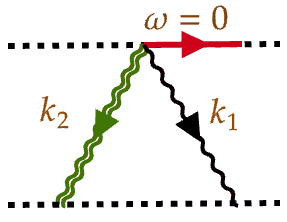}}
\end{minipage}\hspace{0.7 cm}&=i\frac{m_1s_1}{2m_p^2}\times \frac{m_2}{2m_p}\times \frac{m_2s_2}{2m_p}\\ &
\hspace{1 cm}\rmint_{k_{1,2}}(k_1^\mu+k_2^\mu)e^{i(k_1+k_2)\cdot b}\frac{\hat\delta(k_1\cdot v_1+k_2\cdot v_1)\hat\delta(k_1\cdot v_2)\hat\delta(k_2\cdot v_2)}{k_2^2(k_1^2-m^2)}P_{\alpha\beta;\rho\sigma}v_1^\alpha v_1^\beta v_2^\rho v_2^\sigma\,,\\ &=\frac{i m_1s_1m_2^2s_2}{8m_p^4}P_{\alpha\beta;\rho\sigma}v_1^\alpha v_1^\beta v_2^\rho v_2^\sigma\rmint_{q,k_2}\frac{q^\mu\hat\delta(q\cdot v_1)\hat\delta(q\cdot v_2)\hat\delta(k_2\cdot v_2)}{k_2^2[(q-k_2)^2-m^2]}e^{iq\cdot b}\,,\\ &
=\frac{m_1 s_1 m_2^2 s_2}{64\pi^2 m_p^4\gamma\sqrt{\gamma^2-1}}(2\gamma^2-1)\frac{b^\mu}{|b|}\partial_{|b|}\rmint_0^{\infty} dl\,J_{0}(|b|l)\arctan\Big(\frac{l}{m}\Big)\,.
    \end{split}
\end{align}
 \hfsetfillcolor{gray!10}
\hfsetbordercolor{black}
\begin{equation}\label{sa}
\tikzmarkin[disable rounded corners=false]{g}(0.80,-0.55)(-0.4,0.80)  [\Delta p_1^{\mu}]_{(i)}=\frac{m_1 s_1 m_2^2 s_2}{64\pi^2 m_p^4\gamma\sqrt{\gamma^2-1}}(2\gamma^2-1)\frac{b^\mu}{|b|}\partial_{|b|}\rmint_0^{\infty} dl\,J_{0}(|b|l)\arctan\Big(\frac{l}{m}\Big)\,.\tikzmarkend{g} \vspace{0.5cm}
\end{equation}\\\
\textcolor{black}{The integral in \eqref{sa} has a closed-form expression, which can be seen by taking \textbf{logarithmic }representation of  \textbf{ArcTan} which gives,
\begin{align}
    \begin{split}
        [\Delta p_1^{\mu}]_{(i)}&=\frac{m_1 s_1 m_2^2 s_2}{64\pi m_p^4\gamma\sqrt{\gamma^2-1}}(2\gamma^2-1)\frac{b^\mu}{8\pi|b|^2}\\ &\Bigg[2 G_{3,1}^{1,3}\left(\frac{i}{{m\,b}},\frac{1}{2}\Big|
\begin{array}{c}
 1,1,\frac{3}{2} \\
 \frac{3}{2} \\
\end{array}
\right)+2 G_{3,1}^{1,3}\left(-\frac{i}{{m\,b}},\frac{1}{2}\Big|
\begin{array}{c}
 1,1,\frac{3}{2} \\
 \frac{3}{2} \\
\end{array}
\right)+G_{4,2}^{1,4}\left(\frac{i}{{m\,b}},\frac{1}{2}\Big|
\begin{array}{c}
 1,1,\frac{3}{2},\frac{3}{2} \\
 \frac{3}{2},\frac{1}{2} \\
\end{array}
\right)\\ &+G_{4,2}^{1,4}\left(-\frac{i}{{m\,b}},\frac{1}{2}\Big|
\begin{array}{c}
 1,1,\frac{3}{2},\frac{3}{2} \\
 \frac{3}{2},\frac{1}{2} \\
\end{array}
\right)\Bigg]\,.
    \end{split}
\end{align}}
Here have again used that fact that $$P_{\alpha\beta;\rho\sigma}v_1^\alpha v_1^\beta v_2^\rho v_2^\sigma=\frac{2\gamma^2-1}{2}\,.$$
Again, we have a finite massless counterpart of the diagram which gives,
\begin{align}
    \begin{split}
         [\Delta p_1^{\mu}]_{(i)}\Big|_{m\to 0}=-\frac{m_1 s_1 m_2^2 s_2}{64\pi^2 m_p^4\gamma\sqrt{\gamma^2-1}}(2\gamma^2-1)\frac{b^\mu}{|b|^3}\,.\label{4.33a}
    \end{split}
\end{align}
\textbullet $\,\,$  Lastly, there will be another diagram where $h_{\mu\nu}$ in (\ref{vert1}), is replaced by $\varphi$ in the worldline vertex. The worldline vertex under consideration is,
\begin{align}
    \begin{split}
          \mathcal{O}(z):& -\frac{m_1g_1}{2m_p^2}\rmint_{k_1,k_2,\omega}e^{i(k_1+k_2)\cdot b_1}\hat\delta(k_1\cdot v_1+k_2\cdot v_1+\omega)\{(k_{1\rho}+k_{2\rho})v^{\mu}v^{\nu}+2\omega v^{(\mu}\delta_{\rho}^{\nu)}\}\\&\hspace{9 cm}\times\varphi(-k_1)\varphi(-k_2) \eta_{\mu\nu}z^{\rho}(-\omega)\,.
    \end{split}
\end{align}
Therefore, the impulse is given by,
\begin{align}
    \begin{split}
         [\Delta p_1^{\mu}]_{(j)}=\begin{minipage}[h]{0.12\linewidth}
	\scalebox{1.35}{\includegraphics[width=\linewidth]{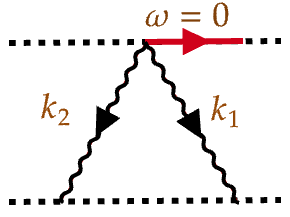}}
\end{minipage}\hspace{0.7 cm}&=i\frac{m_1g_1}{m_p^2}\times \frac{m_2^2 s_2^2}{2m_p^2}\rmint_{q,k_2}\frac{q^\mu\hat\delta(q\cdot v_1)\hat\delta(q\cdot v_2)\hat\delta(k_2\cdot v_2)}{(k_2^2-m^2)[(q-k_2)^2-m^2]}e^{iq\cdot b}\,,\\ &
=\frac{m_1g_1 m_2^2 s_2^2}{8\pi^2 m_p^4\gamma\sqrt{\gamma^2-1}}\frac{b^\mu}{|b|}\partial_{|b|}\rmint_0^\infty dl\,J_{0}(|b|l)\arctan\Big(\frac{l}{2m}\Big)\,.
    \end{split}
\end{align}
Therefore, the contribution to the impulse from the above diagram is given by,\\\
 \hfsetfillcolor{gray!10}
\hfsetbordercolor{black}
\begin{equation}\label{e:barwqp}
\tikzmarkin[disable rounded corners=false]{m}(0.15,-0.6)(-0.1,0.8)  [\Delta p_1^{\mu}]_{(j)}=\frac{m_1g_1 m_2^2 s_2^2}{8\pi^2 m_p^4\gamma\sqrt{\gamma^2-1}}\frac{b^\mu}{|b|}\partial_{|b|}\rmint_0^\infty dl\,J_{0}(|b| l)\arctan\Big(\frac{l}{2m}\Big)\,.\tikzmarkend{m} \vspace{0.5cm}
\end{equation}
\textcolor{black}{Note that, similar to (\ref{sa}), this integral can also be recast in terms of {\bf MeijerG}. } 
The massless limit of  \eqref{4.33a} can be taken, and it gives the following,
\begin{align}
    \begin{split}
         [\Delta p_1^{\mu}]_{(j)}\Big|_{m\to 0}=\frac{m_1g_1 m_2^2 s_2^2}{16 \pi m_p^4\gamma\sqrt{\gamma^2-1}}\frac{b^\mu}{|b|^3}\,.
    \end{split}
\end{align}
\\
Finally, there will be 2PM diagrams coming from the bulk $\varphi^3$  interaction vertex . Below, we will give the details of the contribution to the impulse coming from these vertices. \\
\textbullet $\,\,$ First, we consider the following Feynman diagram.
\begin{align}
    \begin{split}
       [\Delta p_1^{\mu}]_{(k)}&=\begin{minipage}[h]{0.12\linewidth}
	\scalebox{1.4}{\includegraphics[width=\linewidth]{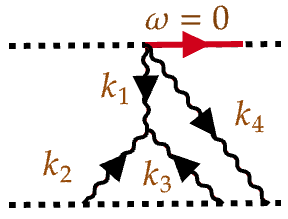}}
\end{minipage}\\&=-i\lambda_3 m_p\Big(\frac{m_1 g_1}{2m_p^2}\Big)\Big(\frac{m_2^3s_2^3}{8m_p^3}\Big)\rmint_{k_i}\hat\delta\Big(\sum_{i=1}^3 k_i\Big)\frac{(k_1+k_4)^\mu\hat\delta(k_1\cdot v_1+k_4\cdot v_1)\hat\delta(k_2\cdot v_2)\hat\delta(k_3\cdot v_2)\hat\delta(k_4\cdot v_2)}{\prod_{i}^4 (k_i^2-m^2)}\\ &\hspace{11 cm}\times e^{i(k_1+k_4)\cdot b_1}e^{i(k_2+k_3)\cdot b_2}e^{-ik_4\cdot b_2}\,.\label{4.51b}
    \end{split}
\end{align}
\vspace{-0.3 cm}
Focusing only on the integral we get,
\begin{align}
    \begin{split}
        [\Delta p_1^{\mu}]_{(k)}^\mu&\sim\rmint_{k_i}\frac{(k_1+k_4)^\mu \hat\delta(k_1\cdot v_1+k_4\cdot v_1)\hat\delta(k_2\cdot v_2)\hat\delta(k_1\cdot v_2)\hat\delta(k_4\cdot v_2)}{(k_1^2-m^2)(k_2^2-m^2)[(k_1+k_2)^2-m^2](k_4^2-m^2)} e^{i(k_1+k_4)\cdot b}\,,\\ &
        \xrightarrow[]{k_1+k_4\to q}\rmint_{k_i,q}\frac{q^\mu\hat\delta(q\cdot v_1)\hat\delta(k_2\cdot v_2)\hat\delta(k_1\cdot v_2)\hat\delta(q\cdot v_2)}{(k_1^2-m^2)(k_2^2-m^2)[(k_1+k_2)^2-m^2][(q-k_1)^2-m^2]}e^{i q\cdot b}\,,\\ &
        =\rmint _q q^\mu \hat{\delta}(q\cdot v_1)\hat\delta(q\cdot v_2)e^{i q \cdot b}\rmint_{k_1,k_2}\frac{\hat\delta(k_1\cdot v_2)\hat\delta(k_2\cdot v_2)}{(k_1^2-m^2)(k_2^2-m^2)[(k_1+k_2)^2-m^2][(q-k_1)^2-m^2]}\,,\\ &
        =\frac{1}{\gamma\sqrt{\gamma^2-1}}\frac{b^\mu}{|b|}\partial_{|b|}\rmint \hat d^2 q\, e^{-i \vec q\cdot \vec b}\, \hat L(\vec q,m)
    \end{split}
\end{align}
where,
\begin{align}
    \begin{split}
        \hat L(\vec q,m)&=\rmint_{\vec k_1,\vec k_2}\frac{1}{(\vec k_1^2+m^2)[(\vec k_1-\vec q)^2+m^2]|\vec k_1|}\arctan\Big(\frac{|\vec k_1|}{2m}\Big)\,,\\ &
        =\rmint_{0}^\infty d\alpha \,e^{-\alpha m^2}\rmint_{\vec k_1}\frac{1}{|\vec k_1|(\vec k_1^2+m^2)} e^{-\alpha(\vec k_1-\vec q)^2}\arctan\Big(\frac{|\vec k_1|}{2m}\Big)\,,\\ &
        =\frac{2\pi}{q} \rmint dk_1 \frac{1}{k_1^2+m^2}\tanh ^{-1}\left(\frac{2 k_1 q}{k_1^2+m^2+q^2}\right)\arctan\Big(\frac{|\vec k_1|}{2m}\Big)\,.
    \end{split}
\end{align}
Therefore, the impulse is (after restoring all the prefractors),\\\\
\hfsetfillcolor{gray!10}
\hfsetbordercolor{black}
\begin{equation}
\tikzmarkin[disable rounded corners=false]{ml}(0,-1)(-0.3,0.8)  [\Delta p_1^{\mu}]_{(k)}=\frac{\lambda_3}{(2\pi)^4} \Big(\frac{m_1 g_1}{2m_p}\Big)\Big(\frac{m_2^3s_2^3}{8m_p^3}\Big)\frac{1}{\gamma\sqrt{\gamma^2-1}}\frac{b^\mu}{|b|}\partial_{|b|}\rmint_0^{\infty} dq \,dk_1 \,\frac{J_{0}(q|b|)}{k_1^2+m^2}\,\textrm{arctanh}\Big(\frac{2 k_1 q}{k_1^2+q^2+m^2}\Big)\arctan\Big(\frac{k_1}{2m}\Big)\label{4.54r}.\tikzmarkend{ml} \vspace{0.5cm}
\end{equation}
The integral in \eqref{4.54r} can be further simplified using the logarithmic representation of ArcTan as shown in \eqref{4.25j}. The massless limit also can be taken as \eqref{4.25j}.
\\\\
\textbullet $\,\,$Another Feynman diagram that we will contribute can be obtained by replacing one scalar propagator in (\ref{4.51b}) with a graviton propagator. The impulse is given by : 
\begin{align}
    \begin{split}
         [\Delta p_1^{\mu}]_{(l)}&=\begin{minipage}[h]{0.12\linewidth}
	\scalebox{1.5}{\includegraphics[width=\linewidth]{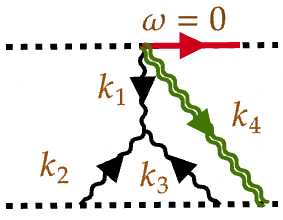}}
\end{minipage}\\ &= \frac{2\gamma^2-1}{2}\frac{\lambda_3}{(2\pi)^4}\Big(\frac{m_1 s_1}{2m_p}\Big)\Big(\frac{m_2^3 s_2^2}{8m_p^3}\Big)\frac{1}{\gamma\sqrt{\gamma^2-1}}\frac{b^\mu}{|b|}\partial_{|b|}\rmint_0^\infty dq  dk_1 \frac{J_0(q|b|)}{k_1^2+m^2}\arctan\Big(\frac{k_1}{2m}\Big)\, \textrm{arctanh}\Big(\frac{2k_1 q}{k_1^2+q^2}\Big)\,.\label{4.55m}
\end{split}
\end{align}
\hfsetfillcolor{gray!10}
\hfsetbordercolor{black}
\begin{equation}
\tikzmarkin[disable rounded corners=false]{mlm}(0.25,-1)(-1.25,0.8)\hspace{-1 cm}  [\Delta p_1^{\mu}]_{(l)}=\frac{2\gamma^2-1}{2}\frac{\lambda_3}{(2\pi)^4}\Big(\frac{m_1 s_1}{2m_p}\Big)\Big(\frac{m_2^3 s_2^2}{8m_p^3}\Big)\frac{1}{\gamma\sqrt{\gamma^2-1}}\frac{b^\mu}{|b|}\partial_{|b|}\rmint_0^\infty dq  dk_1 \frac{J_0(q|b|)}{k_1^2+m^2}\arctan\Big(\frac{k_1}{2m}\Big)\, \textrm{arctanh}\Big(\frac{2k_1 q}{k_1^2+q^2}\Big)\label{4.55r}.\tikzmarkend{mlm} \vspace{0.5cm}
\end{equation}
Again, this integral in \eqref{4.55m} can be further simplified using the logarithmic representation of ArcTan as \eqref{4.25j}.\\
\textbullet $\,\,$ Another Feynman topology contributing to the  impulse at 2PM order is given by: 
\begin{align}
    \begin{split}
         [\Delta p_1^{\mu}]_{(m)}&=\begin{minipage}[h]{0.12\linewidth}
	\scalebox{1.5}{\includegraphics[width=\linewidth]{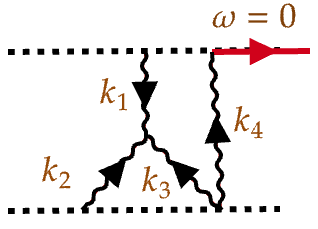}}
\end{minipage}\\ &= -i\lambda_3\Big(\frac{m_1^2 s_1^2}{4m_p}\Big)\Big(\frac{m_2 s_2 m_2 g_2 }{4m_p^3}\Big)\rmint_{k_i}\frac{k_1^\mu \hat\delta(k_1\cdot v_1)\hat\delta(k_2\cdot v_2)\hat\delta(k_3\cdot v_2+k_4\cdot v_2)\hat\delta(k_4\cdot v_1)}{\prod_i (k_i^2-m^2)}\hat\delta^{(4)}(k_1+k_2+k_3)\\ &
\hspace{7 cm}\times e^{i(k_1-k_4)\cdot b_1}e^{i(k_2+k_3+k_4)\cdot b_2}\,,\\ &
        =-i\lambda_3\Big(\frac{m_1^2 s_1^2}{4m_p}\Big)\Big(\frac{m_2 s_2 m_2 g_2 }{4m_p^3}\Big)\rmint_{k_i}\frac{k_1^\mu \hat\delta(k_1\cdot v_1)\hat\delta(k_2\cdot v_2)\hat\delta(-k_1\cdot v_2-k_2\cdot v_2+k_4\cdot v_2)\hat\delta(k_4\cdot v_1)}{(k_1^2-m^2)(k_2^2-m^2)[(k_1+k_2)^2-m^2](k_4^2-m^2)}e^{i(k_1-k_4)\cdot b}\,,\\ &
        \xrightarrow[]{k_1-k_4\to q}-i\lambda_3\Big(\frac{m_1^2 s_1^2}{4m_p}\Big)\Big(\frac{m_2 s_2 m_2 g_2 }{4m_p^3}\Big)\rmint_{k_i,q}\frac{k_1^\mu\hat\delta(k_1\cdot v_1)\hat\delta(k_2\cdot v_2)\hat\delta(q\cdot v_2)\hat\delta(q\cdot v_1)}{(k_1^2-m^2)(k_2^2-m^2)[(k_1+k_2)^2-m^2][(q-k_1)^2-m^2]}e^{iq\cdot b}\,.
    \end{split}\label{4.56e}
\end{align}\\\\
Now, doing the $k_2$ integral we will have (omitting the prefactors),
\begin{align}
    \begin{split}
         [\Delta p_1^{\mu}]_{(m)}\propto \rmint_{k_i,q}\frac{k_1^\mu\hat\delta(k_1\cdot v_1)\hat\delta(q\cdot v_2)\hat\delta(q\cdot v_1)}{(k_1^2-m^2)[(q-k_1)^2-m^2]|\vec k_1|}\arctan\Big(\frac{|\vec k_1|}{2m}\Big)e^{iq\cdot b}\,.
    \end{split}
\end{align}
\textcolor{black}{We see that due to the asymmetric structure of this diagram, it is difficult to obtain a closed-form expression, like the case of $\lambda_4 \varphi^4$ vertex. We first do the $q$ integral. It is clear from the delta function constraints that $q$ and $k_1$ are two-dimensional and three-dimensional vectors, respectively.} Therefore, after performing the $q$ integral, the result will only depend on the second and third components of $\vec k_1$. Therefore, we are left with the following, 
\begin{align}
    \begin{split}
          [\Delta p_1^{\mu}]_{(m)}&\propto 2 \pi \frac{1}{\gamma\sqrt{\gamma^2-1}}\rmint_{k_1} e^{-i ({b^{(2)}} {k_1^{(2)}}+{b^{(3)}}{k_1^{(3)}})} K_0\left(\sqrt{{b^{(2)}}^2+{b^{(3)}}^2} \sqrt{\vec k_1^2-{k_1^{(2)}}^2-{k_1^{(3)}}^2+m^2}\right)\\ &\hspace{7 cm}\times\frac{k_1^\mu}{(k_1^2-m^2)|\vec k_1|}\arctan\Big(\frac{|\vec k_1|}{2m}\Big)\,.\label{4.57mm}
    \end{split}
\end{align}
Now, note that in our parametrisation, the impact parameter takes the form: $b^{\mu}=(0,0,1,0)$. Therefore, restoring the prefactors integral in \eqref{4.57mm} can be recasted as,
\begin{align}
    \begin{split}
         [\Delta p_1^{\mu}]_{(m)}&=\frac{\lambda_3}{(2\pi)^4}\Big(\frac{m_1^2 s_1^2}{4m_p}\Big)\Big(\frac{m_2 s_2 m_2 g_2 }{4m_p^3}\Big)\frac{1}{\gamma\sqrt{\gamma^2-1}}\rmint d^4 k_1 e^{ik_1\cdot b}\,K_{0}\Big(|b||\sqrt{k_{1(1)}^2+m^2}|\Big)\\ &
        \hspace{7 cm}\times\frac{k_1^\mu\hat\delta(k_1\cdot v_1)}{(k_1^2-m^2)|\vec k_1|}\arctan\Big(\frac{|\vec k_1|}{2m}\Big)\,,\\ &
        =\frac{\lambda_3}{(2\pi)^4}\Big(\frac{m_1^2 s_1^2}{4m_p}\Big)\Big(\frac{m_2 s_2 m_2 g_2 }{4m_p^3}\Big)\frac{1}{\gamma\sqrt{\gamma^2-1}}\frac{b^\mu}{|b|}\partial_{|b|}\rmint d^3 k_1 e^{-i \vec k_1\cdot \vec b}K_{0}\Big(|b||\sqrt{k_{1(1)}^2+m^2}|\Big)\\ &
        \hspace{7 cm}\times\frac{1}{(\bar k_1^2+m^2)|\vec k_1|}\arctan\Big(\frac{|\vec k_1|}{2m}\Big)\,.\label{4.58}
    \end{split}
\end{align}
Hence, the contribution has the following form,
\hfsetfillcolor{gray!10}
\hfsetbordercolor{black}
\begin{align}
\begin{split}
\tikzmarkin[disable rounded corners=false]{si}(2.8,-1)(-0.6,1)  [\Delta p_1^{\mu}]_{(m)}&=\frac{\lambda_3}{(2\pi)^4}\Big(\frac{m_1^2 s_1^2}{4m_p}\Big)\Big(\frac{m_2 s_2 m_2 g_2 }{4m_p^3}\Big)\frac{1}{\gamma\sqrt{\gamma^2-1}}\frac{b^\mu}{|b|}\partial_{|b|}\rmint_{-\infty}^{\infty} d^3 k_1 e^{-i \vec k_1\cdot \vec b}K_{0}\Big(|b||\sqrt{k_{1(1)}^2+m^2}|\Big)\\ &\hspace{7 cm}\times \frac{1}{(\bar k_1^2+m^2)|\vec k_1|}\arctan\Big(\frac{|\vec k_1|}{2m}\Big)\label{4.58o}\tikzmarkend{si} \vspace{0.5cm}
\end{split}
\end{align}
where, $\bar k_1=(k_1^{(1)}/\gamma,k_1^{(2)},k_1^{(3)})$. \textcolor{black}{Again, we can see that the integral in \eqref{4.58} has an asymmetry in the different components of the momentum integration and therefore, to the best of our knowledge, does not have a closed-form. For further analysis, one should do the integral componentwise and numerically.} Although, for the massless case, we also do not have any closed-form due to the asymmetry, but one can take a smooth massless limit.
\\\\
\textbullet $\,\,$ Similar to \eqref{4.56e} we will have another contributing diagram where the $k_4$ scalar propagator will be replaced by a graviton propagator. Corresponding contribution to the impulse takes the following form,
\begin{align}
    \begin{split}
         [\Delta p_1^{\mu}]_{(n)} &= \begin{minipage}[h]{0.12\linewidth}
	\scalebox{1.7}{\includegraphics[width=\linewidth]{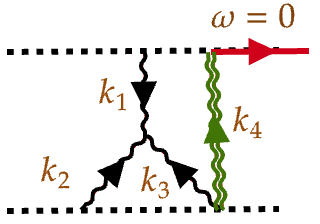}}
\end{minipage} \\&=\frac{\lambda_3}{(2\pi)^4}\Big(\frac{m_1^2 s_1^2}{4m_p}\Big)\Big(\frac{m_2^2 s_2^2  }{4m_p^3}\Big)\frac{2\gamma^2-1}{2\gamma\sqrt{\gamma^2-1}} \frac{b^\mu}{|b|}\partial_{|b|}\rmint d^3 k_1 e^{-i \vec k_1\cdot \vec b}K_{0}\Big(|b||{k_{1(1)}}|\Big)\frac{1}{(\bar k_1^2+m^2)|\vec k_1|}\arctan\Big(\frac{|\vec k_1|}{2m}\Big)\,.
    \end{split}
\end{align}
\hfsetfillcolor{gray!10}
\hfsetbordercolor{black}
\begin{equation}
\tikzmarkin[disable rounded corners=false]{mlmnl}(0.25,-1)(-0.5,0.8)  [\Delta p_1^{\mu}]_{(n)}=\frac{\lambda_3}{(2\pi)^4}\Big(\frac{m_1^2 s_1^2}{4m_p}\Big)\Big(\frac{m_2^2 s_2^2  }{4m_p^3}\Big)\frac{2\gamma^2-1}{2\gamma\sqrt{\gamma^2-1}} \frac{b^\mu}{|b|}\partial_{|b|}\rmint_{-\infty}^{\infty} d^3 k_1 e^{-i \vec k_1\cdot \vec b}K_{0}\Big(|b||{k_{1(1)}}|\Big)\frac{1}{(\bar k_1^2+m^2)|\vec k_1|}\arctan\Big(\frac{|\vec k_1|}{2m}\Big)\label{4.63pp}.\tikzmarkend{mlmnl} \vspace{0.5cm}
\end{equation}
Again, due to the reason mentioned above, this integral does not possess any closed-form expression. One has to do it numerically component-wise. \\\\
\textbullet $\,\,$ We have another type of vertex $\lambda_3 h \varphi^3$, which also contributes to the 2PM diagrams. Calculations are similar to the computations of $\lambda_4 \varphi^4$ vertex \eqref{4.26m}.
\begin{align}
    \begin{split}
         [\Delta p_1^{\mu}]_{(o)}&=\begin{minipage}[h]{0.12\linewidth}
	\scalebox{1.5}{\includegraphics[width=\linewidth]{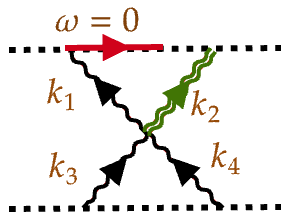}}
\end{minipage}\\ &=-i\lambda_3\Big(\frac{m_1^2 s_1 (m_2 s_2)^2}{16m_p^4}\Big)P^{\alpha}_{\,\,\alpha,\rho\sigma}v_1^\rho v_1^\sigma\rmint_{\{k_i\}}\hat\delta^{(4)}\Big(\sum_{i=1}^{4}k_{i}\Big)\hat\delta(k_1\cdot v_1)\hat\delta(k_2\cdot v_1)\hat\delta(k_3\cdot v_2)\hat\delta(k_4\cdot v_2)\\ &\hspace{5 cm}\times\prod_{j=1,3,4}\frac{1}{(k_j^2-m^2)k_2^2} 
 k_1^{\mu} \,e^{i(k_1+k_2)\cdot b_2}e^{i(k_3+k_4)\cdot b_2}\,.\label{4.55e}
    \end{split}
\end{align}
Now, \eqref{4.55e} has the same form as \eqref{4.26m} except that one denominator is massless. The result is given by,
\hfsetfillcolor{gray!10}
\hfsetbordercolor{black}
\begin{equation}
\tikzmarkin[disable rounded corners=false]{mlmnlo}(0.2,-1)(-0.3,0.8)  [\Delta p_1^{\mu}]_{(o)}=-\frac{\lambda_3}{(2\pi)^3}\Big(\frac{m_1^2 s_1 (m_2 s_2)^2}{16m_p^4}\Big)\frac{1}{\sqrt{\gamma^2-1}}\frac{b^\mu}{|b|}\rmint_0^{\infty} dx {J_{1}(x|b|)}\arctan\Big(\frac{x}{2m}\Big)\arctan\Big(\frac{x}{m}\Big)\,\label{4.63ppl}.\tikzmarkend{mlmnlo} \vspace{0.5cm}
\end{equation}
\textcolor{black}{Like before, although this integral does not possess any closed-form expression, it has a smooth massless limit. In the massless limit, the above diagram can be calculated exactly as},
\begin{align}
    \begin{split}
         [\Delta p_1^{\mu}]_{(o)}\Big|_{m\rightarrow 0}=-{\lambda_3}\Big(\frac{m_1^2 s_1 (m_2 s_2)^2}{1024 \pi m_p^4}\Big)\frac{1}{\sqrt{\gamma^2-1}}\frac{b^\mu}{|b|^3}\,.
        \end{split}
    \end{align} 
\textbullet $\,\, $ There could be another possibility like \eqref{4.19c} where we have one line (graviton) connected with the first worldline and the other three (scalars) are connected with the second worldline. Now, the corresponding contribution to the impulse reads,
\begin{align}
    \begin{split}
         [\Delta p_1^{\mu}]_{(p)}&=\begin{minipage}[h]{0.12\linewidth}
	\scalebox{1.5}{\includegraphics[width=\linewidth]{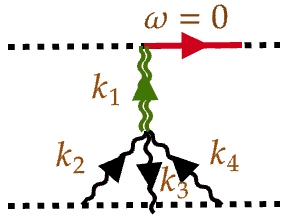}}
\end{minipage}\\ &= -i\lambda_3\Big(\frac{m_1  (m_2 s_2)^3}{16m_p^4}\Big)P^{\alpha}_{\,\,\alpha,\rho\sigma}v_1^\rho v_1^\sigma \rmint_{\{k_i\}}\hat\delta^{(4)}\Big(\sum_{i=1}^{4}k_{i}\Big)\hat\delta(k_1\cdot v_1)\hat\delta(k_2\cdot v_2)\hat\delta(k_3\cdot v_2)\hat\delta(k_4\cdot v_2)\\ &\hspace{4.5 cm}\prod_{j=2}^{4} \frac{1}{(k_j^2-m^2)k_1^2}
\times k_1^{\mu} \,e^{ik_1\cdot b_1}e^{i(k_2+k_3+k_4)\cdot b_2}\,.\label{4.57a}
    \end{split}
\end{align}
Following the derivation of \eqref{4.19c} we will get, \\\\
\hfsetfillcolor{gray!10}
\hfsetbordercolor{black}
\begin{equation}
\tikzmarkin[disable rounded corners=false]{mlmnlolk}(0.6,-1)(-0.3,0.8)  [\Delta p_1^{\mu}]_{(p)}=\frac{\lambda_3}{(2\pi)^3}\Big(\frac{m_1  (m_2 s_2)^3}{16m_p^4}\Big)\frac{1}{\gamma\sqrt{\gamma^2-1}}\frac{b^\mu}{|b|}\partial_{|b|}\rmint_0^\infty dq \,dk_1\,\frac{J_{0}(k_1 |b|)}{k_1^2}\arctan\Big(\frac{q}{2m}\Big)\textrm{arctanh}\Big(\frac{2qk_1}{m^2+q^2+k_1^2}\Big)\,.\label{4.58a}\tikzmarkend{mlmnlolk} \vspace{0.5cm}
\end{equation}
The integral in \eqref{4.58a} can be further simplified using the logarithmic representation of ArcTan. The massless limit can also be taken as \eqref{4.25j}.
\\\\
\textbullet $\,\,$ There will be another another diagram topologically equivalent to \eqref{4.57a} contributing to the impulse where one line (scalar) connects with one worldline and the other three (one graviton and two scalars) connect with the other worldline. The corresponding contribution is,
\begin{align}
    \begin{split}
      [\Delta p_1^{\mu}]_{(q)} &=\begin{minipage}[h]{0.12\linewidth}
	\scalebox{1.4}{\includegraphics[width=\linewidth]{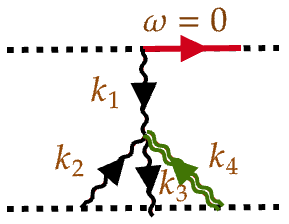}}
\end{minipage}\\ &= -i\lambda_3\Big(\frac{m_1s_1 m_2 (m_2 s_2)^2}{16m_p^4}\Big)P^{\alpha}_{\,\,\alpha,\rho\sigma}v_2^\rho v_2^\sigma \rmint_{\{k_i\}}\hat\delta^{(4)}\Big(\sum_{i=1}^{4}k_{i}\Big)\hat\delta(k_1\cdot v_1)\hat\delta(k_2\cdot v_2)\hat\delta(k_3\cdot v_2)\hat\delta(k_4\cdot v_2)\\ & 
\hspace{5 cm}\times\prod_{j=2}^{4}\frac{1}{(k_j^2-m^2)k_1^2} k_1^{\mu} \,e^{ik_1\cdot b_1}e^{i(k_2+k_3+k_4)\cdot b_2}\,,
    \end{split}
\end{align}
Similarly, following the derivation of \eqref{4.19c} we have,\\\\
\hfsetfillcolor{gray!10}
\hfsetbordercolor{black}
\begin{equation}
\tikzmarkin[disable rounded corners=false]{mlmnlol}(0.2,-1)(-0.,0.8)[\Delta p_1^{\mu}]_{(q)} =\frac{\lambda_3}{(2\pi)^3}\Big(\frac{m_1s_1 m_2 (m_2 s_2)^2}{16m_p^4}\Big)\frac{1}{\gamma\sqrt{\gamma^2-1}}\frac{b^\mu}{|b|}\partial_{|b|}\rmint_0^\infty dq \,dk_1\,\frac{J_{0}(k_1 |b|)}{k_1^2+m^2}\arctan\Big(\frac{q}{m}\Big)\textrm{arctanh}\Big(\frac{2qk_1}{m^2+q^2+k_1^2}\Big).\label{4.66t}\tikzmarkend{mlmnlol} \vspace{0.5cm}
\end{equation}
The integral in \eqref{4.66t} can be further simplified using the logarithmic representation of ArcTan. The massless limit can also be taken as \eqref{4.25j}.
\\
\textbullet $\,\,$ So far, the diagrams that contributed to the impulse at 2PM order require expanding the point particle action upto $\mathcal{O}({z})\,.$ But there is another diagram we need to consider to complete the computation of impulse at 2PM, but that requires expanding the point particle action upto $\mathcal{O}({z}^2)$. 
\begin{align}
    \begin{split}
       [\Delta p_1^{\mu}]_{(r)} =\begin{minipage}[h]{0.12\linewidth}
	\vspace{4pt}
	\scalebox{1.3}{\includegraphics[width=\linewidth]{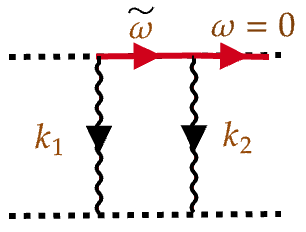}}
\end{minipage}\hspace{0.7 cm}&= -im_1\Big(\frac{s_1m_2 s_2}{2\sqrt{2}m_p^2}\Big)^2\rmint_{k_i,\tilde\omega}\hat\delta(k_1\cdot v_2)\hat\delta(k_2\cdot v_2)\hat\delta(-k_1\cdot v_1+\tilde\omega)\hat\delta(-k_2\cdot v_1-\tilde\omega+\omega)\\ &\hspace{0 cm}\times (2\tilde\omega v_1^{{\rho}}-k_1^{\rho}) (\frac{1}{2}k_{2\rho}k_{2}^\mu+\tilde\omega k_{2}^\mu v_{1\rho}-\omega k_{2\rho}v_{1}^\mu-\omega\,\tilde\omega\delta^\mu_{\rho})\frac{e^{i(k_1+k_2)\cdot b}}{(k_1^2-m^2)(k_2^2-m^2)\tilde\omega^2}\Big |_{\omega=0},\\ &
        =im_1\Big(\frac{s_1m_2 s_2}{4m_p^2}\Big)^2\textcolor{black}{\rmint_{k_1,k}\frac{(k-k_1)^{\mu}\hat\delta(k_1\cdot v_2)\hat\delta(k\cdot v_2)\hat\delta(k\cdot v_1)e^{ik\cdot b}}{(k_1^2-m^2)[(k-k_1)^2-m^2](k_1\cdot v_1)^2}k_1\cdot(k-k_1)}\,.\label{4.6 mm}
    \end{split}
\end{align}
One can write down the integral as ,
\begin{align}
    \begin{split}
       [\Delta p_1^{\mu}]_{(r)} &=im_1\Big(\frac{s_1m_2 s_2}{4m_p^2}\Big)^2\rmint \hat d^4k\,\hat\delta(k\cdot v_1)\hat\delta(k\cdot v_2)e^{ik\cdot b}\rmint \hat d^4 k_1\,\hat\delta(k_1\cdot v_2)\frac{k_1\cdot(k-k_1)(k-k_1)^{\mu}}{(k_1^2-m^2)[(k-k_1)^2-m^2](k_1\cdot v_1)^2}\,,\\ &
        =im_1\Big(\frac{ s_1m_2 s_2}{4m_p^2}\Big)^2\rmint \hat d^4k\,\hat\delta(k\cdot v_1)\hat\delta(k\cdot v_2)e^{ik\cdot b}\mathcal{K}^{\mu}\label{4.37a}
    \end{split}
\end{align}
where $\mathcal{K}^{\mu}(k,v_i)$ can be reduced by Passarino-Veltman reduction as,
\begin{align}
    \begin{split}
        \mathcal{K}^{\mu}=a_1 k^{\mu}+a_2 v_1^{\mu}+a_3 v_2^{\mu}\,.\label{4.38a}
    \end{split}
\end{align}
It is evident that with the support $k\cdot v_1 =0\,\text{and } k\cdot v_2=0$, one can write $ v_{2\mu}\mathcal{K}^{\mu}=0$, implying, $a_2\gamma+a_3=0$. Hence, we left with the following,
\begin{align}
    \begin{split}
\mathcal{K}^{\mu}=a_2(v_1^{\mu}-\gamma v_2^{\mu})+a_1k^\mu\,.
    \end{split}
\end{align}
From \eqref{B.22aa} it is clear that the $a_1$ coefficient is non-zero for 
and is given by,
\begin{align}
    \begin{split}
     a_1=   \frac{1}{4}(-\vec k^2-2m^2)\hat\chi_k(|\vec k|,m)
    \end{split}
\end{align}
where, $\hat\chi_{k}(k,m)$ is defined in \eqref{B.16ab}. Similarly,
$a_2$ can be fixed by contracting both side by $v_1^{\mu}$,
\begin{align}
    \begin{split}
        a_2&=\frac{1}{\gamma^2-1}\rmint \hat d^{4}k_1 \frac{\hat\delta(k_1\cdot v_2)\,k_1\cdot(k-k_1)}{(k_1^2-m^2)[(k-k_1)^2-m^2](k_1\cdot v_1+i\epsilon)}\,,\\ &
        =\frac{1}{2(\gamma^2-1)}\rmint \hat d^4 k_1\hat\delta(k_1\cdot v_2)\Big[-\frac{1}{(k_1^2-m^2)(k_1\cdot v_1+i\epsilon)}-\frac{1}{(k-k_1)^2-m^2](k_1\cdot v_1+i\epsilon)}\\\ &
        \hspace{3.3 cm} \frac{k^2}{(k_1^2-m^2)[(k-k_1)^2-m^2](k_1\cdot v_1+i\epsilon)}
      -\frac{2m^2}{(k_1^2-m^2)[(k-k_1)^2-m^2](k_1\cdot v_1+i\epsilon)}  \Big]\,.\label{3.13}
    \end{split}
\end{align}
The first two terms will not contribute. Now doing a change of variable, $k_1-k\rightarrow -q$, $a_2$ can be written as,
\begin{align}
    \begin{split}
        a_2=-\frac{1}{2(\gamma^2-1)}\rmint \hat d^4q \frac{(k^2-2m^2)\hat\delta(q\cdot v_2)}{(q^2-m^2)[(q-k)^2-m^2](q\cdot v_1-i\epsilon)}\,.\label{3.14}
    \end{split}
\end{align}
Now adding (\ref{3.13}) and (\ref{3.14}) and using the fact,
\begin{align}
    \begin{split}
       - i\hat \delta(x)=\frac{1}{x+i\epsilon}-\frac{1}{x-i\epsilon}
    \end{split}
\end{align}
one gets,
\begin{align}
    \begin{split}
        a_2=\frac{-i}{4(\gamma^2-1)}\rmint \hat d^4 q \frac{(k^2-2m^2)\hat\delta(q\cdot v_1)\hat\delta(q\cdot v_2)}{(q^2-m^2)[(q-k)^2-m^2]}\,.
    \end{split}
\end{align}
Thereafter, $\Delta p_1^{\mu}$ can be written as,
\begin{align}
    \begin{split}
         \hspace{-1 cm}[\Delta p_1^{\mu}]_{(r)} &= -m_1\Big(\frac{s_1m_2 s_2}{8\sqrt{2}\sqrt{\gamma^2-1}m_p^2}\Big)^2\rmint \hat d^4 k\,\hat d^4 q\,\hat\delta(k\cdot v_1)\hat\delta(k\cdot v_2)\hat\delta(q\cdot v_1)\hat\delta(q\cdot v_2)\,e^{i k\cdot b}\,\frac{k^2-2m^2}{(q^2-m^2)[(q-k)^2-m^2]}\\ &
\hspace{5 cm}\times (v_1^{\mu}-\gamma v_2^{\mu})+im_1\Big(\frac{s_1m_2 s_2}{4m_p^2}\Big)^2\rmint_{k}\hat\delta(k\cdot v_1)\hat\delta(k\cdot v_2)e^{ik\cdot b}k^{\mu}a_1\,,
         \\&
        =-\frac{m_1}{(2\pi)^4}\Big(\frac{s_1m_2 s_2}{8\sqrt{2}m_p^2}\Big)^2\frac{1}{(\gamma^2-1)^3}\rmint  d^2 k\,e^{-ik\cdot b}(-k^2-2m^2)\rmint  d^2 q\,\frac{1}{(q^2+m^2)[(q-k)^2+m^2]}(v_1^{\mu}-\gamma v_2^{\mu})\\ &
      \hspace{5 cm}  -m_1\Big(\frac{s_1m_2 s_2}{8m_p^2}\Big)^2\frac{1}{\pi^{3/2}(\gamma^2-1)^{3/2}} \frac{b^\mu}{|b|}\Big(m K_1(2 |b| m)\Big)
        \,.
    \end{split}
\end{align}
The $`q$' integral can be done using dimensional regularisation with $d=2-\epsilon$. We need to be cautious about the UV (or IR) divergences that may appear while doing the integral. The integral can be written as,
\begin{align}
    \begin{split}
      I&=  \rmint {d}^{2-\epsilon}q \frac{1}{(q^2+m^2)[(q-k)^2+m^2]}\,,
      \\ &
       \xrightarrow[]{\epsilon \rightarrow 0} (2\pi)\frac{2 \log \left(\sqrt{\frac{k^2}{4 m^2}+1}+\frac{k}{2 m}\right)}{m\,k  \sqrt{\frac{k^2}{4 m^2}+1}}\,=(2\pi)\frac{4[\log(k+\sqrt{k^2+4m^2})-\log(2m)]}{k\sqrt{k^2+4m^2}}
    \end{split}
\end{align}
Now, the integral $\Delta p_1^{\mu}$ can be expressed as,
\hfsetfillcolor{gray!10}
\hfsetbordercolor{black}
\begin{align}
\begin{split} \label{eee22}
\tikzmarkin[disable rounded corners=false]{e11}(6,-0.79)(-0.15,0.7)  [\Delta p_1^{\mu}]_{(r)} =&-\frac{m_1}{(2\pi)^2}\Big(\frac{s_1m_2 s_2}{8\sqrt{2}\,m_p^2}\Big)^2\frac{1}{(\gamma^2-1)^3}(v_1^\mu-\gamma v_2^\mu)\rmint_{0}^{\infty}{d}k\,k\, J_{0}(k|b|)(k^2+2m^2)
      \\&\hspace{6cm} \Bigg( \frac{4[\log(k+\sqrt{k^2+4m^2})-\log(2m)]}{k\sqrt{k^2+4m^2}}\Bigg)\\ &
      \hspace{0 cm}-m_1\Big(\frac{s_1m_2 s_2}{8m_p^2}\Big)^2\frac{1}{\pi^{3/2}\gamma^3(\gamma^2-1)^{3/2}} \frac{b^\mu}{|b|}\Big(m K_1(2 |b| m)\Big)\,.
\tikzmarkend{e11}
\end{split}
\end{align}\\\\
In the massless limit we have,
\begin{align}
    \begin{split}
     [\Delta p_1^{\mu}]_{(r)} \Big|_{m\to 0}&= -\frac{m_1}{(2\pi)^2}\Big(\frac{s_1m_2 s_2}{4\sqrt{2}\,m_p^2}\Big)^2\frac{1}{(\gamma^2-1)^3}(v_1^\mu-\gamma v_2^\mu)\Bigg(\rmint_{0}^{\infty}dk \, k\,J_{0}(k|b|) \log(2 k)-4\frac
     {\log(2\epsilon)J_{1}(|b|\Lambda)}{|b|}\Bigg)\\ &
       \hspace{3 cm}-m_1\Big(\frac{s_1m_2 s_2}{8m_p^2}\Big)^2\frac{1}{\pi^{3/2}(\gamma^2-1)^{3/2}} \frac{b^\mu}{|b|^2}
    \end{split}
\end{align}
\vspace{-0 cm}
where $\epsilon$ and $\Lambda$ are the IR and UV cut-off respectively. The second term is a divergent one. In fact, UV/IR  divergence is mixed, i.e. it is divergent in both the limit  ${\epsilon}\rightarrow 0 \,\& \,\Lambda\rightarrow \infty\,.$ So we ignore this in the classical limit. The finite part takes the following form,
\begin{align}
    \begin{split}
    [\Delta p_1^{\mu}]_{(r)} \Big|_{m\to 0}\sim N_1\frac{(v_1^{\mu}-\gamma v_2^{\mu})}{4(\gamma^2-1)^3} \frac{1}{|b|^2}\textcolor{black}{+N_2 \frac{b^\mu}{(\gamma^2-1)^{3/2}|b|^2}}\,.
    \end{split}
\end{align}
The total impulse (due to the scalar field) at 2PM order is the sum of  (\ref{e:barwq244}), (\ref{e:barwq248}), (\ref{e11}), (\ref{e:barwq246}), (\ref{e:barwqpp}), (\ref{e:barwq24}), (\ref{sa}), (\ref{e:barwqp}), (\ref{4.54r}), (\ref{4.55r}), (\ref{4.58o}), (\ref{4.63pp}), (\ref{4.63ppl}), (\ref{4.58a}), (\ref{4.66t}) and (\ref{eee22}) as well as the terms that come from interchanging the worldline one and two. 
\begin{align}
\begin{split}
     \Delta p_1^{\mu}\Big|^{\textrm{2PM}, \textrm{Total}}_{\textrm{scalar}}= & \sum_{\upsilon=c}^{r}[\Delta p_1^{\mu}]_{(\upsilon)}\,.
     \end{split}
\end{align}
\vspace{0 cm}
Finally, collecting all individual expressions we get,\\
\hfsetfillcolor{white!10}
\hfsetbordercolor{white}
\begin{align}
\begin{split} \label{e22}
\tikzmarkin[disable rounded corners=false]{ma1}(-0.10,-1)(-1.3,1)
\hspace{-1 cm}\Delta p_1^{\mu}\Big|^{\textrm{2PM},\textrm{total}}_{\textrm{scalar}}=&\frac{m^2\, m_1m_2^2 s_2}{32\pi^2\,m_p^4\,\sqrt{\gamma^2-1}}\frac{b^\mu}{|b|}\partial_{|b|}\Big(s_2 I_2(m,|b|)+s_1 \bar{I}_2(m,|b|)\Big)\\&+\frac{m_1 m_2^2 s_2}{8\,m_p^4}\frac{b^{\mu}}{|b|}\partial_{|b|}\Big(s_2 I_{4}(m,|b|)+\frac{s_1 }{16 \pi^2 \sqrt{\gamma^2-1}}\bar{I}_4(m,|b|)\Big)\\ &+\frac{m_1 m_2^2 s_2}{8\pi^2m_p^4\gamma\sqrt{\gamma^2-1}}\frac{b^\mu}{|b|}\partial_{|b|}\Bigg[ \frac{s_1\,(2\gamma^2-1) }{8}\rmint dl\,J_{0}(|b|l)\arctan\Big(\frac{l}{m}\Big)
        +g_1 s_2\rmint_0^\infty dl\,J_{0}(|b|l)\arctan\Big(\frac{l}{2m}\Big)\Bigg]\\&-\frac{m_1}{(2\pi)^2}\Big(\frac{s_1m_2 s_2}{4\sqrt{2}\,m_p^2}\Big)^2\frac{1}{(\gamma^2-1)^3}(v_1^\mu-\gamma v_2^\mu)\rmint_{0}^{\infty}{d}k\,k\, J_{0}(k|b|)(k^2+2m^2)     \\&\hspace{6cm}\times \Bigg( \frac{\log(k+\sqrt{k^2+4m^2})-\log(2m)}{k\sqrt{k^2+4m^2}}\Bigg)\\ &
      \hspace{0 cm}-m_1\Big(\frac{s_1m_2 s_2}{8m_p^2}\Big)^2\frac{1}{\pi^{3/2}(\gamma^2-1)^{3/2}} \frac{b^\mu}{|b|}\Big(m K_1(2 |b| m)\Big)\\&+\frac{\lambda_4\, m_1s_1 (m_2s_2)^2}{128\pi^3m_p^4\sqrt{\gamma^2-1}}\frac{b^\mu}{|b|}\partial_
{|b|}\Bigg[m_1s_1\,I_3(m,|b|)+\frac{m_2 s_2}{4\gamma}\Big(2(1-\log(3m^2))K_{0}(|b|m)-\\&\hspace{9cm}\rmint dk_1\frac{J_{0}(k_1 |b|)}{k_1^2+m^2}\Theta(k_1,m)\Big)\Bigg]\\&+\frac{\lambda_3}{(2\pi)^4} \Big(\frac{m_2^3s_2^3}{16m_p^4}\Big)\frac{1}{\sqrt{\gamma^2-1}}\frac{b^\mu}{|b|}\partial_{|b|}\Bigg[\Big(m_1 g_1 I_{5}(m,|b|)+\frac{(2\gamma^2-1)\,m_1 s_1}{2}\bar{I}_{5}(m,|b|)\Bigg]\\&+\frac{\lambda_3}{(2\pi)^3}\Big(\frac{m_1 (m_2 s_2)^2}{16m_p^4}\Big)\frac{1}{\sqrt{\gamma^2-1}}\frac{b^\mu}{|b|}\partial_{|b|}\Bigg[-m_1s_1 I  _{6}(m,|b|)+\frac{m_2 s_2}{\gamma}\bar{I}_{6}(m,|b|)+\frac{m_2 s_1}{\gamma}\bar{I}_{6}(m,|b|)\Bigg]\\&+\frac{\lambda_3}{(2\pi)^4}\Big(\frac{m_1^2 s_1^2}{4m_p}\Big)\Big(\frac{m_2 s_2}{4m_p^3}\Big)\frac{1}{\gamma\,\sqrt{\gamma^2-1}}\frac{b^\mu}{|b|}\partial_{|b|}\Bigg[m_2g_2\,I_{7}(m,|b|)+\Big(\frac{m_2 s_2\,(2\gamma^2-1) }{2}\Big)\bar{I}_{7}(m,|b|)\Bigg]+(1\leftrightarrow 2)\nonumber
\tikzmarkend{ma1}
\end{split}
\end{align}
\newpage
where, \begin{align}
    \begin{split}  
    &I_2(m,|b|)=\rmint _{0}^{\infty}dx\,\frac{J_{0}(|b|x)}{x^2}\arctan\Big(\frac{x}{2m}\Big)\,,\quad 
    \bar{I}_{2}(m,|b|)=\rmint_{0}^\infty dx\frac{J_{0}(b|x|)}{x^2+m^2}\arctan\Big(\frac{x}{m}\Big)\,,\\&
    I_{3}(m,|b|)=\rmint_{0}^\infty dx\, \frac{J_{0}(x|b|)}{x}\arctan^2\Big(\frac{x}{2m}\Big)\,,\quad
    \bar{I}_{4}(m,|b|)=\rmint_0^\infty dx \frac{x^2}{x^2+m^2}J_{0}(x|b|)\arctan\Big(\frac{x}{m}\Big)\,,\\&
    I_{5}(m,|b|)=\rmint_0^{\infty} dx \,dz \,\frac{J_{0}(x|b|)}{z^2+m^2}\,\textrm{arctanh}\Big(\frac{2 x\,z}{x^2+z^2+m^2}\Big)\arctan\Big(\frac{z}{2m}\Big)\,,\\&
    \bar{I}_{5}(m,|b|)=\rmint_0^\infty dx  dz \frac{J_0(x|b|)}{z^2+m^2}\arctan\Big(\frac{z}{2m}\Big)\, \textrm{arctanh}\Big(\frac{2x\,z}{x^2+z^2}\Big)\,,\\&
    I_{6}(m,|b|)=\rmint_0^{\infty} dx \frac{J_{0}(x|b|)}{x}\arctan\Big(\frac{x}{2m}\Big)\arctan\Big(\frac{x}{m}\Big)\,,\\&
    \bar{I}_{6}(m,|b|)=\rmint_0^\infty dx \,dz\,\frac{J_{0}(z |b|)}{z^2}\arctan\Big(\frac{x}{2m}\Big)\textrm{arctanh}\Big(\frac{2x\,z}{m^2+x^2+z^2}\Big)\,,\\&
    \bar{\bar{I}}_6(m,|b|)=\rmint_0^\infty dx\,dz\,\frac{J_{0}(z |b|)}{z^2+m^2}\arctan\Big(\frac{x}{m}\Big)\textrm{arctanh}\Big(\frac{2x\,z}{m^2+x^2+z^2}\Big)\,,\\&
   I_{7}(m,|b|)= \rmint_{-\infty}^{\infty} d^3 z e^{-i \vec z\cdot \vec b}K_{0}\Big(|b||\sqrt{z_{(1)}^2+m^2}|\Big)\frac{1}{(\bar z^2+m^2)|\vec z|}\arctan\Big(\frac{|\vec z|}{2m}\Big)\,,\\&\bar{I}_{7}(m,|b|)=\rmint_{-\infty}^{\infty} d^3 z e^{-i \vec z\cdot \vec b} K_{0}\Big(|b||z_{(1)}|\Big)\frac{1}{(\vec{z}^2+m^2)|\vec z|}\arctan\Big(\frac{|\vec z|}{2m}\Big)
    \end{split} 
\end{align}
and $I_{4}(m,|b|)$ is defined in (\ref{4.37}). \textcolor{black}{As discussed previously, to the best of our knowledge these integrals do not possess any closed-form expression (except $I_5(m,|b|),\bar{I}_5(m,|b|), \bar{I}_6(m,|b|)$ and $\bar{\bar{I}}_6(m,|b|)$ as they can be done to some extent using method discussed around (\ref{4.25j}). They can be evaluated numerically and possess smooth behaviour w.r.t. $|b|\,.$ They also admit smooth massless limits on a case-by-case basis.} \par
Note that the expression for the total impulse will also have a contribution from the pure Einstein-Hilbert part. However, its result is well known in the literature \cite{Mogull:2020sak}. Hence, instead of reproducing it here, we only focused on the new contributions from various vertices involving the scalar field upto 2PM order. We have also shown the corresponding results for the massless case. The fall-off w.r.t $|b|$ is analogous to the gravitational case. \\\\
Before we end this section, although we have computed the impulse up to 2PM, we have not discussed its connection with the scattering amplitude. \textcolor{black}{We elaborate on this connection in the Appendix~(\ref{sec7}).}
\section{Computation of waveform}\label{sec5}
In this section, we will compute the frequency domain gravitational waveform due to different field configurations. In the wave zone, $f_{\varphi,h}(k)$ is defined by the one point function as follows,
 \hfsetfillcolor{gray!10}
\hfsetbordercolor{black}
\begin{equation}\label{e:bar}\begin{split}
\tikzmarkin[disable rounded corners=false]{p}(0.8,-0.58)(-0.05,0.55) &  f_{h}(k):=\frac{1}{4\pi m_p}\epsilon^{\mu}\epsilon^{\nu}k^2\langle h_{\mu\nu}(k)\rangle\Big|_{k^2\rightarrow 0}\,, \\ &
        f_{\varphi}(k) := (k^2-m^2)\frac{\langle \varphi(k)\rangle}{m_p}\Big|_{k^2\rightarrow m^2}\tikzmarkend{p}\\
        \end{split}
\end{equation}
\\
where $k^\mu=\Omega\, n^\mu$ describes the on-shell momentum of the graviton/scalar. As, on-shell graviton is massless then, $n_\mu n^\mu=0$  and for the scalar $n_\mu n^\mu=\frac{m^2}{\Omega^2}$. $n^\mu$ can be parameterized as, 
\begin{align}
    \begin{split}
n^\mu\Big|_{g,h}=\Big(1,\sqrt{1-\frac{m_s^2}{\Omega^2}}\boldsymbol{\hat{x}}\Big)
    \end{split}
\end{align}
where,
\begin{align}
    \begin{split}
\boldsymbol{\hat{x}}^{\mu}=e_{1}^{\mu}\cos\theta+\sin(\theta)(e_{2}^{\mu}\cos\phi+e_3^{\mu}\sin\phi),\,e_{i}^{\mu}=(0,\hat\xi_{i})\,.
    \end{split}
\end{align}
with $\hat \xi\in (\hat i,\hat j,\hat k)$. In the parametrization mentioned above, the $n$ can be written as follows,
\begin{align}
    \begin{split}
        n=(1,\cos(\theta),\sin(\theta)\cos(\phi),\sin(\theta)\sin(\phi))\,.
    \end{split}
\end{align}
First, we will compute the corrections to the waveform from the scalar degrees of freedom. For the sake of simplicity, we first take the scalar field to be massless, and in the next section, we will discuss the massive integrals and how the integrals get complicated in the massive case. It has been shown in \cite{Mogull:2020sak} that in a two-body scattering problem, the scattering amplitude with an on-shell external graviton (or scalar) is related to the WQFT correlator. In general, the scattering amplitude has the following schematic form.
\begin{align}
    \begin{split}
        \mathcal{M}[g,\varphi]\equiv \begin{minipage}[h]{0.12\linewidth}
	\scalebox{2.5}{\includegraphics[width=\linewidth]{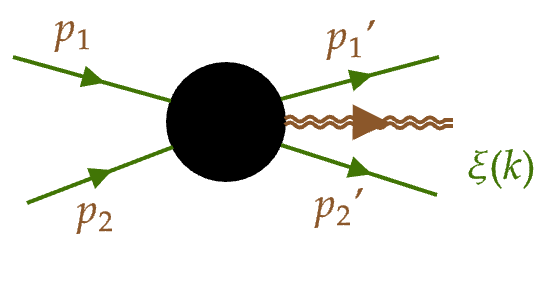}}
\end{minipage}\hspace{4 cm}
    \end{split}
\end{align}
Using the result established in \cite{Mogull:2020sak}, one can show that the connection between the S-matrix element $\langle\phi_1,\phi_2\Big |\,\mathcal{S}\,\Big |\phi_1,\phi_2,\chi\rangle$ with the WQFT correlator,
\begin{align}
    \begin{split}
     (k^2-m_{\chi}^2)   \langle \chi(k)\rangle \propto \rmint_{k_1,k_2} d\mu_{1,2}(k_i,k)\lim_{\hbar\rightarrow 0}\mathcal{M}[g,\varphi](p_i,p_i',k),\,\chi\in (h_{\mu\nu},\varphi).\label{5.5m}
    \end{split}
\end{align}
where, in \eqref{5.5m} $k_i=p_i-p_i'$ and $d\mu_{1,2}$ is the measure of integration depends on the interaction.\\
\subsection{Scalar waveform}
Now we will initiate the computation of the scalar waveform. We will compute it upto 2PM order. Next, we list all the diagrams contributing upto this order and evaluate the corresponding expression.\\\\
\subsubsection{Scalar waveform at 1PM}
In this subsection we list down the diagrams that contributes to the 1PM scalar waveform.\\\\
\textbullet $\,\,$ We start by considering the self-interaction vertices and their contribution to the waveform. First, we will consider the $\frac{\lambda_3}{\textcolor{black}{3!}} m_p\,\varphi^3$ vertex. Its contribution to the one-point function has the following form and comes at 1PM order \footnote{\textcolor{black}{Again, note that the combinatorial factor associated with this diagram is 3!. We have multiplied it by that. For the subsequent diagrams, we will also multiply by the suitable combinatorial factors from the beginning.}}. 
\begin{align}
    \begin{split} \label{wave0}
        k^2\Big\langle\varphi(k)\Big\rangle\equiv \begin{minipage}[h]{0.12\linewidth}
	\scalebox{1.45}{\includegraphics[width=\linewidth]{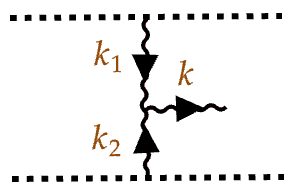}}
\end{minipage}\hspace{1.2 cm}=\lambda_3\Big(\frac{m_1s_1m_2s_2}{4m_p}\Big)\rmint \frac{d\mu_{1,2}(k)}{k_1^2k_2^2}\,.
    \end{split}
\end{align}
where the integral measure is given by,
\begin{align}
    \begin{split}
        \rmint d\mu_{1,2}(k)=\rmint_{k_1,k_2}e^{ik_1\cdot b_1}e^{ik_2\cdot b_2}\hat\delta(k_1\cdot v_1)\hat\delta(k_2\cdot v_2)\hat\delta^{(4)}(k_1+k_2-k)\,.
    \end{split}
\end{align}
Now, our focus is to compute the time domain waveform, which is the Fourier transform of the frequency domain waveform. Hence, we have to do a Fourier transformation of (\ref{wave0}). This gives the following, 
\begin{align}
    \begin{split}
        f_{\varphi}(x)&= \lambda_3\Big(\frac{m_1s_1m_2s_2}{4m_p^2}\Big)\rmint_{\Omega}e^{-ik\cdot x}\rmint_{k_i}e^{ik_1\cdot b_1}e^{ik_2\cdot b_2}\frac{\hat\delta(k_1\cdot v_1)\hat\delta(k_2\cdot v_2)}{k_1^2k_2^2}\hat\delta^{(4)}(k_1+k_2-k)\,,\\ &
     =   \lambda_3\Big(\frac{m_1s_1m_2s_2}{4m_p^2}\Big) \rmint_{\Omega}e^{-ik\cdot (x-b_1)}\underbrace{\rmint_{k_2}e^{-ik_2\cdot b}\frac{\hat\delta(k\cdot v_1-k_2\cdot v_1)\hat\delta(k_2\cdot v_2)}{k_2^2(k-k_2)^2}}_{J_{(0)}(k)}\,.
     \label{5.47 m}
    \end{split}
\end{align}
The integral in \eqref{5.47 m} can be evaluated using the results derived in \eqref{A.2} and \eqref{A.8}.  

\hfsetfillcolor{gray!10}
\hfsetbordercolor{black}
\begin{equation}\label{wave1}\begin{split}
\tikzmarkin[disable rounded corners=false]{u}(0.2,-0.60)(-0.3,0.80)  f_{\varphi}(x)&= \lambda_3\Big(\frac{m_1s_1m_2s_2}{4m_p^2}\Big) \rmint_{\Omega}e^{-ik\cdot (x-b_1)}\,J_{(0)}(k),\\ &
        =-\frac{\lambda_3}{(2\pi)^3}\Big(\frac{m_1s_1m_2s_2}{4m_p^2}\Big) \frac{|b|}{4\gamma}\rmint_{0}^{1}dy\,\frac{1}{\Bar\Delta(y)}\sqrt{\frac{l^2}{|b|^2\bar\Delta^2}+1},\,l:=n\cdot(x-b_1+y\,b)
\tikzmarkend{u}\\
        \end{split}
\end{equation}
\vspace{0.5 cm}\\
where, $\Bar\Delta$ is defined in \eqref{A.4} and \eqref{A.6}. We have restored the factors of $\pi$ that come from the integration measures and the delta functions. So at 1PM order, (\ref{wave1}) gives the entire contribution to the scalar waveform. Next, we will extend our study to a 2PM order.
\\\\
\subsubsection{Scalar waveform at 2PM}
In this subsection we intend to compute the 2PM scalar waveform coming from purely scalar sector and scalar-graviton interaction sector.\\\\
\textbullet $\,\,$ \textit{We start with the self-interacting vertex, namely,  $\frac{\lambda_4}{\textcolor{black}{4!}}\varphi^4$. We will show that it doesn't contribute to the waveform.} To show that, we first write down the one-point function for this case. 
\begin{align}
    \begin{split}\label{5.11 mn}
        k^2\Big\langle\varphi(k)\Big\rangle\equiv    \begin{minipage}[h]{0.12\linewidth}
	\vspace{4pt}
	\scalebox{1.7}{\includegraphics[width=\linewidth]{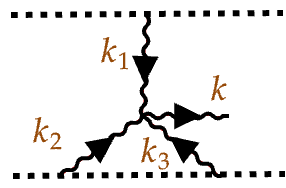}}
\end{minipage}\hspace{1.4 cm}=\rmint \frac{d\mu_{1,2}(k)}{k_1^2 k_2^2 k_3^2}\,.
    \end{split}
\end{align}
where, in this case, the integral measure has the following form
\begin{align}
    \begin{split}
       \rmint d\mu_{1,2}(k)=\rmint_{k_1,k_2,k_3}e^{i k_1\cdot b_1}e^{ik_2\cdot b_2}e^{i k_3 \cdot b_2}\hat\delta(k_1\cdot v_1)\hat\delta(k_2\cdot v_2)\hat\delta(k_3\cdot v_2)\hat\delta^{(4)}(k_1+k_2+k_3-k)\,.
    \end{split}
\end{align}
Hence, the corresponding contribution to the waveform has the following form,
\begin{align}
    \begin{split}
  \label{4PHI}      f_{\varphi}(x)&=\rmint_{\Omega}e^{-ik\cdot x}\rmint_{k_i}e^{ik_1\cdot b_1}e^{i k_2\cdot b_2}e^{i k_3\cdot b_2}\frac{\hat\delta(k_1\cdot v_1)\hat\delta(k_2\cdot v_2)\hat\delta(k_3\cdot v_2)}{k_1^2 k_2^2 k_3^2}\hat\delta^{(4)}(k_1+k_2+k_3-k)\,,\\ &
=\rmint_{\Omega}e^{-ik\cdot x}\rmint_{k_2,k_3}\frac{\hat\delta((k-k_2-k_3)\cdot v_1)\hat\delta(k_2\cdot v_2)\hat\delta{(k_3\cdot v_2)}}{k_2^2 k_3^2 (k-k_2-k_3)^2}\,e^{i (k-k_2-k_3)\cdot b_1}e^{i(k_2+k_3)\cdot b_2} \,,\\ &
\xrightarrow[]{q=k_2+k_3}\rmint_{\Omega}e^{-i k\cdot x}\rmint_{k_3,q} \frac{\hat\delta(k\cdot v_1-q\cdot v_1)\hat\delta(k_3\cdot v_2)\hat\delta(k_3\cdot v_2-q\cdot v_2)}{k_3^2 (q-k)^2(q-k_3)^2}e^{i (k-q)\cdot b_1}e^{i(q-k_3)\cdot b_2}e^{ik_3\cdot b_2}\,.
    \end{split}
\end{align}
We have the following integral to be solved, which can be systematically done by introducing $\alpha$ parametrization,
\begin{align}
    \begin{split}
        f_{\varphi}(x)&=\rmint_{\Omega}e^{-i k\cdot x+i k\cdot b_{1}}\rmint_{k_3,q} \frac{\hat\delta(k\cdot v_1-q\cdot v_1)\hat\delta(k_3\cdot v_2)\hat\delta(q\cdot v_2)}{k_3^2 \, (q-k_3)^2(q-k)^2}e^{-i q\cdot b}\,,\\ &
        =\int_{\Omega}e^{-i k\cdot(x-b_1)}\int_{0}^\infty d\alpha\, d\beta\, d\gamma\, \exp\Big[-i\alpha (q-k)^2-i\beta(q-k_3)^2-i\gamma k_3^2 \Big]\\ &
        \hspace{4 cm}\times  \hat\delta(k\cdot v_1-q\cdot v_1)\hat\delta(k_3\cdot v_2)\hat\delta(q\cdot v_2) e^{-iq\cdot b}\,,\\ &
=\int_{0}^{\infty}d\hat\alpha\,d\hat\beta\,d\hat\gamma \int_{\Omega}e^{-i\Omega \,n\cdot (x-b_1)}\int_{\vec k_3,\vec q}\hat\delta(\Omega\,n\cdot v_1+\vec q\cdot \vec v_1)\,\exp\Big[i\hat\alpha(\vec q^2-2\Omega \vec q\cdot \vec n)\\ &
\hspace{5 cm}+i\hat\beta (\vec q^2-2\vec q\cdot \vec k_3+\vec k_3^2)+{i\hat\gamma \vec k_3^2}\Big]e^{i\vec q\cdot \vec b}
    \end{split}
\end{align}
Now doing the $\Omega $ integral we have,
\begin{align}
    \begin{split}
        f_{\varphi}(x)&=\frac{1}{n\cdot v_1}\int_{0}^{\infty}d\hat\alpha\,d\hat\beta\,d\hat\gamma \int_{\vec k_3,\vec q} \exp\Big[i\frac{\vec q\cdot \vec v_1}{n\cdot v_1}n\cdot \tilde x\Big]\exp\Big[i(\hat\alpha+\hat\beta)\,\vec q^2+2i\hat\alpha \Big(\frac{\vec q\cdot \vec v_1}{n\cdot v_1}\Big)(\vec q\cdot \vec n)+i\vec q\cdot \vec b\Big]\\ &
        \hspace{5 cm}\times \exp\Big[i(\hat\beta+\hat\gamma)\vec k_3^2-2i\hat\beta \vec k_3\cdot \vec q\Big]\,,\Tilde{x}\equiv x-b_1
    \end{split}
\end{align}
Now we do the $\vec k_3$ integral by dimensional regularisation,
\begin{align}
    \begin{split}
        f_{\varphi}(x)&=\frac{1}{n\cdot v_1}\int_{0}^{\infty}d\hat\alpha\,d\hat\beta\,d\hat\gamma \int_{\vec q}\exp\Big[i\frac{\vec q\cdot \vec v_1}{n\cdot v_1}n\cdot \tilde x\Big]\exp\Big[i(\hat\alpha+\hat\beta)\,\vec q^2+2i\hat\alpha \Big(\frac{\vec q\cdot \vec v_1}{n\cdot v_1}\Big)(\vec q\cdot \vec n)+i\vec q\cdot \vec b\Big]\\ &
    \hspace{4 cm} \times e^{-i\frac{\pi}{4}+i\frac{\epsilon}{2}}\pi^{\frac{3}{2}-\epsilon}(\hat\beta+\hat\gamma)^{-\frac{3}{2}+\epsilon}\exp\Big(-i \frac{\hat\beta^2 \vec q^2}{\hat\beta +\hat\gamma}\Big)\\ &
    =\frac{1}{n\cdot v_1}e^{-i\frac{\pi}{4}+i\frac{\epsilon}{2}}\pi^{\frac{3}{2}-\epsilon}\int_{0}^{\infty}d\hat\alpha\,d\hat\beta\,d\hat\gamma \,(\hat\beta+\hat\gamma)^{-\frac{3}{2}+\epsilon}\int d^3q \exp\Big(i\lambda_1 \vec q^2+2i \lambda_2 (\vec q\cdot \vec v_1)(\vec q\cdot \vec n)+i\vec q\cdot \vec \lambda_3\Big)\label{5.16k}
    \end{split}
\end{align}
where,
\begin{align}
    \begin{split}
&\lambda_1(\hat\alpha,\hat\beta,\hat\gamma)=\hat\alpha+\hat\beta -\frac{\hat\beta^2}{\hat\beta +\hat\gamma}\\ &
\lambda_2(\hat\alpha,\hat\beta,\hat\gamma)=\frac{\hat\alpha }{n\cdot v_1}\\ &
\vec \lambda_3=\frac{n\cdot \tilde x}{n\cdot v_1}\vec v_1+\vec b\equiv \Upsilon(x)\vec v_1+\vec b
    \end{split}
\end{align}
The non-triviality comes in the $q$ integral due to the presence of $\vec {q} \cdot \vec n\,\vec {q} \cdot \vec {v_1} $ term. For the sake of simplicity, we choose $ n=(1,1,0,0)$ and the $\vec q$ integral can be done component wise,
\begin{align}
    \begin{split}
        f_{\varphi}(x)&=\frac{1}{n\cdot v_1}e^{-i\frac{\pi}{4}+i\frac{\epsilon}{2}}\pi^{\frac{3}{2}-\epsilon}\int_{0}^{\infty}d\hat\alpha\,d\hat\beta\,d\hat\gamma \,(\hat\beta+\hat\gamma)^{-\frac{3}{2}+\epsilon}\int dq_{(1)}\exp\Big(i(\lambda_1+2\lambda_2\gamma\beta\cos\epsilon)q_{(1)}^2+i q_{(1)}\Upsilon(x)\,\gamma\beta\Big)\\ &
        \hspace{3 cm }\times \int dq_{(2)}\exp\Big(i\lambda_1 q_{(2)}^2+iq_{(2)}b\Big)\int dq_{(3)} \exp(i\lambda_1 q_{(3)}^2)\\ & =  \frac{1}{n\cdot v_1}e^{-i\frac{\pi}{4}+i\frac{\epsilon}{2}}\pi^{\frac{3}{2}-\epsilon}\int_{0}^{\infty}d\hat\alpha\,d\hat\beta\,d\hat\gamma \,(\hat\beta+\hat\gamma)^{-\frac{3}{2}+\epsilon}\Bigg(\sqrt{\frac{\pi}{\lambda_1 +2\lambda_2\gamma\beta}}\\ &
        \hspace{3 cm}\times\exp\Big(i\frac{\pi}{4}-i\frac{\Upsilon(x)^2\gamma^2\beta^2}{4\lambda_1 +8\lambda_2\gamma\beta}\Big)\Bigg)\times \Bigg(\sqrt{\frac{\pi}{\lambda_1}}\exp\Big(i\frac{\pi}{4}-i\frac{b^2}{4\lambda_1}\Big)\Bigg)\times \sqrt{\frac{\pi}{\lambda_1}}\exp\Big(i\frac{\pi}{4}\Big)\\ &
        =\frac{\pi^3}{n\cdot v_1}\int_0^\infty d\hat\alpha \, d\hat\beta\,d\hat\gamma\,(\hat\beta+\hat\gamma)^{-3/2}\frac{1}{\lambda_1\sqrt{\lambda_1+2\lambda_2\gamma\beta}}\exp\Big(-i\frac{b^2}{4\lambda_1}-i\frac{\Upsilon(x)^2\gamma^2\beta^2}{4\lambda_1 +8\lambda_2\gamma\beta}\Big)\label{5.18k}
    \end{split}
\end{align}
The integral in \eqref{5.18k} does not admit any closed form. One has to perform the remaining integrals numerically and see whether there is any finite contribution.\par  However, we can extract some information by doing an asymptotic analysis,
\begin{align}
    \begin{split}
        \textrm{Reg.}|f_{\varphi}(x)|\le \frac{\pi^3}{n\cdot v_1}\textrm{Reg.}\int_0^\infty d\hat\alpha \, d\hat\beta\,d\hat\gamma\,(\hat\beta+\hat\gamma)^{-3/2}\frac{1}{\lambda_1\sqrt{\lambda_1+2\lambda_2\gamma\beta}}\label{5.19k}
    \end{split}
\end{align}
Now, expanding the integrand \eqref{5.19k} around two limit ($0,\infty$) we get,
\begin{align}
     \textrm{Reg.}|f_{\varphi}(x)|\le 0\implies  \textrm{Reg.}|f_{\varphi}(x)|\to 0
\end{align}
To make this statement more concrete, we make an analysis by using the method of region in Appendix~(\ref{App2}). But a more precise numerical analysis is required to see whether there is any finite contribution from this integral, which we leave for future investigations.  In a similar fashion, one can, in principle, compute the waveform corresponding to $\lambda_3h\varphi^3$ vertex. 
\\\\
\textbullet $\,\,$ The simplest 2PM contribution comes from quadratic scalar field coupling in the worldline.
\begin{align}
    \begin{split}
        k^2\Big\langle\varphi(k)\Big\rangle\equiv \begin{minipage}[h]{0.12\linewidth}
	\scalebox{1.6}{\includegraphics[width=\linewidth]{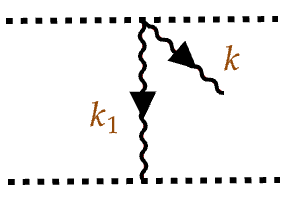}}
\end{minipage}\hspace{1 cm}=\Big(\frac{m_1g_1}{2m_p^2}\Big)\times \Big(\frac{m_2s_2}{2m_p}\Big)\rmint d\mu_{1,2}(k)\frac{1}{k_1^2}\label{5.7mm}
    \end{split}
\end{align}
where the integral measure has the following form,
\begin{align}
    \begin{split}
        d\mu_{1,2}(k)=\rmint_{k_1}e^{-ik_1\cdot b}e^{ik\cdot b_1}\hat\delta(k_1\cdot v_2)\hat\delta(k\cdot v_1-k_1\cdot v_1)\,.
    \end{split}
\end{align}
Therefore, the  time domain waveform has the following form,
\begin{align}
    \begin{split}
        f_{\varphi}(x)&=\Big(\frac{m_1g_1}{2m_p^2}\Big)\times \Big(\frac{m_2s_2}{2m_p}\Big)\frac{1}{n\cdot v_1}\rmint_{\Omega} e^{-ik\cdot (x-b_1)}\rmint_{k_1} e^{-i k_1\cdot b}\frac{\hat\delta(k_1\cdot v_2)\hat\delta\Big(\Omega-\frac{k_1\cdot v_1}{n\cdot v_1}\Big)}{k_1^2}\,,\\ &
         =  \Big(\frac{m_1g_1}{2m_p^2}\Big)\times \Big(\frac{m_2s_2}{2m_p}\Big)\frac{1}{n\cdot v_1}\rmint_{k_1}e^{-ik_1\cdot w_1}\frac{\hat\delta(k_1\cdot v_2)}{k_1^2},\,\textrm{with}, \, w_1\equiv \frac{n\cdot (x-b_1)}{n\cdot v_1}v_1+b,\nonumber
         \end{split}
\end{align}
\begin{align}
\begin{split}
  \hspace{-3 cm}  & 
      = -\Big(\frac{m_1g_1}{2m_p^2}\Big)\times \Big(\frac{m_2s_2}{2m_p}\Big)\frac{1}{n\cdot v_1}\frac{1}{4\pi|\vec w_1|}\,.
    \end{split}
\end{align}
Therefore the final result of the integrals looks,\\\\
\hfsetfillcolor{gray!10}
\hfsetbordercolor{black}
\begin{equation}\label{e:ba}\begin{split}
\tikzmarkin[disable rounded corners=false]{q}(0.2,-0.6)(-0.1,1) f_{\varphi}(x)&=-\Big(\frac{m_1g_1}{2m_p^2}\Big)\times \Big(\frac{m_2s_2}{2m_p}\Big)\frac{1}{n\cdot v_1}\frac{1}{4\pi|\vec w_1|}\,.
\tikzmarkend{q}\\
        \end{split}
\end{equation}\\
\textbullet $\,\,$ Another 2PM contribution comes from the following:
\begin{align}
    \begin{split}
        k^2\Big\langle\varphi(k)\Big\rangle\Big|_{k^2\rightarrow 0}\equiv \begin{minipage}[h]{0.12\linewidth}
	\scalebox{1.5}{\includegraphics[width=\linewidth]{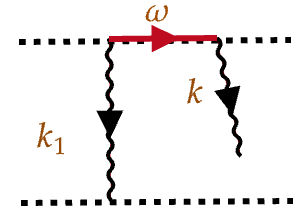}}
\end{minipage}\hspace{1 cm}=m_1\Big(\frac{s_1}{2m_p}\Big)^2 \Big(\frac{m_2s_2}{2m_p}\Big)\rmint d\mu_{1,2}(k)\frac{\{2\omega (v_1)_{\rho}-(k_1)_{\rho}\}\{-2\omega (v_1)_{\rho}+ k_{\rho}\}}{\omega^2 k_1^2}\label{5.10 m}
    \end{split}
\end{align}
where, the measure $d\mu_{1,2}(k)$ has the following form,
\begin{align}
    \begin{split}
        \rmint d\mu_{1,2}(k)=\rmint_{k_1,\omega}e^{i(k-k_1)\cdot b_1}e^{i k_1\cdot b_2}\hat\delta(k_1\cdot v_1-\omega)\hat\delta(k \cdot v_1-\omega)\hat\delta(k_1\cdot v_2)\,.
    \end{split}
\end{align}
Now it's contribution to the time domain waveform is,
\begin{align}
    \begin{split}
        f_{\varphi}(x)& = -m_1\Big(\frac{s_1}{2m_p}\Big)^2 \Big(\frac{m_2s_2}{2m_p}\Big) \rmint_{\Omega} e^{-ik\cdot x} \rmint d\mu_{(a)}(k) \frac{k\cdot k_1}{{\omega}^2\,k_1^2}\,,\\ &
        =-m_1\Big(\frac{s_1}{2m_p}\Big)^2 \Big(\frac{m_2s_2}{2m_p}\Big)\rmint_{\Omega} e^{-ik\cdot x}\rmint_{k_1} e^{-ik_1\cdot b}e^{i k\cdot b_1}\frac{k\cdot k_1\hat\delta(k_1\cdot v_2)}{k_1^2\,(k_1\cdot v_1+i\epsilon)^2}\hat\delta[(k-k_1)\cdot v_1]\,.
    \end{split}
\end{align}
Therefore, the whole integral can be written as,
\begin{align}
\begin{split}
      f_{\varphi}(x)&= -m_1\Big(\frac{s_1}{2m_p}\Big)^2 \Big(\frac{m_2s_2}{2m_p}\Big)\rmint_{\Omega}e^{-ik\cdot (x-b_1)}\rmint_{\omega,k_1}e^{-i k_1\cdot b}\hat{\delta}(k_1\cdot v_2)\hat{\delta}(\Omega\, n\cdot v_1-\omega)\hat\delta(k_1\cdot v_1-\omega)\frac{\Omega(n\cdot k_1)}{\omega^2 k_1^2}\,,\\ &
       = -m_1\Big(\frac{s_1}{2m_p}\Big)^2 \Big(\frac{m_2s_2}{2m_p}\Big)\frac{1}{(n\cdot v_1)^2}\rmint_{\Omega}e^{-ik\cdot (x-b_1)}\rmint_{k_1}e^{-ik_1\cdot b}\hat\delta(k_1\cdot v_2)\hat{\delta}(\Omega-\frac{k_1\cdot v_1}{n\cdot v_1})\frac{n\cdot k_1}{(k_1\cdot v_1+i\epsilon)k_1^2}\,,\\ &
     = -m_1\Big(\frac{s_1}{2m_p}\Big)^2 \Big(\frac{m_2s_2}{2m_p}\Big)\frac{n^{\mu}}{(n\cdot v_1)^2}\rmint_{k_1}e^{-i k_1\cdot w_1}\hat{\delta}(k_1\cdot v_2)\frac{k_{1\mu}}{(k_1\cdot v_1+i\epsilon)k_1^2},\,\, w_1\equiv\frac{n\cdot (x-b_1)}{n\cdot v_1}v_1+b \,.\label{7.6}
      \end{split}
\end{align}
The integral in (\ref{7.6}) can be easily done from the frame of the second particle as follows,
\begin{align}
    \begin{split}
        \mathcal{J}^{\mu}:=\rmint_{k_1}e^{-i k_1\cdot w_1}\frac{k_1^{\mu}\,\hat\delta(k_1\cdot v_2)}{(k_1\cdot v_1+i\epsilon)k_1^2}&=-i\rmint d\tau \,\theta(\tau)\,\rmint_{k_1}\hat{\delta}(k_1\cdot v_2) e^{-ik_1\cdot (w_1-\tau\, v_1)}\frac{k_1^{\mu}}{k_1^2}\,,\\ &
        =-i\rmint d\tau\,\theta(\tau)\rmint_{k_1}\hat\delta{(k_1^{0})} \exp[{-ik_1\cdot \underbrace{(w_1-\tau\, v_1)}_{\tilde{w}_1}}]\frac{k_1^{\mu}}{k_1^2}\,,\\ &
        \xrightarrow[]{k_1^0\rightarrow 0}-\rmint d\tau\, \theta(\tau)\rmint_{\vec k_1}\frac{k_1^i}{-\vec k_1^2}\,e^{i \vec{k}_1\cdot \vec {\tilde w}_1}\,,\\ &
        =-\rmint_{-\infty}^{\infty}d\tau \, \theta(\tau)\frac{(\vec w_1-\tau\,\vec v_1)^{i}}{|\vec w_1-\tau\,\vec v_1|^{3}}\,.
    \end{split}
\end{align}
Now form (\ref{7.6}), it is clear that we need to compute the following quantity,
\begin{align}
    \begin{split}
        n_{\mu}\mathcal{J}^{\mu}\rightarrow n_{i}\mathcal{J}^{i}=\rmint_{-\infty}^{\infty}d\tau\, \theta(\tau)\frac{\vec n\cdot (\vec w_1-\tau\, \vec v_1)}{|\vec w_1-\tau \vec v_1|^3}\,.
    \end{split}
\end{align}
To proceed further, we have to properly parameterize the impact parameter $b_1,\,b_2$. As we are in the frame of the second particle, it is convenient to choose $b_1=(0,0,b,0)$ and $b_2=0$.
\begin{align}
    \begin{split}
       \vec n\cdot (\vec w_1-\tau \vec v_1)&\rightarrow \Big[\frac{n\cdot (x-b_1)}{n\cdot v_1}-\tau\Big]\vec n\cdot \vec v_1+\vec n\cdot \vec b\\ &
       =(u_1-\tau)\gamma\tilde\beta+\chi,\,\,\text{with},\chi:=b\sin\theta\cos\varphi,\,\tilde\beta:=\beta \cos\theta
    \end{split}
\end{align}
and,
\begin{align}
    \begin{split}
        |\vec w_1-\tau \vec v_1|^3\rightarrow (\gamma^2-1)^{3/2}\Big[\tau^2-u_1^2-2\tau u_1+\frac{|b|^2}{\gamma^2-1}\Big]^{3/2}.
    \end{split}
\end{align}
Hence, 
\begin{align}
    \begin{split}
    n_{\mu}\mathcal{J}^{\mu}=\frac{\gamma  \tilde{\beta } \left(2 u_1^2-\frac{|b|^2}{\left(\gamma ^2-1\right)^2}\right)+\chi  \left(\sqrt{\frac{|b|^2}{\left(\gamma ^2-1\right)^2}-u_1^2}+u_1\right)}{(\gamma^2-1)^{3/2}\left(\frac{|b|^2}{\left(\gamma ^2-1\right)^2}-2 u_1^2\right) \sqrt{\frac{|b|^2}{\left(\gamma ^2-1\right)^2}-u_1^2}}
    \end{split}
\end{align}
where, we define, $u_i=\frac{n\cdot (x-b_1)}{n\cdot v_i}$.Therefore, restoring the factors of $\pi$ from delta functions and integration measures, the contribution to the waveform from the particular diagram in  (\ref{5.10 m}) has the following,\\
\hfsetfillcolor{gray!10}
\hfsetbordercolor{black}
\begin{align}\begin{split}\label{5.24 mmm}
\tikzmarkin[disable rounded corners=false]{r}(0.25,-0.75)(-0.3,1.3) f_{\varphi}(x)= -\frac{m_1}{(2\pi)^2}\Big(\frac{s_1}{2m_p}\Big)^2 \Big(\frac{m_2s_2}{2m_p}\Big)\frac{1}{\gamma^2(1-\tilde\beta)^2(\gamma^2-1)^{3/2}}\frac{\gamma  \tilde{\beta } \left(2 u_1^2-\frac{|b|^2}{\left(\gamma ^2-1\right)^2}\right)+\chi  \left(\sqrt{\frac{|b|^2}{\left(\gamma ^2-1\right)^2}-u_1^2}+u_1\right)}{\left(\frac{|b|^2}{\left(\gamma ^2-1\right)^2}-2 u_1^2\right) \sqrt{\frac{|b|^2}{\left(\gamma ^2-1\right)^2}-u_1^2}}\,.
\tikzmarkend{r}\\
 \end{split}
\end{align}\\
\textbullet $\,\,$ Finally, another 3-point scalar-graviton interaction vertex involving derivative contributes to 2PM radiation. It is of the following form: $h^{\mu\nu}\partial_{\mu}\varphi\partial_{\nu} \varphi$  As we are computing the one-point function of the scalar field, it would be useful to partially integrate over the interaction Lagrangian and separate one of the scalar fields as,
\begin{align}
    \begin{split}
        S_{\textrm{int.}}=\frac{1}{m_p}\rmint d^4x \,h^{\mu\nu} \partial_{\mu}\varphi\partial_{\nu}\varphi=-\frac{1}{m_p}\rmint d^4 x \Big[\underbrace{h^{\mu\nu}\partial_{\mu}\partial_{\nu}\varphi}_{\textrm{Term I}}+\underbrace{\partial_{\mu}h^{\mu\nu}\partial_{\nu}\varphi}_{\textrm{Terrm II}}\Big]\,\varphi.\label{5.22 m}
    \end{split}
\end{align}
Now, the contribution to the scalar one-point function from Term I is given by,
\begin{align}
    \begin{split}
        k^2\Big\langle \varphi(k)\Big\rangle\equiv \begin{minipage}[h]{0.12\linewidth}
	\scalebox{1.6}{\includegraphics[width=\linewidth]{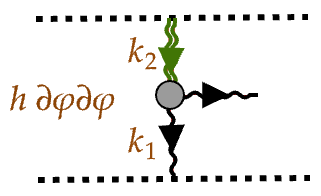}}
\end{minipage}\hspace{1 cm}=\Big(\frac{m_1m_2s_2}{2m_p^3}\Big)\rmint d\mu_{1,2}(k)\frac{v_1^\alpha v_1^\beta\,P_{\mu\nu;\alpha\beta}k_2^\mu k_2^\nu}{k_1^2k_2^2}\label{5.23 m}
    \end{split}
\end{align}
where the integral measure has the following form,
\begin{align}
    \begin{split}
     \rmint   d\mu_{1,2}(k)=\rmint_{k_1,k_2}e^{ik_1\cdot b_1}e^{ik_2 \cdot b_2}\hat\delta(k_1\cdot v_1)\hat\delta(k_2\cdot v_2)\hat\delta^{(4)}(k_1+k_2-k)\,.
    \end{split}
\end{align}
Hence, the time domain waveform has the following form,
\begin{align}
    \begin{split}
        f_{\varphi}(x)\Big|_1=\Big(\frac{m_1m_2s_2}{2m_p^3}\Big)\rmint_{\Omega}e^{-ik\cdot x}\rmint_{k_1,k_2}e^{ik_1\cdot b_1}e^{ik_2\cdot b_2}\hat\delta(k_1\cdot v_1)v_1^{\alpha}v_1^{\beta}\hat\delta(k_2\cdot v_2)\frac{k_2^{\mu}k_2^{\nu}P_{\mu\nu,\alpha\beta}}{k_1^2k_2^2}\delta^{(4)}(k_1+k_2-k)\,.\label{5.25mmm}
    \end{split}
\end{align}
The integral in \eqref{5.25mmm} can be done using the partial fraction approach, where we can separate the denominator with undefined momentum as,
\begin{align}
    \begin{split}
        \frac{1}{k_1^2 k_2^2}=-\frac{1}{2}\frac{1}{k_1^2(k_1\cdot k)}-\frac{1}{2}\frac{1}{k_2^2(k_2\cdot k)}\,.
    \end{split}
\end{align}
Therefore, the waveform can be reduced to two independent momentum integrals,
\begin{align}
    \begin{split} \label{e:::}
        f_{\varphi}(x)\Big|_{1}=\Big(\frac{m_1m_2s_2}{4 m_p^3}\Big)\Big(f_{\varphi}(x)\Big|_{1}^{(1)}+f_{\varphi}(x)\Big|_{1}^{(2)}\Big)\,.
    \end{split}
\end{align}
where,
\begin{align}
    \begin{split} \label{5.38nm}
        f_{\varphi}(x)\Big|_{1}^{(1)}&=v_1^{\alpha}v_1^{\beta}P_{\mu\nu;\alpha\beta}\rmint_{\Omega}e^{-ik\cdot x}\rmint_{k_1,k_2}e^{ik_1\cdot b_1}e^{ik_2\cdot b_2}\frac{k_2^\mu k_2^\nu\hat\delta(k_1\cdot v_1)\hat\delta(k_2\cdot v_2)}{k_1^2(k_1\cdot k)}\delta^{(4)}(k_1+k_2-k)\,,\\ &
=v_1^{\alpha}v_1^{\beta}P_{\mu\nu;\alpha\beta}\rmint_{\Omega}e^{-ik\cdot (x-b_2)}\rmint_{k_1}e^{ik_1\cdot b}\frac{\hat\delta(k_1\cdot v_1)\hat\delta(\Omega-k_1\cdot v_2)}{k_1^2(k_1\cdot n)\Omega}(k-k_1)^{\mu}(k-k_1)^{\nu}\,,\\ &
=v_1^{\alpha}v_1^{\beta}P_{\mu\nu;\alpha\beta}\rmint_{k_1}e^{-i k_1\cdot \tilde w_1}\frac{\hat\delta(k_1\cdot v_1)}{k_1^2 (k_1\cdot n)(k_1\cdot v_2)}\Big[n^\mu n^\nu (k_1\cdot v_2)^2-2 (k_1\cdot v_2)k_1^{(\mu}n^{\nu)}+k_1^{\mu}k_1^{\nu}\Big]\,.
    \end{split}
\end{align}
The integral of concern is the following,
\begin{align}
    \begin{split}
        I^{\mu\nu}&=\rmint_{k_1}\frac{e^{-ik_1\cdot \tilde w_1}\hat\delta(k_1\cdot v_1)}{k_1^2(k_1\cdot n)(k_1\cdot v_2)}k_1^{\mu}k_1^{\nu},\, \tilde w_1=\frac{n\cdot (x-b_2)}{n\cdot v_2}v_2-b\,.\label{5.29mmm}
    \end{split}
    \end{align}
{\color{black}One can see that because of the delta function constraint, one can replace  $\tilde w_1$ with $\hat w_2:=-b+u_2 (v_2-\gamma v_1)$. Also, $I^{\mu\nu}$ is orthogonal to $\hat w_2$ and $v_1$. Therefore it can be written in term of a basis plane orthogonal to  $\hat w_2$ and $v_1$.} $I^{\mu\nu}$ is orthogonal to $\hat w_2$ and $v_1$.  Therefore,
    \begin{align}
        \begin{split}
I^{\mu\nu}&=\Pi^{\mu}_{\alpha}\,\Pi^{\nu}_{\beta}\Big[c_{vv} v_2^{\alpha}v_2^{\beta}+c_{nn} n^{\alpha}n^{\beta}+c_{nv} v_2^{(\alpha}n^{\beta)}\Big]\,.
        \end{split}
    \end{align}
One can notice that $I^{\mu\nu}$ is symmetric under the exchange of $v_2 \leftrightarrow n$ implies $c_{nn}=c_{vv}$. Now, the integral can be done using the integral reduction technique by Passarino and Veltman, redefining the basis and writing the following ansatz, 
\begin{align}
        \begin{split}
\label{Ansatz2}I^{\mu\nu}&=c_{a}\Pi_{1}^{\mu\nu}+ 2c_{b}\Big(\Pi_{1}.v_{2}\Big)^{(\mu}\Big(\Pi_{1}.n\Big)^{\nu)}
\end{split}
    \end{align}
where $c_{a}$ and $c_{b}$ are two new constants and $\Pi_{1}^{\mu\nu}= |w_{1}|^{2}\,P_{1}^{\mu\nu}\,+ w_{1}^{\mu}\,w_{1}^{\nu}$ is the projection operator orthogonal to $w_{1}^{\mu}$ and $v_{1}^{\mu}$. One can evaluate the constants from the following two equations:
\begin{align}
\label{NV2}  &  n\cdot I\cdot v_2=\rmint_{k_1}e^{-ik_1\cdot \hat w_1}\frac{\hat\delta(k_1\cdot v_1)}{k_1^2}= c_{a}\Big(n\cdot \Pi\cdot v_{2}\Big)+ c_{b}\,\Big[\Big(n\cdot \Pi\cdot v_{2}\Big)\,\Big(v_{2}\cdot \Pi\cdot n\Big)+ \Big(v_{2}\cdot \Pi\cdot v_{2}\Big)\,\Big(n\cdot \Pi\cdot n\Big)\Big]\,,\\ &
\label{NIN}  n\cdot I \cdot n= n_\mu \rmint_{k_1}e^{-ik_1\cdot \hat w_2}\frac{k_1^\mu\hat\delta(k_1\cdot v_1)}{k_1^2 (k_1\cdot v_2)}= c_{a}\Big(n\cdot \Pi\cdot n\Big)+ 2c_{b}\Big(n\cdot \Pi\cdot v_{2}\Big)\,\Big(n\cdot \Pi\cdot n\Big)\,.
\end{align}
The first equality of (\ref{NIN}) gives,
\begin{align}
\begin{split}
   n_\mu \rmint_{k_1}e^{-ik_1\cdot \hat w_2}\frac{k_1^\mu\hat\delta(k_1\cdot v_1)}{k_1^2 (k_1\cdot v_2)}&\to -i n_\mu\, \rmint d\tau\,\theta(\tau)\rmint_{ k_1} e^{-i k_1\cdot (\hat{w}_2-\tau v_2)}\frac{{k}_1^\mu}{k_1^2}\hat\delta(k_1\cdot v_1)\,,\\ &
   =-i\, \grave{n}_i\rmint d\tau\,\theta(\tau)
   \rmint_{\bar k_1}e^{i \bar k_1\cdot \grave{\underbar{$w$}}_2}\frac{\bar k_1^i}{-\bar k_1^2}\,,\\ &
   =-\grave n_i\rmint d\tau \,\theta(\tau)\frac{\grave{\underbar{$w$}}(\tau)^i}{|\grave{\underbar{$w$}}(\tau)|^3},\,\grave{\underbar{$w$}}\equiv \hat{w}_2-\tau v_2.
   \end{split}
\end{align}
where,
\begin{align}
    \begin{split}
        \grave n_i\equiv \Big\{\gamma(n^{(1)}-\beta n^{(0)}),n^{(2)},n^{(3)}\Big\}\,.
    \end{split}
\end{align}
Comparing with the second equality of (\ref{NIN}), we get,\\
\begin{align}\begin{split}\label{m:}
  &c_a\to -\frac{-(n\cdot \Pi\cdot v_2) (n\cdot I\cdot n) (v_2\cdot \Pi\cdot n)+2(n\cdot \Pi\cdot v_2)(n\cdot \Pi\cdot n) (n\cdot I\cdot v_2)-(v_2\cdot \Pi\cdot v_2) (n\cdot \Pi\cdot n)(n\cdot I \cdot n)}{-2 (n\cdot \Pi\cdot v_2)^2+(n\cdot \Pi\cdot v_2) (v_2\cdot \Pi\cdot n)+(v_2\cdot \Pi\cdot v_2) (n\cdot \Pi\cdot n)},\\&
 c_b\to \frac{(n\cdot \Pi\cdot n) (n\cdot I\cdot v_2)-(n\cdot \Pi\cdot v_2) (n\cdot I\cdot n)}{(n \cdot \Pi\cdot n) \left(-2 (n\cdot \Pi\cdot v_2)^2+(n\cdot \Pi\cdot v_2) (v_2\cdot \Pi\cdot n)+(v_2\cdot \Pi\cdot v_2) (n\cdot \Pi\cdot n)\right)}.\\
\end{split}
\end{align}\\
Next, we compute the integral of the R.H.S of the first equality of (\ref{NV2}) to get,
\begin{eqnarray}
\label{CAINT} \rmint_{k_1}e^{-ik_1\cdot \hat w_1}\frac{\hat\delta(k_1\cdot v_1)}{k_1^2} =-\frac{1}{4\pi}\frac{1}{|\grave{w}_2|}
\end{eqnarray}
where $\grave{w}_2$ is a three vector defined as,
\begin{align}
    \grave{w}_2=\Big\{\gamma(\hat{w}_2^{(1)}-\beta \hat{w}_2^{(0)}),\hat{w}_2^{(2)},\hat{w}_2^{(3)}\Big\}\,.
\end{align}
Thus, the integral $I^{\mu\nu}$ can be finally written in a concise form from the ansatz (\ref{Ansatz2}) as,
\begin{align}
\begin{split} 
\hspace{-1 cm}\label{IMUNU}I^{\mu\nu} =-\Big[\frac{-(n\cdot \Pi\cdot v_2) (n\cdot I\cdot n) (v_2\cdot \Pi\cdot n)+2(n\cdot \Pi\cdot v_2)(n\cdot \Pi\cdot n) (n\cdot I\cdot v_2)-(v_2\cdot \Pi\cdot v_2) (n\cdot \Pi\cdot n)(n\cdot I \cdot n)}{-2 (n\cdot \Pi\cdot v_2)^2+(n\cdot \Pi\cdot v_2) (v_2\cdot \Pi\cdot n)+(v_2\cdot \Pi\cdot v_2) (n\cdot \Pi\cdot n)}\Big]\,\Pi^{\mu\nu} \\
+ 2\,\Big[\frac{(n\cdot \Pi\cdot n) (n\cdot I\cdot v_2)-(n\cdot \Pi\cdot v_2) (n\cdot I\cdot n)}{(n \cdot \Pi\cdot n) \left(-2 (n\cdot \Pi\cdot v_2)^2+(n\cdot \Pi\cdot v_2) (v_2\cdot \Pi\cdot n)+(v_2\cdot \Pi\cdot v_2) (n\cdot \Pi\cdot n)\right)}\Big]\Big(\Pi \cdot v_{2}\Big)^{(\mu}\Big(\Pi\cdot n\Big)^{\nu)}\,.
\end{split}
\end{align}
Once we have the closed-form expression for $I^{\mu\nu}\,,$ one can write down the exact expression for the first part of the waveform after restoring the factor of $\pi$.
\hfsetfillcolor{gray!10}
\hfsetbordercolor{black}
\begin{align}\begin{split}\label{e:}
\tikzmarkin[disable rounded corners=false]{ss}(0.1,-0.5)(-0.2,0.73)
f_{\varphi}(x)\Big|_{1}^{(1)}=-\,\frac{v_1^\alpha v_1^\beta}{4\pi^2}P_{\mu\nu;\alpha\beta}\Big[n^{\mu}n^\nu I^{\rho\sigma}v_{2\rho}v_{2\sigma}-2v_{2\rho}I^{\rho(\mu}n^{\nu)}+I^{\mu\nu}\Big].\tikzmarkend{ss}\\
\end{split}
\end{align}
Now we focus on the other part.
\begin{align}
    \begin{split}
        f_{\varphi}(x)\Big|_{1}^{(2)}&=-\,v_1^\alpha v_1^\beta P_{\mu\nu;\alpha\beta}\rmint_{\Omega}e^{-ik\cdot x}\rmint_{k_1,k_2}e^{ik_1\cdot b_1}e^{ik_2\cdot b_2}\frac{k_2^\mu k_2^\nu \hat\delta(k_1\cdot v_1)\hat\delta(k_2\cdot v_2)}{k_2^2 (k_2\cdot k)}\hat\delta^{(4)}(k_1+k_2-k)\,,\\ &
        =-\,v_1^\alpha v_1^\beta P_{\mu\nu;\alpha\beta}\rmint_{\Omega}e^{-i \Omega \,n \cdot x}\rmint_{k_2}e^{i (k-k_2)\cdot b_1} e^{ik_2\cdot b_2} \frac{k_2^\mu k_2^\nu \,\hat\delta(k_2\cdot v_2)\hat\delta(\Omega\,n\cdot v_1-k_2\cdot v_1)}{k_2^2 (k_2\cdot n)\Omega}\,,\\ &
        =-\,v_1^\alpha v_1^\beta P_{\mu\nu;\alpha\beta} \underbrace{\rmint_{k_2} e^{-i k_2 \cdot w_1}k_2^\mu k_2^\nu\, \frac{\hat\delta(k_2\cdot v_2)}{k_2^2(k_2\cdot n)(k_2\cdot v_1)}}_{\bar{I}^{\mu\nu}} \textrm{with},\,w_1\equiv \frac{n\cdot (x-b_1)}{n\cdot v_1}v_1+b\,.\label{5.43mm}
    \end{split}
\end{align}
The integral in \eqref{5.43mm} can be similar to \eqref{5.29mmm} and is given by,
\begin{align}
    \begin{split} \label{e::}
\tikzmarkin[disable rounded corners=false]{s}(0.2,-0.50)(-0.1,0.60)f_{\varphi}(x)\Big|_{1}^{(2)}=-\,v_1^{\alpha}v_1^\beta P_{\mu\nu;\alpha\beta} \bar I^{\mu\nu}.\tikzmarkend{s}\\
    \end{split}
\end{align}
where $\bar I^{\mu\nu}$ is defined (\ref{5.43mm}) and can be calculated as \eqref{5.29mmm}. So adding (\ref{e:}) and  (\ref{e::}) we get the full expression for $f_{\phi}(x)\Big|_1$ mentioned in (\ref{e:::}).\\\\
Now, we deal with the Term II of the interaction Lagrangian of \eqref{5.22 m}. The corresponding contribution to the waveform is given by,
\begin{align}
    \begin{split} \label{ex6}
        f_{\varphi}(x)\Big|_{2}=\Big(\frac{m_1 m_2 s_2}{2m_p^3}\Big)v_1^\alpha v_1^\beta P_{\mu\nu;\alpha\beta}\rmint_{\Omega} e^{-ik\cdot x}\rmint_{k_1,k_2}e^{ik_1\cdot b_1}e^{ik_2\cdot b_2}\frac{k_1^\mu k_2^\nu\hat\delta(k_1\cdot v_1)\hat\delta(k_2\cdot v_2)}{k_1^2 k_2^2}\hat\delta^{(4)}(k_1+k_2-k)\,.
    \end{split}
\end{align}
Now, doing the integration over $k_1$ using the delta function, we will get,
\begin{align}
    \begin{split}
         f_{\varphi}(x)\Big|_{2}=\Big(\frac{m_1 m_2 s_2}{2m_p^3}\Big)v_1^\alpha v_1^\beta P_{\mu\nu;\alpha\beta}\rmint_{\Omega}e^{-ik\cdot (x-b_1)}\rmint_{k_2}e^{-i k_2 \cdot b}\frac{(k-k_2)^\mu \,k_2^\nu}{k_2^2\,(k-k_2)^2}\hat\delta(k\cdot v_1-k_2\cdot v_1)\hat\delta(k_2\cdot v_2)\,.\label{5.46 m}
    \end{split}
\end{align}
Again this can be split into two parts and can be dealt with separately. 
\begin{align}
    \begin{split} \label{ex3}
    f_{\varphi}(x)\Big|_{2}=\Big(\frac{m_1 m_2 s_2}{2m_p^3}\Big)\Big(f_{\varphi}(x)\Big|^{(1)}_{2}+f_{\varphi}(x)\Big|^{(2)}_{2}\Big)\,.
     \end{split}
\end{align}
The first part gives,
\begin{align}
    \begin{split}
        f_{\varphi}(x)\Big|_{2}^{(1)}=\underbrace{(v_1\cdot P\cdot v_1)_{\mu\nu}\,n^{\mu}}_{T_{\nu}}\rmint_{\Omega}e^{-ik\cdot (x-b_1)}\,\Omega\underbrace{\rmint_{k_2}\frac{k_2^\nu}{k_2^2(k-k_2)^2}\hat\delta(k\cdot v_1-k_2\cdot v_1)\hat\delta(k_2\cdot v_2)}_{J_{(1)}^\nu}\,.\label{5.57 m}
    \end{split}
\end{align}
The integral in \eqref{5.57 m} can be done using the results in \eqref{A.9 m}. We know that $J_{(1)}^{\mu}$ has two parts. We first concentrate on the first part, i.e. $\mathcal{J}^{\nu}$ in \eqref{A.9 m}. Therefore, the first term of $f_{\varphi}(x)\Big|^{(1)}_{2}$ takes the following form:
\begin{align}
    \begin{split} \label{ex}
        \widetilde {f_{\varphi}(x)}\Big|_{2}^{(1)}&=T_{\nu}\rmint_{\Omega}e^{-ik\cdot (x-b_1)}\Omega\,\mathcal{J}^{\nu}(\Omega)\,,\\ &
        =i\,\frac{\partial}{\partial l}\,\Big(f_1+f_2+f_3\Big)
    \end{split}
\end{align}
where, $f_i$'s are defined in \eqref{A.19 m}. Apart from the $\mathcal{J}^{\nu}$ part we have another part in form of $J_{(1)}^\nu$, as shown in (\ref{A.9 m}), which gives,
\begin{align}
    \begin{split} \label{ex1}
      \widetilde{ \widetilde{f_{\varphi}(x)}}\Big|_{2}^{(1)}&= \frac{|b| \,n^\nu}{\gamma}T_{\nu}\rmint_{\Omega}e^{-ik\cdot (x-b_1)}\,\Omega^2 \rmint_{0}^1dy\,y\,\,e^{-iyk\cdot b}\frac{K_1(|b|\Delta(k,y))}{4\pi\Delta(k,y)}\,, \\ &
       =\frac{|b|^2 \,T\cdot n}{4\gamma}\rmint_0^\infty dy\,y\rmint d\Omega\, e^{-i\Omega l}\,\Omega^2\, \rmint_{0}
^\infty \frac{dt}{t^2}\,\exp\Big(-t-\frac{\Omega^2|b|^2\bar\Delta(y)^2}{4t}\Big)\,,\\ &
=\frac{\sqrt{\pi} \,T\cdot n}{|b|^3\gamma}\rmint_{0}^1dy\,\frac{y}{\bar\Delta(y)^5}\rmint_0^\infty dt\,t^{-1/2}(|b|^2\bar\Delta^2-2l^2t)\exp\Big(-\frac{l^2 t}{|b|^2\bar\Delta^2}-t\Big)\,,\\ &
=\frac{{\pi} \,T\cdot n}{|b|^3\gamma}\rmint_0^1 dy\,\frac{y}{\Bar \Delta}\times \Bigg[\frac{b^2\Bar\Delta^2}{\sqrt{\frac{l^2}{|b|^2 \Bar\Delta ^2}+1}}-l^2\frac{1}{ \left(\frac{l^2}{|b|^2 \Bar\Delta ^2}+1\right)^{3/2}}\Bigg],\,l:=n\cdot (x-b_1-yb)\,.
\end{split}
\end{align}
Then adding (\ref{ex}) and (\ref{ex1}) we get the full expression for $f_{\phi}(x)\Big|_{2}^{(1)}\,.$
The second part in \eqref{ex3} is given by,
\begin{align}
    \begin{split}
    f_{\varphi}(x)\Big|_{2}^{(2)}=(v_1\cdot P\cdot v_1)_{\mu\nu}\rmint_{\Omega}e^{-ik\cdot (x-b_1)}\rmint_{k_2}e^{-ik_2\cdot b}\frac{k_2^\mu\,k_2^\nu}{k_2^2(k-k_2^2)}\hat\delta(k\cdot v_1-k_2\cdot v_1)\hat\delta(k_2\cdot v_2)\,.\label{5.50 m}
    \end{split}
\end{align}
Now, we will rewrite \eqref{5.50 m} to match with \eqref{5.25mmm}. To do this, we introduce an auxiliary variable $k_1$ rewrite (\ref{5.50 m}) using a four-dimensional delta function in the following way.

\begin{align}
\begin{split}f_{\varphi}(x)\Big|_{2}^{(2)}=(v_1\cdot P\cdot v_1)_{\mu\nu}\rmint_{\Omega}e^{-ik\cdot x}\rmint_{k_2}e^{ik_1\cdot b_1}e^{ik_2\cdot b_2}\frac{k_2^\mu\,k_2^\nu}{k_2^2 k_1^2}\hat\delta(k_1\cdot v_1)\hat\delta(k_2\cdot v_2)\hat\delta^{(4)}(k_1+k_2-k)\,.\end{split}
\end{align}
Then we can proceed just like the case of (\ref{5.38nm}). Finally we get, \\\\
\hfsetfillcolor{gray!10}
\hfsetbordercolor{black}
\begin{equation}\label{ex4}\begin{split}
\tikzmarkin[disable rounded corners=false]{v}(0.3,-0.5)(-0.5,1) f_{\varphi}(x)\Big|_{2}^{(2)}&
      =(v_1\cdot P\cdot v_1)_{\mu\nu}\Big[n^{\mu}n^\nu I^{\rho\sigma}v_{2\rho}v_{2\sigma}-2v_{2\rho}I^{\rho(\mu}n^{\nu)}+I^{\mu\nu}+\bar I^{\mu\nu}\Big].
\tikzmarkend{v}\\
        \end{split}
\end{equation}\\
Here $I^{\mu\nu}$ and $\bar{I}^{\mu\nu}$ are defined in (\ref{IMUNU}) and (\ref{5.43mm}) respectively. Then  adding (\ref{ex4}) with (\ref{ex}) and (\ref{ex1}) we get the entire expression for $f_{\phi}(x)\Big|_2$ defined in (\ref{ex3}). Finally, adding (\ref{ex3}) with (\ref{5.25mmm}) we get the entire contribution to the waveform coming from this 3-point vertex at 2PM order. \\\\
\textbullet $\,\,$ Apart from the above-mentioned diagrams, there will be three more diagrams coming from the $\lambda_3\varphi^3$ vertex, which will contribute to the scalar wave form.
\begin{align}
    \begin{split}
        f(x) &\equiv\begin{minipage}[h]{0.12\linewidth}
	\vspace{4pt}
	\scalebox{1.3}{\includegraphics[width=\linewidth]{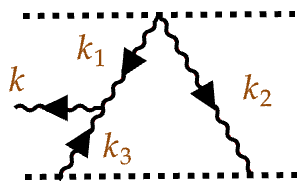}}
\end{minipage}\\ &=\lambda_3\frac{m_1 g_1 m_2^2 s_2^2}{8m_p^3}\rmint e^{-ik\cdot x}\rmint_{k_i}e^{i(k_1+k_2)\cdot b_1}e^{i(k_3-k_2)\cdot b_2}\frac{\hat\delta(k_1\cdot v_1+k_2\cdot v_1)\hat\delta(k_2\cdot v_2)\hat\delta(k_3\cdot v_2)}{k_1^2 k_2^2 k_3^2}\hat\delta^{(4)}(k-k_1-k_3)\,,\\ &
        =\lambda_3\frac{m_1 g_1 m_2^2 s_2^2}{8m_p^3}\rmint_{k}e^{-ik\cdot (x-b_1)}\rmint_{k_2} e^{ik_2\cdot b}\frac{\hat\delta(k_2\cdot v_2)}{k_2^2}\rmint_{k_3}\hat\delta(k_3\cdot v_2)\hat\delta(k_3\cdot v_1-k\cdot v_1-k_2\cdot v_1)\frac{e^{-ik_3\cdot b}}{k_3^2(k_3-k)^2}.\label{5.62x}
    \end{split}
\end{align}
The $k_3$ integral can be further simplified using Feynman parametrization.
\begin{align}
    \begin{split}
        I_{k_3}&=\rmint_{k_3}\hat\delta(k_3\cdot v_2)\hat\delta(k_3\cdot v_1-k\cdot v_1-k_2\cdot v_1)\frac{e^{-ik_3\cdot b}}{k_3^2(k_3-k)^2},\quad \textrm{with}\quad k^2=0\,,\\ &
        =\rmint_0^1 dy\, e^{-iy k\cdot b}\rmint_{0}^\infty dt\,t\,\rmint_{\tilde k_3}\exp\Big[i \tilde k_3\cdot b-t\, \tilde k_3^2-t\Sigma(y,k,k_2\cdot v_1)^2\Big]\,,\\ &
    =\frac{|b|}{\gamma}\rmint_0^1dy\,e^{-iyk\cdot b}\frac{K_1\Big[|b|\Sigma(k,y,k_2\cdot v_1)\Big]}{4\pi \Sigma(k,y,k_2\cdot v_1)}\,.
    \end{split}
\end{align}
 Note that in our velocity parametrization: $\frac{k_2\cdot v_1}{\gamma\beta}=-k_2^{(1)}$, which implies the function $\Sigma$ depends only one component of $k_2$. Hence we can do the other two components integral of $k_2$, which reads,
 \begin{align}
     \begin{split}
         f(x)&\sim\frac{|b|}{4\pi\gamma}\rmint_0^1 dy \rmint_{k}e^{-ik\cdot (x-b_1+yb)}\rmint dz K_0(|z||b|)\frac{K_1\Big[|b|\Sigma(k,y,k_2\cdot v_1)\Big]}{\Sigma(k,y,k_2\cdot v_1)}\\ &
         =\frac{|b|}{4\pi\gamma}\rmint_{0}^1 dy\rmint_{-\infty}^{\infty} d \Omega\rmint_{-\infty}^\infty dz \,e^{-i \Omega \lambda(x,y,b)}K_{0}(|z||b|)\frac{K_1\Big[|b|\Sigma(\Omega,y,z)\Big]}{\Sigma(\Omega,y,z)},\,\lambda\equiv n\cdot(x-b_1+yb)\label{6.30d}
     \end{split}
 \end{align}
where,
\begin{align}
    \begin{split}
        \Sigma=\sqrt{\frac{\Omega^2\Big[y^2(n\cdot v_2)^2+(1-y)^2(n\cdot v_1)^2+2y(1-y)\gamma (n\cdot v_2)(n\cdot v_1)\Big]}{\gamma^2-1}+z^2-2z\,\Omega\,\Big(\frac{y}{\beta}n\cdot v_2+\frac{1-y}{\gamma\beta}n\cdot v_1\Big)}\,.
    \end{split}
\end{align}
The integral in \eqref{6.30d} can be further simplified by using the integral representation of Bessel functions,
\begin{align}
    \begin{split}
        f(x)&\sim  \frac{|b|}{4\pi\gamma}\rmint_{0}^1 dy\rmint_{-\infty}^{\infty} d \Omega\rmint_{-\infty}^\infty dz \,e^{-i \Omega \lambda(x,y,b)}K_{0}(|z||b|)\frac{K_1\Big[|b|\Sigma(\Omega,y,z)\Big]}{\Sigma(\Omega,y,z)}\,,\\ &
      =  \frac{|b|^2}{16\pi\gamma}\rmint_{0}^1 dy\rmint_{-\infty}^{\infty} d \Omega\rmint_{-\infty}^\infty dz \,e^{-i \Omega \lambda(x,y,b)}\rmint dt_1\, \frac{1}{2t_1}\exp\Big(-t_1-\frac{z^2 |b|^2}{4t_1}\Big)\rmint \frac{dt_2}{t_2^2}\exp\Big(-t_2-\frac{|b|^2\Sigma^2}{4t_2}\Big)\,,\\ &
      =\frac{|b|^2}{32\pi \gamma}\rmint_0^1 dy \rmint_{-\infty}^{\infty} d\Omega e^{-i \Omega\lambda(x,y,b)} \rmint_0^\infty \frac{dt_1 dt_2}{t_1 t_2^2}e^{-(t_1+t_2)}\rmint_{-\infty}^{\infty}dz \exp\Big(-\frac{z^2|b|^2}{4t_1}-\frac{(z^2-2z\theta_1+\theta_2)|b|^2}{4t_2}\Big)\,.
    \end{split}
\end{align}
After doing the $z$ integral we left with,
\begin{align}
    \begin{split}
        f(x)&\sim \frac{|b|^2}{32\pi \gamma}\int_0^1 dy \rmint_{-\infty}^{\infty} d\Omega e^{-i \Omega\lambda(x,y,b)} \rmint_0^\infty \frac{dt_1 dt_2}{t_1 t_2^2}e^{-(t_1+t_2)}\frac{2 \sqrt{\pi } \exp\Big({\frac{|b|^2 \theta _1^2 t_1}{4 t_2^2+4 t_1 t_2}}-\frac{\theta_2|b|^2}{4t_2}\Big)}{b \sqrt{\frac{1}{t_2}+\frac{1}{t_1}}}\,,\\ &
        =\frac{|b|\textcolor{black}{\sqrt{\pi}}}{32\pi \gamma}\rmint_0^1 dy\rmint_0^\infty \frac{dt_1 dt_2}{t_1t_2^2}\frac{e^{-(t_1+t_2)}}{\sqrt{\frac{1}{t_1}+\frac{1}{t_2}}}\rmint_{-\infty}^{\infty} d\Omega\,e^{-i\Omega \lambda}\exp\Big[-\Omega^2\Big(\frac{\hat{\theta}_2|b|^2}{4t_2}-\frac{|b|^2\hat\theta_1^2t_1}{4t_2^2+4t_1t_2}\Big)\Big]\,.\\ &
       \label{6.33d}
    \end{split}
\end{align}
In \eqref{6.33d} the $\Omega$ integral is convergent if,
$$\hat{\theta}_2>\hat{\theta}_1^2\frac{t_1}{t_1+t_2}\implies \underbrace{(\hat\theta_2-\hat\theta_1^2)}_{-y^2}t_1+\hat\theta_2 t_2>0\,,$$
which sets a bound on the $t_1,\, t_2$ integral. The region of integral lies in $t_1< \frac{\hat\theta_2}{y^2}\,t_2$. Now within this region one can do the $\Omega$ integral which reads,
\begin{align}
\begin{split}
      f(x)&   =\frac{|b|}{16 \gamma}\rmint_0^1 dy\rmint_0^\infty \frac{dt_2}{t_2^2}\rmint_0^{\frac{\hat\theta_2}{y^2}t_2}\frac{ dt_1}{t_1}\frac{e^{-(t_1+t_2)}}{\sqrt{\frac{1}{t_1}+\frac{1}{t_2}}}\frac{1}{\sqrt{\frac{\hat{\theta}_2|b|^2}{4t_2}-\frac{|b|^2\hat\theta_1^2t_1}{4t_2^2+4t_1t_2}}}\exp\Big[-\frac{\lambda^2}{4}\frac{1}{\frac{\hat{\theta}_2|b|^2}{4t_2}-\frac{|b|^2\hat\theta_1^2t_1}{4t_2^2+4t_1t_2}}\Big]\\ &\hspace{8 cm}+ \textrm{pure divergence coming from $\Omega$ integral.}\\ &
        \xrightarrow[]{\textrm{finite part}}\frac{1}{8 \gamma}\rmint_0^1 dy\rmint_0^\infty dt_2 \rmint_{0}^{\frac{\hat\theta_2}{y^2}t_2}dt_1\frac{e^{(-t_1-t_2)}}{t_2\sqrt{t_1}\sqrt{(\hat\theta_2-\hat\theta_1^2)t_1+\hat\theta_2 t_2}}\exp\Big[-\frac{\lambda^2}{|b|^2}\frac{t_2(t_1+t_2)}{(\hat\theta_2-\hat\theta_1^2)t_1+\hat\theta_2 t_2}\Big]
    \end{split}
\end{align}
where,
\begin{align}
    \begin{split}
        &\theta_1(y)=\Omega\Big(\frac{y}{\beta}n\cdot v_2+\frac{1-y}{\gamma\beta}n\cdot v_1\Big):=\Omega\, \hat\theta_1\,,\\ &
        \theta_2(y)=\frac{\Omega^2}{{\gamma^2-1}}\Big({y^2(n\cdot v_2)^2+(1-y)^2(n\cdot v_1)^2+2y (1-y)\gamma(n\cdot v_2)(n\cdot v_1)}\Big):=\Omega^2\,\hat\theta_2\,.
    \end{split}
\end{align}
Restoring the prefactors, the waveform has the following form,\\\\
\hfsetfillcolor{gray!10}
\hfsetbordercolor{black}
\begin{align}
    \begin{split}
    \tikzmarkin[disable rounded corners=false]{e35}(0.35,-1)(-0.2,1.1)  
    f(x)&=\frac{\lambda_3}{(2\pi)^5}\frac{\sqrt{\pi}m_1 g_1 m_2^2 s_2^2}{64\gamma m_p^3}\rmint_0^1 dy\rmint_0^\infty dt_2 \rmint_{0}^{\frac{\hat\theta_2}{y^2}t_2}dt_1\frac{e^{-(t_1+t_2)}}{t_2\sqrt{t_1}\sqrt{(\hat\theta_2-\hat\theta_1^2)t_1+\hat\theta_2 t_2}}\\ &\hspace{8 cm}\times\exp\Big[-\frac{\lambda^2}{|b|^2}\frac{t_2(t_1+t_2)}{(\hat\theta_2-\hat\theta_1^2)t_1+\hat\theta_2 t_2}\Big]\,.\label{5.70l}
           \tikzmarkend{e35}
    \end{split}
\end{align}\\\\
The integral in \eqref{5.70l}, to the best of our knowledge, does not have any closed-form and one can in principle do the integral numerically while doing further investigations.\\
\textbullet $\,\,$ Another waveform diagram has the following form:
\begin{align}
    \begin{split}
        f(x)&\sim\begin{minipage}[h]{0.12\linewidth}
	\vspace{4pt}
	\scalebox{1.7}{\includegraphics[width=\linewidth]{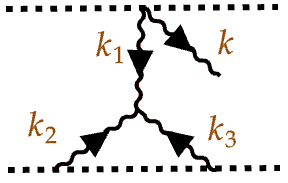}}
\end{minipage}\\ &=\lambda_3\frac{m_1 g_1 m_2^2 s_2^2}{8m_p^3} \rmint_{k}e^{-ik\cdot x}\rmint_{k_i}\hat\delta^{(4)}\Big(\sum_ik_i\Big)\frac{\hat\delta(k_1\cdot v_1+k\cdot v_1)\hat\delta(k_2\cdot v_2)\hat\delta(k_3\cdot v_2)}{k_1^2 k_2^2 k_3^2}e^{i(k+k_1)\cdot b_1}e^{i(k_2+k_3)\cdot b_2}\,,\\ &
        =\lambda_3\frac{m_1 g_1 m_2^2 s_2^2}{8m_p^3}\rmint_{\Omega}e^{-i\Omega\,n\cdot(x-b_1)}\rmint_{k_1,k_2}\frac{\hat\delta(k\cdot v_1+k_1\cdot v_1)\hat\delta(k_2\cdot v_2)\hat\delta(k_1\cdot v_2)}{k_1^2 k_2^2 (k_1+k_2)^2}e^{ik_1\cdot b}\,,\\ &
        =\lambda_3\frac{m_1 g_1 m_2^2 s_2^2}{8m_p^3}\frac{1}{n\cdot v_1}\rmint_{k_1,k_2}e^{-i k_1\cdot \delta_1}\frac{\hat\delta(k_1\cdot v_2)\hat\delta(k_2\cdot v_2)}{k_1^2 k_2^2 (k_1+k_2)^2}\,,\delta_1\equiv \frac{n\cdot(x-b_1)}{n\cdot v_1}v_1-b\,,\\ &
        =\lambda_3\frac{m_1 g_1 m_2^2 s_2^2}{64m_p^3(n\cdot v_1)}\rmint_{\vec k_1}\frac{e^{i \vec k_1\cdot \vec \delta_1}}{|\vec k_1|^3}\,.\label{6.35}
    \end{split}
\end{align}
The integral in \eqref{6.35} can be done using the dimensional regularisation and we get,\vspace{0.5 cm}
\hfsetfillcolor{gray!10}
\hfsetbordercolor{black}
\begin{align}
    \begin{split}
    \tikzmarkin[disable rounded corners=false]{e34}(0.4,-1)(-0.2,0.99)  
           f(x)=\frac{\lambda_3}{\textcolor{black}{(2\pi)^3}}\frac{m_1 g_1 m_2^2 s_2^2}{8m_p^3} \frac{1}{32\pi^2(n\cdot v_1)}\Big[{-\log (\vec\delta_1 ^2/\Lambda^2)-\gamma_E +\log (4)}\Big]\,.\label{5.72l}
           \tikzmarkend{e34}
    \end{split}
\end{align}
\\\\
\textit{Last but not the least, the total contribution to the scalar waveform at 2PM order is sum of (\ref{e:ba}), (\ref{5.24 mmm}), (\ref{5.25mmm}),(\ref{ex6}), \eqref{5.70l} and \eqref{5.72l}.} \textcolor{black}{Also, note that we need to replace worldline 1 with worldline 2 in all of our results and add them to get the total contribution, as the final result has to be symmetric under this exchange. }
\subsection{Gravitational waveform at 2PM: contribution due to extra scalar DOF}
Now, we will also spell out the correction to the gravitational waveform due to the presence of the extra scalar field. The leading order correction comes from bulk scalar graviton interaction with interacting action:
\begin{align}
    S_{int}=\frac{1}{m_p}\rmint d^4x \,h^{\mu\nu}\partial_{\mu}\varphi\partial_{\nu}\varphi\,.
\end{align}
The corresponding graviton one-point function has the following form:

\begin{align}
    \begin{split}
        k^2\Big\langle h_{\mu\nu}(k)\Big\rangle\Big|_{k^2\rightarrow 0}= \begin{minipage}[h]{0.12\linewidth}
	\vspace{4pt}
	\scalebox{1.4}{\includegraphics[width=\linewidth]{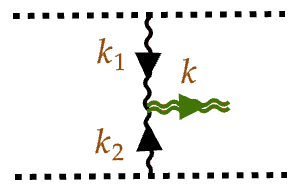}}
\end{minipage}\hspace{1 cm}=\Big(\frac{m_1 s_1 m_2 s_2}{2m_p}\Big)^2\rmint d\mu_{1,2}(k) \frac{k_1^\mu k_2^\nu} {k_1^2 k_2^2}\,.\label{5.63a}
    \end{split}
\end{align}
where the integral measure takes the form,
\begin{align}
    \begin{split}
        \rmint d\mu_{1,2}(k)= \rmint_{k_1,k_2}e^{ik_1\cdot b_1} e^{i k_2\cdot b_2}\hat\delta(k_1\cdot v_1)\hat\delta(k_2\cdot v_2)\hat\delta^{(4)}(k_1+k_2-k)
    \end{split}
\end{align}
Therefore, the correction to the time domain waveform for this particular interaction is given by,
\begin{align}
    \begin{split}
        f_{h}&=\Big(\frac{m_1 s_1 m_2 s_2}{2m_p}\Big)^2\frac{1}{4\pi m_p}\epsilon^\mu \epsilon^\nu \rmint_{\Omega}e^{-ik\cdot x}\rmint d\mu_{1,2}(k)\,\frac{k_{1\mu}k_{2\nu}}{k_1^2\,k_2^2}\,,\\ &
        =\Big(\frac{m_1 s_1 m_2 s_2}{2m_p}\Big)^2\frac{1}{4\pi m_p}\epsilon^\mu \epsilon^\nu \rmint_{\Omega}e^{-ik\cdot x}\rmint_{k_2}e^{i(k-k_2)\cdot b_1}e^{ik_2\cdot b_2}\frac{(k-k_2)_{\mu}k_{2\nu}}{k_2^2(k-k_2)^2}\,\hat\delta\Big(k\cdot v_1-k_2\cdot v_1\Big)\hat\delta(k_2\cdot v_2)\,.\label{5.55mm}
    \end{split}
\end{align}
The total waveform in \eqref{5.55mm} can be separated into two parts, and the parts will be treated separately,
\begin{align}
    \begin{split}
        f_h(x)\Big|_{1}=\Big(\frac{m_1 s_1 m_2 s_2}{2m_p}\Big)^2\frac{\epsilon^\mu \epsilon^\nu}{4\pi m_p}\rmint_{\Omega}e^{-ik\cdot (x-b_1)}k_{\mu}\rmint_{k_2}e^{-ik_2\cdot b}\frac{k_{2\nu}}{k_2^2(k-k_2)^2}\,\hat\delta\Big(k\cdot v_1-k_2\cdot v_1\Big)\hat\delta(k_2\cdot v_2)\,.\label{5.56 mm}
    \end{split}
\end{align}
It is evident from \eqref{5.56 mm} that $f_h\Big|_{1}\sim \epsilon\cdot k$ and we know that $\epsilon\cdot k=0$. Therefore, the first part will not contribute to the waveform. Now, the second part of the waveform takes the following form,
\begin{align}
    \begin{split}
        f_{h}(x)\Big|_{2}=-\Big(\frac{m_1 s_1 m_2 s_2}{2m_p}\Big)^2\frac{\epsilon^\mu \epsilon^\nu}{4\pi m_p}\rmint_{\Omega}e^{-ik\cdot x}\rmint_{k_1,k_2}\frac{k_{2\mu}k_{2\nu}}{k_1^2k_2^2}\hat\delta(k_1\cdot v_1)\hat\delta(k_2\cdot v_2)\hat\delta^{(4)}(k_1+k_2-k)\,.
    \end{split}
\end{align}
Restoring the factors of $\pi$ from the delta functions and the integration  measures, this integral is the same as \eqref{5.25mmm} and results in,
\\\\
\hfsetfillcolor{gray!10}
\hfsetbordercolor{black}
\begin{equation}\label{e}
\begin{split}
\tikzmarkin[disable rounded corners=false]{w}(0.3,-0.5)(-0.5,1)  f_{h}\Big|_{2}(x)=\Big(\frac{m_1 s_1 m_2 s_2}{2m_p}\Big)^2\frac{1}{2(2\pi)^2 \pi m_p }\epsilon^{\mu}\epsilon^{\nu}(I_{\mu\nu}+\bar I_{\mu\nu}).
\tikzmarkend{w}\\
        \end{split}
\end{equation}\\
where we again used the fact that $n\cdot \epsilon=0$ and $I^{\mu}$ and $\bar{I}^{\mu\nu}$ are defined in (\ref{IMUNU}) and (\ref{5.43mm}) respectively.
\section{Towards massive waveform}\label{sec6}
As mentioned in the previous section, we could get exact results only when we set the mass of the scalar field to zero. In this section, we will discuss how the radiation integrals become more complicated when we make the scalar field massive. Then, we provide an approximate way to evaluate those integrals and get an analytic result. \par 
For massive scalar, the radiated momentum has to be parametrized as follows,
\begin{align}
    k^{\mu}=\Omega n^\mu,\, n^\mu=(1,\sqrt{1-\frac{m^2}{\Omega^2}}\boldsymbol{\hat{x}})\,.\label{6.1 m}
\end{align}
\textit{From \eqref{6.1 m} it is clear that the unit vector towards the direction of the observed scalar depends on the observed frequency $\Omega$. We need to integrate over all possible frequency $\Omega$ for computing the time domain waveform. There lies the difficulty due to the presence of complicated phase factors in the time-domain waveform integrand. }
\subsection*{Radiation integrands for massive scalar waveform}
Before proceeding further, we first list all the relevant massive integrands having a massless counterpart, which we evaluated in the previous section.
\begin{itemize}
  \item The massive integrated corresponding to $ \lambda_3 \varphi^3$ in \eqref{wave0} vertex has the following form:
    \begin{align}
        \begin{split}
          f_{\varphi}\propto \rmint_{\Omega}e^{-ik\cdot x}\rmint_{k_i}e^{ik_1\cdot b_1}e^{ik_2\cdot b_2}\frac{\hat\delta(k_1\cdot v_1)\hat\delta(k_2\cdot v_2)}{(k_1^2-m^2)(k_2^2-m^2)}\hat\delta^{(4)}(k_1+k_2-k)\,. \label{6.11 m}
        \end{split}
    \end{align}
\item The massive integrated corresponding to $\lambda_4 \varphi^4$ in \eqref{5.11 mn} vertex has the following form:
\begin{align} 
    \begin{split} \label{6.11 mm}
        f_{\varphi}\propto \rmint_{\Omega}e^{-ik\cdot x}\rmint_{k_i}e^{ik_1\cdot b_1}e^{i k_2\cdot b_2}e^{i k_3\cdot b_2}\frac{\hat\delta(k_1\cdot v_1)\hat\delta(k_2\cdot v_2)\hat\delta(k_3\cdot v_2)}{(k_1^2-m^2)(k_2^2 -m^2)(k_3^2-m^2)}\hat\delta^{(4)}(k_1+k_2+k_3-k)\,.
    \end{split} 
\end{align}
\item  
Massive integral corresponding to worldline radiation in \eqref{5.24 mmm} \footnote{Another way to approximate this integral is to take a large velocity limit. We have discussed that in detail in Appendix~(\ref{app22}). Although this approximation helps get an approximate closed-form result for this diagram, it is unclear whether a large velocity approximation will also be useful for all other massive integrals. Hence, we focused on the stationary-phase method in the main text.}:
\begin{align}\label{6.4 mm}
f_{\varphi}(x)&\propto\rmint_{\Omega}e^{-ik\cdot (x-b_1)}\rmint_{\omega,k_1}e^{-i k_1\cdot b}\hat{\delta}(k_1\cdot v_2)\hat{\delta}[\Omega\, n(m,\Omega)\cdot v_1-\omega]\hat\delta(k_1\cdot v_1-\omega)\frac{\Omega(n\cdot k_1)}{\omega^2 (k_1^2-m^2)}\,.
\end{align}
     \item The massive integrated corresponding to the diagram \eqref{5.22 m} coming from the derivative interaction term $\rmint d^4x h^{\mu\nu}\partial_{\mu}\varphi\partial_\nu \varphi$ has the following form:
    \begin{align}
        \begin{split}
            f_{\varphi}\propto \rmint_{\Omega}e^{-ik\cdot x}\rmint_{k_1,k_2}e^{ik_1\cdot b_1}e^{ik_2\cdot b_2}\hat\delta(k_1\cdot v_1)v_1^{\alpha}v_1^{\beta}\hat\delta(k_2\cdot v_2)\frac{k_2^{\mu}k_2^{\nu}P_{\mu\nu,\alpha\beta}}{k_1^2(k_2^2-m^2)}\delta^{(4)}(k_1+k_2-k) \,.\label{6.10 m}
        \end{split}
    \end{align}
\end{itemize}
\subsection{Stationary Phase (SP) approximation: the need and general procedure}
We mentioned in the previous subsection how the massive integrals get complicated. Now, we will evaluate the integrals mentioned above using the \textit{stationary phase approximation} for large observation distance (and time) $|x|\rightarrow \infty$. This will help us to get an approximate analytical result. Note that the idea behind the stationary phase approximation stems from the fact that sinusoids with rapidly varying phases interfere destructively. However, they can be added constructively if they have the same phases. Physically, in a scattering scenario, we expect a burst signal when the emitted radiation gets detected around some frequencies. Hence, we believe that using an approximation method such as the stationary phase method to evaluate these integrals that will give the waveform is reasonable. 
We have to evaluate the following type of integrals.
\begin{align}
\begin{split}
        f_{\varphi}\propto \rmint_{\Omega}\exp\Big({-i\underbrace{\Omega \,n(\Omega)\cdot (x-b_1)}_{l(\Omega)}}\Big)g(\Omega)\label{6.13 m}
    \end{split}
\end{align}
The stationary point can be obtained by,
\begin{align}
    \begin{split}
        \frac{d}{d\Omega}\Big(\Omega \,n(\Omega)\cdot (x-b_1) \Big)\Big |_{\Omega_0}=0\,.\label{6.14 m}
    \end{split}
\end{align}
Therefore, the integral can be approximated to,
\begin{align}
    \begin{split}
        f_{\varphi}\propto g(\Omega_0) \,e^{-i l(\Omega_0)}e^{\frac{i\pi}{4}}\sqrt{\frac{\pi}{|l''(\Omega_0)|}}+\mathcal{O}\Big(\frac{1}{|x-b_1|}\Big)
    \end{split}
\end{align}
\subsection{Dealing the massive integrals via SP approximation}
We start with the integral mentioned in (\ref{6.11 m}) we get, 
\begin{align}
    g(\Omega_0)=\rmint_{k_2}e^{-ik_2\cdot b}\frac{\hat\delta(k_0\cdot v_1-k_2\cdot v_1)\hat\delta(k_2\cdot v_2)}{[(k_2-k_0)^2-m^2](k_2^2-m^2)},\,k_0\equiv \Omega_0 \,n({\Omega_0})\,.
\end{align}
Then the final integral over $k_2$ can be done using Feynman parametrization in the following way, 
\begin{align}
    \begin{split}
        g(\Omega_0)&=\rmint_{0}^1 dy\rmint e^{-ik_2\cdot b}\frac{\hat\delta(k_2\cdot v_1-k_0\cdot v_1)\hat\delta(k_2\cdot v_2)}{[(k_2-yk_0)^2-(1-y+y^2)m^2]^2},\,k_0^2=m^2\,,\\ &
        \xrightarrow[]{k_2-yk_0\rightarrow \bar q}\rmint_0^1 dy e^{-iy\,k_0\cdot b}\rmint \frac{e^{-i \bar q\cdot b}}{[\bar q^2-(1-y+y^2)m^2]^2}\hat\delta(\bar q\cdot v_1-(1-y)k_0\cdot v_1)\hat\delta(\bar q\cdot v_2+y k_0\cdot v_2)\,,\\ &
        =\frac{1}{\gamma}\rmint_0^1\,dy\,e^{-iy\,k_0\cdot b}\rmint \frac{dt}{4\pi}\exp\Big(-\frac{|b|^2}{4t}-t\Delta(k_0,y)^2-t(1-y+y^2)m^2\Big)\,,\\ &
        =\frac{b}{4\pi \gamma}\rmint_0^1
dy\,e^{-iy\,k_0\cdot b}\frac{K_1\Big( b\sqrt{\Delta(y)^2+(1-y+y^2)m^2}\Big)}{\sqrt{\Delta(y)^2+(1-y+y^2)m^2}}\,.
\end{split}
\end{align}
Then, the contribution to the waveform takes the following form,
\hfsetfillcolor{gray!10}
\hfsetbordercolor{black}
\begin{equation}\label{e}\begin{split}
\tikzmarkin[disable rounded corners=false]{y1}(0.3,-0.5)(-0.5,1)  f_{\varphi}\propto \frac{b}{4\pi\gamma}\sqrt{\frac{\pi}{|l''(\Omega_0)|}}\rmint_0^1 dy\,e^{-i\sigma(b,\Omega_0,m,y)}\frac{K_1\Big( b\sqrt{\Delta(y)^2+(1-y+y^2)m^2}\Big)}{\sqrt{\Delta(y)^2+(1-y+y^2)m^2}},\,\sigma=y k_0\cdot b+l(\Omega_0)-\frac{\pi}{4}\,.
\tikzmarkend{y1}\\
        \end{split}
\end{equation}
\\
Next, we consider the integral in \eqref{6.4 mm} and apply  the method of stationary phase to it. It gives the following,
\begin{align}
    \begin{split}
        f(\varphi_0)\propto g(\Omega_0)e^{-il(\Omega_0)}e^{\frac{i\pi}{4}}\sqrt{\frac{\pi}{|l''(\Omega_0)|}}
    \end{split}
\end{align}
where $g(\Omega_0)$ is given by,
\begin{align}
    \begin{split}
        g(\Omega_0)&=\frac{1}{\Omega_0^2 \,\Big(n(\Omega_0)\cdot v_1 \Big)^2}\rmint_{k_1}e^{-ik_1\cdot b}\frac{-k_1\cdot k_0}{(k_1^2-m^2)}\hat\delta(k_1\cdot v_1-k_0\cdot v_1)\hat\delta(k_1\cdot v_2)\,,\\ &
        =-\frac{k_0^\mu}{\Omega_0^2 \,\Big(n(\Omega_0)\cdot v_1 \Big)^2}\underbrace{\rmint_{k_1}e^{-ik_1\cdot b}\frac{k_1^\mu}{k_1^2-m^2}\hat\delta(k_1\cdot v_1-k_0\cdot v_1)\hat\delta(k_1\cdot v_2)}_{g^\mu(\Omega_0)}\,.
    \end{split}
\end{align}
Here the $g_\mu(\Omega_0)$ can be evaluated using Passarino-Veltman reduction.
\begin{align}
    \begin{split}
        g^{\mu}(\Omega_0)= \lambda_b b^\mu+\lambda_1 v_1^\mu+\lambda_2 v_2^\mu\,.
    \end{split}
\end{align}
To extract the constants, one needs to contract the index structure in RHS with LHS, which gives the following:\\\\
\textbullet $\,\,$ Contracting with $b^\mu$ gives,
\begin{align}
    \begin{split}
        -\lambda_b \,|b|^2&=\rmint_{k_1} e^{-i k_1\cdot b} \frac{k_1\cdot b}{k_1^2-m^2}\hat\delta(k_1\cdot v_1-k_0\cdot v_1)\hat\delta(k_1\cdot v_2)\,,\\ &
        =i\lim_{\kappa\to 1}\frac{\partial}{\partial \kappa}\rmint_{0}^\infty dl\,l\,\frac{J_{0}(\kappa\,l|b|)}{l^2+\hat m^2},\,\hat m^2\equiv m^2+\Big(\frac{k_0\cdot v_1}{\gamma\sqrt{\gamma^2-1}}\Big)^2\,.\label{6.35 m}
    \end{split}
\end{align}
\textbullet $\,\,$ Contracting with $v_1^\mu$ we will get,
\begin{align}
    \begin{split}
        \lambda_1+\gamma\lambda_2&= \rmint_{k_1} e^{-i k_1\cdot b} \frac{k_1\cdot v_1}{k_1^2-m^2}\hat\delta(k_1\cdot v_1-k_0\cdot v_1)\hat\delta(k_2\cdot v_2)\,,\\ &
        =k_0\cdot v_1\rmint_{k_1}e^{-ik_1\cdot b}\frac{\hat\delta(k_1\cdot v_1-k_0\cdot v_1)\hat\delta(k_1\cdot v_2)}{k_1^2-m^2}\,,\\ &
        =k_0\cdot v_1\rmint_0^\infty dl\,l\,\frac{J_{0}(l|b|)}{l^2+\hat m^2
        }\,.\label{6.36 m}
    \end{split}
\end{align}
\textbullet $\,\,$ Contracting with $v_2^\mu$ we will get,
\begin{align}
    \begin{split}
        \gamma \lambda_1+\lambda_2=0,\,\textrm{as}\,k_1\cdot v_2=0\,.\label{6.37 m}
    \end{split}
\end{align}
\textcolor{black}{Solving \eqref{6.36 m} and \eqref{6.37 m} we will get,
\begin{align}
    \begin{split}
      &  \lambda_1=\frac{1}{1-\gamma^2}\,k_0\cdot v_1\rmint_0^\infty dl\,l\,\frac{J_{0}(l|b|)}{l^2+\hat m^2}=\frac{1}{1-\gamma^2}\,(k_0\cdot v_1)K_0\left(|b| \hat{m}\right)\,,\\ &
      \lambda_2=\frac{\gamma}{\gamma^2-1} k_0\cdot v_1\,\rmint_0^\infty dl\,l\,\frac{J_{0}(l|b|)}{l^2+\hat m^2}=\frac{\gamma}{\gamma^2-1} (k_0\cdot v_1)\,K_0\left(|b| \hat{m}\right)
    \end{split}
\end{align}
and, from \eqref{6.35 m} we will get,
\begin{align}
    \lambda_b=\frac{i}{|b|}\rmint_0^\infty dl\,l^2\,\frac{J_1(l|b|)}{l^2+\hat m^2}=\frac{i}{|b|}\hat{m}\, K_1\left(|b| \hat{m}\right)\,.\label{6.19ab}
\end{align}
}
Therefore, it's contribution to the  waveform looks like,
\\ 
\hfsetfillcolor{gray!10}
\hfsetbordercolor{black}
\begin{align}
    \begin{split}
    \tikzmarkin[disable rounded corners=false]{e3}(0.4,-1)(-0.2,0.99)  
           f(x)\propto -\frac{1}{\Omega_0^2 \,\Big(n(\Omega_0)\cdot v_1 \Big)^2} \,e^{-il(\Omega_0)}e^{\frac{i\pi}{4}}\sqrt{\frac{\pi}{|l''(\Omega_0)|}}\,k_0\cdot[\lambda_b b+\lambda_1 v_{1}+\lambda_2 v_{2}]\,.
           \tikzmarkend{e3}
    \end{split}
\end{align}\\
Now, we analyze the integral, which comes from the derivative interaction as mentioned in \eqref{6.10 m}. It takes the following form,
\begin{align}
    \begin{split}
       f_{\varphi}\propto (v_1\cdot P\cdot v_1)_{\mu\nu} \rmint_{\Omega}\exp\Big(-i\Omega\,n(\Omega)\cdot (x-b_1)\Big)J_{(2)}^{\mu\nu}(\Omega)\,,
    \end{split}
\end{align}
where,
\begin{align}
    \begin{split}
        J_{(2)}^{\mu\nu}(\Omega)=\rmint_{q}\hat\delta(q\cdot v_1-k\cdot v_1)\hat\delta(q\cdot v_2)\frac{q^\mu q^\nu}{(q-k)^2(q^2-m^2)}e^{-iq \cdot b}\,.\label{6.20 m}
    \end{split}
\end{align}
The integral over $q$ in \eqref{6.20 m} has been done in \eqref{A.22 m}. Then using the method of stationary phase, we get,
\begin{align}
    \begin{split}
        f_{\varphi}\propto (v_1\cdot P\cdot v_1)_{\mu\nu}\sqrt{\frac{\pi}{|l''(\Omega_0)|}}e^{-il(\Omega_0)}e^{\frac{i\pi}{4}}\,J^{\mu\nu}_{(2)}(\Omega_0)\,.
    \end{split}
\end{align}
where $l(\Omega)$ and the saddle point $\Omega_0$ is defined in \eqref{6.13 m} and \eqref{6.14 m} respectively.\\\\
\textbullet $\,\,$ Apart from the integrals listed in (\ref{6.11 m}), (\ref{6.4 mm}) and (\ref{6.10 m}), one can possibly have one term proportional to scalar mass $m$  contributing to the waveform at 2PM order. 
\textcolor{black}{$$S_{int}=-\frac{1}{2}\frac{m^2}{m_{p}}\rmint d^4x\,h\varphi^2\,.$$}
The corresponding contribution to the scalar one-point function is , 
\begin{align}
   \begin{split}
      k^2\Big\langle \varphi(k)\Big\rangle=m^2\Big(\frac{m_1m_2s_2}{8m_p^3}\Big)\rmint d\mu_{1,2}(k)\frac{v_1^\alpha v_1^\beta\,P_{\mu\nu;\alpha\beta}\,\eta^{\mu\nu}}{k_1^2(k_2^2-m^2)}\,.\label{5.23 m}
    \end{split}
\end{align}
where, the integral measure has the following form,
\begin{align}
    \begin{split}
        d\mu_{1,2}(k)&=\rmint_{k_1,k_2}\hat{\delta}^{(4)}(k_1+k_2-k)e^{ik_1\cdot b_1+ik_2\cdot b_2 }\hat\delta(k_1\cdot v_1)\hat\delta(k_2\cdot v_2)\,,\\ &
        =\rmint_{k_2}e^{ik\cdot b_1}e^{-ik_2\cdot b}\hat\delta(k\cdot v_1-k_2\cdot v_1)\hat\delta(k_2\cdot v_2)\,.
    \end{split}
\end{align}
Therefore the corresponding contrubution to scalar waveform has the following form,
\begin{align}
    \begin{split}
        f_{\varphi}(x)&\propto\begin{minipage}[h]{0.12\linewidth}
	\vspace{4pt}
	\scalebox{1.5}{\includegraphics[width=\linewidth]{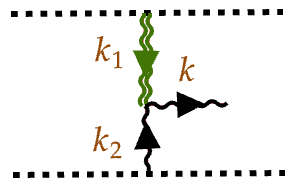}}
\end{minipage}\\ &=-m^2\Big(\frac{m_1m_2s_2}{4m_p^3}\Big)\rmint_{\Omega}e^{-ik\cdot(x-b_1)}\rmint_{k_2}e^{-ik_2\cdot b}\frac{\hat\delta(k_2\cdot v_2)\hat\delta(k\cdot v_1-k_2\cdot v_1)}{(k-k_2)^2(k_2^2-m^2)}\,,\\ &
        =-m^2\Big(\frac{m_1m_2s_2}{4m_p^3}\Big)\frac{|b|}{4\pi\gamma}\sqrt{\frac{\pi}{|l''(\Omega_0)|}}\rmint_0^1 dy\,e^{-i\sigma(|b|,\Omega_0,m,y)}\frac{K_1\Big(|b|\sqrt{\Delta(y)^2+(1-y)^2m^2}\Big)}{\sqrt{\Delta(y)^2+(1-y)^2m^2}}\,.\label{6.26f}
    \end{split}
\end{align}
One can see that the contribution to the waveform from \eqref{6.26f} identically vanishes when one takes the scalar to be massless. Hence, the contribution to the massive scalar waveform takes the form: \\\\
\hfsetfillcolor{gray!10}
\hfsetbordercolor{black}
\begin{align}
    \begin{split}
    \tikzmarkin[disable rounded corners=false]{e33}(0.1,-1)(-0.1,0.9)  
           f(x)\propto -m^2\Big(\frac{m_1m_2s_2}{8m_p^3}\Big)\frac{|b|}{4\pi\gamma}\sqrt{\frac{\pi}{|l''(\Omega_0)|}}\rmint_0^1 dy\,e^{-i\sigma(|b|,\Omega_0,m,y)}\frac{K_1\Big(|b|\sqrt{\Delta(y)^2+(1-y)^2m^2}\Big)}{\sqrt{\Delta(y)^2+(1-y)^2m^2}}\,.
           \tikzmarkend{e33}
    \end{split}
\end{align}\\
\textbullet $\,\,$ The massive counterpart of \eqref{5.62x} will take the form.
\begin{align}
    \begin{split}
        f(x)&\propto \rmint_{k}e^{-ik\cdot (x-b_1)}\rmint_{k_2} e^{ik_2\cdot b}\frac{\hat\delta(k_2\cdot v_2)}{k_2^2-m^2}\rmint_{k_3}\hat\delta(k_3\cdot v_2)\hat\delta(k_3\cdot v_1-k\cdot v_1-k_2\cdot v_1)\frac{e^{-ik_3\cdot b}}{(k_3^2-m^2)[(k_3-k)^2-m^2]}\,,\\ &
        =e^{-il(\Omega_0)}e^{i\frac{\pi}{4}}\sqrt{\frac{\pi}{|l''(\Omega_0)|}}g(\Omega_0)
    \end{split}
\end{align}
where,
\begin{align}
    \begin{split}
        g(\Omega_0)=\rmint_{k_2}\frac{e^{i k_2\cdot b}\,\hat\delta(k_2\cdot v_2)}{k_2^2-m^2}I_{k_3}(m,k_2,b,k_0)\,.
    \end{split}
\end{align}
Now, $I_{k_3}(m,k_2,b,k_0)$ takes the form,
\begin{align}
    \begin{split}
        I_{k_3}(m,k_2,b,k_0)&=\frac{|b|}{\gamma}\rmint_0^1dy\,e^{-iyk_0\cdot b}\frac{K_1\Big[|b|\sqrt{\Sigma(k_0,y,k_2\cdot v_1)^2+(1-y+y^2)m^2}\Big]}{4\pi\sqrt{ \Sigma(k_0,y,k_2\cdot v_1)^2+(1-y+y^2)m^2}}\,.\\ &
    \end{split}
\end{align}
Therefore, 
\begin{align}
    \begin{split}
        g(\Omega_0)=\frac{|b|}{\gamma}\rmint_0^1 dy\,e^{-iy k_0\cdot b}\rmint_{-\infty}^{\infty} dz\, K_{0}(\sqrt{z^2+m^2}|b|)\frac{K_1\Big[|b|\sqrt{\Sigma(k_0,y,z)^2+(1-y+y^2)m^2}\Big]}{4\pi\sqrt{ \Sigma(k_0,y,z)^2+(1-y+y^2)m^2}}
    \end{split}
\end{align}
Therefore the full waveform proportional to,
\hfsetfillcolor{gray!10}
\hfsetbordercolor{black}
\begin{align}
    \begin{split}
    \tikzmarkin[disable rounded corners=false]{e32}(0.4,-1)(-1.3,0.99)  
        \hspace{-1 cm}   f(x)\propto e^{-il(\Omega_0)}e^{i\frac{\pi}{4}}\sqrt{\frac{\pi}{|l''(\Omega_0)|}}\frac{|b|}{\gamma}\rmint dy\,e^{-iy k_0\cdot b}\rmint_{-\infty}^{\infty} dz\, K_{0}(\sqrt{z^2+m^2}|b|)\frac{K_1\Big[|b|\sqrt{\Sigma(k_0,y,z)^2+(1-y+y^2)m^2}\Big]}{4\pi\sqrt{ \Sigma(k_0,y,z)^2+(1-y+y^2)m^2}}.\,
           \tikzmarkend{e32}
    \end{split}
\end{align}
\textbullet $\,\,$ The massive counter part of \eqref{6.35} takes the form.
\begin{align}
    \begin{split}
        f(x)&\propto e^{-il(\Omega_0)}e^{i\frac{\pi}{4}}\sqrt{\frac{\pi}{|l''(\Omega_0)|}}\rmint_{k_1,k_2}\frac{\hat\delta(k_0\cdot v_1+k_1\cdot v_1)\hat\delta(k_2\cdot v_2)\hat\delta(k_1\cdot v_2)}{(k_1^2-m^2) (k_2^2-m^2) [(k_1+k_2)^2-m^2]}e^{ik_1\cdot b}\,
    \end{split}
\end{align}
After integrating over the loop momenta, the full waveform takes the following form,\\
\hfsetfillcolor{gray!10}
\hfsetbordercolor{black}
\begin{align}
    \begin{split}
    \tikzmarkin[disable rounded corners=false]{e31}(0.5,-1)(-0.1,0.99)  
           f(x)\sim e^{-il(\Omega_0)}e^{i\frac{\pi}{4}}\sqrt{\frac{\pi}{|l''(\Omega_0)|}} \rmint_0^\infty dl\,l\,\frac{J_{0}(l|b|)}{(l^2+\hat{m}^2)\sqrt{l^2+\hat{m}^2-m^2}}\arctan\Big(\frac{l^2+\hat{m}^2-m^2}{2m}\Big).\,
           \tikzmarkend{e31}
    \end{split}
\end{align}\\
\textbullet $\,\,$ The massive counter part of \eqref{5.11 mn} which is written in \eqref{6.11 mm} can be casted again by stationary phase approximation and is given by,
\begin{align}
    \begin{split}
        f_{\varphi}(x)&\propto e^{-il(\Omega_0)}e^{i\frac{\pi}{4}}\sqrt{\frac{\pi}{|l''(\Omega_0)|}}\rmint_{k_3,q} \frac{\hat\delta(k\cdot v_1-q\cdot v_1)\hat\delta(k_3\cdot v_2)\hat\delta(q\cdot v_2)}{(k_3^2-m^2) \, [(q-k_3)^2-m^2][(q-k)^2-m^2]}e^{-i q\cdot b}\\ &
        = e^{-il(\Omega_0)}e^{i\frac{\pi}{4}}\sqrt{\frac{\pi}{|l''(\Omega_0)|}}\int_0^\infty d\hat\alpha\,d\hat\beta\,d\hat\gamma\int_{\vec k_3,\vec q}\hat\delta(\Omega_0\,n\cdot v_1+\vec q\cdot \vec v_1)\,\exp\Big[i\hat\alpha(\vec q^2-2\Omega_0 \vec q\cdot \vec n)\\ &
\hspace{5 cm}+i\hat\beta (\vec q^2-2\vec q\cdot \vec k_3+\vec k_3^2)+{i\hat\gamma \vec k_3^2}+i(\hat\beta+\hat\gamma)m^2\Big]e^{i\vec q\cdot \vec b}
    \end{split}
\end{align}
Now from the delta function constraint we can do the integral over $q_{(1)}$ and we left with,
\begin{align}
    \begin{split}
       f_{\varphi}(x)&\sim e^{-il(\Omega_0)}e^{i\frac{\pi}{4}}\sqrt{\frac{\pi}{|l''(\Omega_0)|}}\frac{1}{\gamma\sqrt{\gamma^2-1}}\int_0^\infty d\hat\alpha\,d\hat\beta\,d\hat\gamma \,e^{i(\hat\beta+\hat\gamma) m^2}\exp\Big[i(\hat\alpha+\hat\beta)\Big(\frac{\Omega_0 n\cdot v_1}{\gamma\beta}\Big)^2+2i\hat\alpha\frac{\Omega_0^2n\cdot v_1}{\gamma\beta}\Big]\\ &
       \hspace{3 cm}\times  e^{-i\frac{\pi}{4}}\pi^{\frac{3}{2}}(\hat\beta+\hat\gamma)^{-\frac{3}{2}}\int_{\tilde q}\exp\Big(-i \frac{\hat\beta^2 \vec q^2}{\hat\beta +\hat\gamma}\Big)\exp\Big(i(\hat\alpha +\hat\beta)\tilde q^2+i\tilde q\cdot b\Big)\\ &
       = \pi^{\frac{5}{2}}e^{-il(\Omega_0)}\sqrt{\frac{\pi}{|l''(\Omega_0)|}}\frac{1}{\gamma\sqrt{\gamma^2-1}}\int_0^\infty d\hat\alpha\,d\hat\beta\,d\hat\gamma \,e^{i(\hat\beta+\hat\gamma) m^2}\exp\Big[i(\hat\alpha+\hat\beta)\Big(\frac{\Omega_0 n\cdot v_1}{\gamma\beta}\Big)^2+2i\hat\alpha\frac{\Omega_0^2n\cdot v_1}{\gamma\beta}\\ &
       \hspace{3 cm}-i\frac{\hat\beta^2}{\hat\beta+\hat\gamma}\frac{\Omega_0^2(n\cdot v_1)^2}{\gamma^2(\gamma^2-1)}\Big]  (\hat\beta+\hat\gamma)^{-\frac{3}{2}}\frac{e^{\frac{-ib^2}{4\lambda_1}}}{\lambda_1},\,\textrm{with},\,\lambda_1\equiv \hat\alpha+\hat\beta-\frac{\hat\beta^2}{\hat\beta+\hat\gamma}\label{6.36f}
    \end{split}
\end{align}
The integral in \eqref{6.36f}, to the best of our knowledge, does not have any closed form and should be done numerically. One can analogously find the contribution from $\lambda_3 h\varphi^3$ interaction vertex.
\\
\subsection*{Radiation integrals for gravitational waveform due to massive scalar}
Finally we will also discuss diagrams which contributes to the gravitational waveform due the presence of massive scalar.\\
\textbullet $\,\,$ The correction to the gravitational waveform comes from bulk scalar graviton interaction with interacting action:
\begin{align}
    S_{int}=-\frac{m^2}{2m_p}\rmint d^4x \,h\varphi^2\,.
\end{align}
Note that this diagram does not have any massless couterpart as the vertex is proportional to the mass of the scalar field. It's contribution to the  graviton one-point function takes the following form:
\begin{align}
    \begin{split}
        k^2\Big\langle h_{\mu\nu}(k)\Big\rangle\Big|_{k^2\rightarrow 0}= \begin{minipage}[h]{0.12\linewidth}
	\vspace{4pt}
	\scalebox{1.5}{\includegraphics[width=\linewidth]{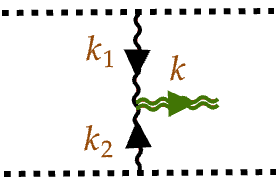}}
\end{minipage}\hspace{1.4 cm}=\rmint d\mu_{1,2}(k) \frac{\eta^{\mu\nu}} {(k_1^2-m^2) (k_2^2-m^2)}\,,
    \end{split}
\end{align}
where the integral measure takes the form,
\begin{align}
    \begin{split}
        \rmint d\mu_{1,2}(k)= \rmint_{k_1,k_2}e^{ik_1\cdot b_1} e^{i k_2\cdot b_2}\hat\delta(k_1\cdot v_1)\hat\delta(k_2\cdot v_2)\hat\delta^{(4)}(k_1+k_2-k)\,.
    \end{split}
\end{align}

Therefore, the correction to the time domain waveform for this particular interaction is given by,
\begin{align}
    \begin{split}
        f_{h}&=\frac{1}{4\pi m_p}\epsilon^\mu \epsilon^\nu \rmint_{\Omega}e^{-ik\cdot x}\rmint d\mu_{1,2}(k)\,\frac{\eta_{\mu\nu}}{(k_1^2-m^2)\,(k_2^2-m^2)}\,.\\ &
       \label{6.38a}
    \end{split}
\end{align}
But as $\epsilon$ is null, i.e. $\epsilon^2=0$, the index structure of the above integral make sure it's contribution to the  waveform vanishes.\\\\
\textbullet $\,\,$ Next we focus on the contribution  to the gravitational waveform from massive scalar field through derivative interaction. The massless counterpart of it is shown in \eqref{5.63a}. The time domain waveform has the following form,
\begin{align}
    \begin{split}
          f_{h}(x)&
        \propto\frac{1}{4\pi m_p}\epsilon^\mu \epsilon^\nu \rmint_{\Omega}e^{-ik\cdot x}\rmint_{k_2}e^{i(k-k_2)\cdot b_1}e^{ik_2\cdot b_2}\frac{(k-k_2)_{\mu}k_{2\nu}}{(k_2^2-m^2)[(k-k_2)^2-m^2]}\,\hat\delta\Big(k\cdot v_1-k_2\cdot v_1\Big)\hat\delta(k_2\cdot v_2)\,,\\ &
=-\frac{1}{4\pi m_p}\epsilon^\mu \epsilon^\nu \rmint_{\Omega}e^{-ik\cdot x}\rmint_{k_2}e^{i(k-k_2)\cdot b_1}e^{ik_2\cdot b_2}\frac{k_{2\mu}k_{2\nu}}{(k_2^2-m^2)[(k-k_2)^2-m^2]}\,\hat\delta\Big(k\cdot v_1-k_2\cdot v_1\Big)\hat\delta(k_2\cdot v_2)\,,\\ &
=-\frac{1}{4\pi m_p}\epsilon\cdot \hat J_{(2)}\cdot \epsilon\,.
    \end{split}
\end{align}
where,
\begin{align}
    \begin{split}
        \hat J_{\mu\nu}^{(2)}=J_{\mu\nu}^{(2)}[(1-y)^2\to (1-y+y^2)].
    \end{split}
\end{align}
where $J_{\mu\nu}^{(2)}$ is defined in (\ref{A.22 m}).\\\\
\textbullet $\,\,$
The other diagram corresponding to the graviton radiation from worldline-1 with bulk $\lambda \varphi^3$ vertex is given by,

\begin{align}
    \begin{split}
        k^2\Big\langle h_{\mu\nu}(k)\Big\rangle\Big|_{k^2\rightarrow 0}= \begin{minipage}[h]{0.12\linewidth}
	\vspace{4pt}
	\scalebox{1.6}{\includegraphics[width=\linewidth]{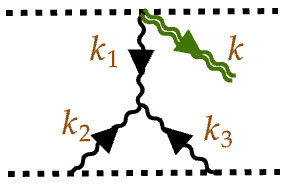}}
\end{minipage}\hspace{1.4 cm}=\rmint d\mu_{1,2}(k)\frac{1}{(k_1^2-m^2)(k_2^2-m^2)(k_3^2-m^2)}\,,
    \end{split}
\end{align}
where the measure looks like,
\begin{align}
    \begin{split}
        d\mu_{1,2}(k)=\rmint_{k_i}\hat\delta^{(4)}\Big(\sum_{i}k_i\Big)\hat\delta(k_1\cdot v_1+k\cdot v_1)\hat\delta(k_2\cdot v_2)\hat\delta(k_3\cdot v_2)e^{i(k+k_1)\cdot b_1}e^{i(k_2+k_3)\cdot b_2}\,.
    \end{split}
\end{align}
Therefore, omitting the prefactors the corresponding waveform takes the form,
\begin{align}
    \begin{split}
        f(x)&\sim \rmint_{\Omega}e^{-ik\cdot x}\rmint_{k_i}\frac{\hat\delta^{(4)}\Big(\sum_{i}k_i\Big)\hat\delta(k_1\cdot v_1+k\cdot v_1)\hat\delta(k_2\cdot v_2)\hat\delta(k_3\cdot v_2)}{(k_1^2-m^2)(k_2^2-m^2)(k_3^2-m^2)}e^{i(k+k_1)\cdot b_1}e^{i(k_2+k_3)\cdot b_2}\,,\\ &
        =\rmint_{\Omega}e^{-i\Omega n\cdot (x-b_1)}\rmint_{k_1,k_2}\frac{\hat\delta(k_1\cdot v_1+k\cdot v_1)\hat\delta(k_2\cdot v_2)\hat\delta(k_1\cdot v_2)}{(k_1^2-m^2)(k_2^2-m^2)[(k_1+k_2)^2-m^2]}e^{ik_1\cdot b}\,,\\ &
        =-\frac{4\pi(2\pi)^2}{(n\cdot v_1)|\vec \delta_1|}\rmint_{0}^{\infty} dk_1 \frac{\sin{(k_1|\vec \delta_1|)}}{(k_1^2+m^2)}\arctan\Big(\frac{|\vec k_1|}{2m}\Big)\,.
    \end{split}
\end{align}
Now, restoring the prefactors the waveform can be recasted as,
\hfsetfillcolor{gray!10}
\hfsetbordercolor{black}

\begin{align}
    \begin{split}
    \tikzmarkin[disable rounded corners=false]{e311}(0.25,-1)(-0.3,1.3) 
        f(x)=-\frac{\lambda_3 }{2(2\pi)^3}\Big(\frac{m_1 s_1m_2^2 s_2^2}{4m_p^3}\Big)\frac{1}{(n\cdot v_1)|\vec \delta_1|}\rmint_{0}^{\infty} dy \frac{\sin{(y|\vec \delta_1|)}}{(y^2+m^2)}\arctan\Big(\frac{y}{2m}\Big)\,.
        \tikzmarkend{e311}
    \end{split}
\end{align}
\\
\section{Discussions and outlook}\label{sec8}
Inspired by the recent developments in scattering amplitude techniques in QFT, we consider a scalar-tensor theory of gravity. After briefly discussing the novel WQFT formalism \cite{Mogull:2020sak}, we compute the two main observables in a scattering event for scalar-tensor theory. We compute the impulse and waveform coming from the scalar contribution to the gravitational field by computing the one-point correlators for the fields (scalar and graviton) and worldline degrees of freedom. \textit{To the best of our knowledge, this is the first study regarding applying WQFT to compute impulse and waveform for a non-GR theory, namely the Scalar-Tensor theory. Also, note that this study helps us extend the computation of the gravitational waveform in the post-Minkwoskian regime for a theory beyond GR.} Below, we summarize the main findings of our paper.
\begin{itemize}
    \item First we compute the impulse $\Delta p^\mu$ in the massive scalar tensor theory with scalar potential $V(\varphi)\sim \lambda_3 \varphi^3+\lambda_4 \varphi^4$ up to 2PM. We mainly concentrate on the corrections to the impulse coming from the scalar degree of freedom. Also, we have provided the corresponding expressions when we take the massless limit. \textcolor{black}{Note that to the best of our knowledge, some of the massive integrals as encountered in this paper do not possess any closed-form expression. They can be evaluated numerically and can be shown to have a smooth behaviour w.r.t. $|b|.$ Also, they possess smooth massless limits as discussed in the main text. Hence, these massless expressions serve as a consistency check of our results.}  The fall-off in that case w.r.t. $|b|$ is analogous to the gravitational case.

\item Next, we compute the radiation integrals coming from the field one point function $\langle\chi(k)\rangle$ and time domain waveform in the massless case for the scalar waveform as well as the correction due to gravitational waveform due to the presence of bulk scalar interaction vertices. Note that total waveform will also have contributions from the GR term, which is already known in the literature. As discussed in the introduction (\ref{intro}), the simplest way to add extra degrees of freedom is to introduce a scalar field. Further motivation was provided for considering the Scalar-Tensor theory there. Keeping this in mind, we have focused only on the correction due to the presence of extra scalar degrees of freedom. To the best of our knowledge, this is the first study of PM waveform for a Scalar-Tensor theory. 
\item Eventually, we proceed to the compute the waveform where the scalar field has non-zero mass. In this case, we find that the analytical (exact) computation of waveform becomes significantly difficult due to the presence of a complicated phase structure. \textit{We propose a procedure to handle those integrals using the stationary phase approximation.} As we expect a burst kind of signal from scattering events, this is a reasonable approximation one can make. 
    \item Apart from that, we would like to emphasize that we show different approaches to computing different kinds of Feynman integrals and discuss the underlying subtleties. Some of these integrals do not appear in the corresponding GR computation. We hope that this will help while exploring WQFT methods for other non-GR theories of gravity. 
\end{itemize}
The connection to the scattering amplitude is trivial in this process. One need not compute the effective action by integrating out the different energy modes, similar to the effective field theory techniques \cite{Bhattacharyya:2023kbh}. To this end, one eventually encounters the loop integrals that one will achieve in scattering processes with intermediate loops.
To the best of our knowledge, we tried to compute diagrams of radiation and impulse with massive propagators in several places, which is new to the literature of WQFT for scalar-tensor theories. It will also be quite an interesting follow-up to incorporate the spin of each black hole by considering finite-size effects in the worldline action. In fact, WQFT of $\mathcal{N}=1 $ \textit{supersymmetric spinning particles} exist in literature \cite{Jakobsen:2021lvp}. Eventually, as a follow-up, one can also try to calculate 3PM three-body radiation. As an advantage of this procedure, one can do this without taking the classical limit ($\hbar\rightarrow 0$) explicitly in each diagram computation. Last but not least, as the connection to the scattering amplitude is apparent, it is important to investigate the Double-copy setup to make it clear in this formalism. We hope to report on some of these issues in the near future. 


\section*{Acknowledgments}
We would like to thank Alok Laddha for his insightful comments on some parts of the draft. We also thank Abhishek Chowdhuri for collaborating in the initial stage of the work. D.G. would like to thank Samim Akhtar for his insightful comments on some parts of the draft. A.B. would like to thank the speakers of the workshop “Testing Aspects of General Relativity-II” (11-13th April 2023) and “New insights into particle physics from quantum information and gravitational waves” (12-13th June 2023) at Lethbridge University, Canada, funded by McDonald Research Partnership-Building Workshop grant by McDonald Institute for useful discussions. S.G (PMRF ID: 1702711) and S.P (PMRF ID: 1703278) are supported by the Prime Minister’s Research Fellowship of the Government of India. A.B also like to thank the Department of Physics and Astronomy of the University of Lethbridge, especially Saurya Das and  FISPAC Research Group, Department of Physics, University of Murcia, especially Jose J. Fernández-Melgarejo for hospitality during the course of this work. S.G and S.P acknowledge the support from the International Centre for Theoretical Sciences (ICTS) during the course of the work while attending a school there. A.B is supported by the Core Reserach Grant (CRG/2023/005112), Mathematical Research Impact Centric Support Grant (MTR/2021/000490) by the Department of Science and Technology Science and Engineering Research Board (India) and the Relevant Research Project grant (202011BRE03RP06633-BRNS) by the Board Of Research In Nuclear Sciences (BRNS), Department of Atomic Energy (DAE), India. A.B also acknowledge the associateship program of the Indian Academy of Science (IASc), Bengaluru.

\appendix
\section{Sketching the derivation of the worldline action} \label{appnew}
In our case, the matter Lagrangian has the following form:
\begin{align}
    \begin{split}
        \mathcal{L}=g^{\mu\nu}\partial_\mu\phi_i^{\dagger}\partial_\nu\phi_i-m_i(\varphi)^2\phi_i^{\dagger}\phi_i
    \end{split}
\,.\end{align}
We start by representing the partition function (in Euclidean signature)  using the Schwinger proper time parametrization.
\begin{align}
    \Gamma[g,\varphi]=\log\Big[\rmint \mathcal{D}[\phi,\phi^{\dagger}]\,e^{-S}\Big] &=-\log\Big[\det(\nabla_{\mu}\nabla^\mu+m(\varphi)^2)\Big]\,,\\ &
    =-\textrm{Tr}\log\Big[\nabla_{\mu}\nabla^\mu+m(\varphi)^2\Big]\,,\\ &
    =\rmint_0^\infty \frac{dT}{T}\rmint \frac{d^4 k}{(2\pi)^4}\exp\Big[-\frac{1}{2}eT\Big(g_{\mu\nu}k^\mu k^\nu+m(\varphi)^2\Big)\Big]
\end{align}
where, $e$ is the einbein. Now we convert the result into the path integral over $x(\tau)$.
\begin{align}
    \begin{split}
        \Gamma[g,\phi]=\rmint_0^\infty \frac{dT}{T}\mathcal{N}(T)\rmint \mathcal{D}[x]\exp\Big[-\rmint d\tau (\frac{1}{2e}g_{\mu\nu}\dot x^\mu\dot x^\nu+\frac{e}{2}m(\varphi)^2)\Big]\,.\label{D.5}
    \end{split}
\end{align}
The result in \eqref{D.5} is a one-dimensional field theory of $x^\mu(\tau)$. For the  case of gravitational field, the integral measure becomes metric dependent which causes the existence of Lee-Yang ghost. However in classical limit the ghost fields are irrelevant and can be ignored as shown in \cite{Mogull:2020sak}.
\section{Impulse via Eikonal: a connection to scattering amplitude}\label{sec7}
In section~(\ref{sec1}) we computed impulse up to 2PM. Now we discuss it's connection with scattering amplitude. It is shown in \cite{Amati:1987wq, Amati:1990xe} that the Eikonal phase $\chi$ is related to 4-point scattering amplitude as,
\begin{align}
    \begin{split}
e^{i\chi}&\equiv 1+\frac{i}{4m_1 m_2}\rmint e^{iq\cdot b}\hat\delta(q\cdot v_1)\hat\delta(q\cdot v_2)\lim_{\hbar\rightarrow 0}\mathcal{M}_4(\phi_1,\phi_2\rightarrow \phi_1,\phi_2)\,,\\ &
=1+\frac{i}{4m_1 m_2}\rmint_{q_{\perp}}e^{i q_{\perp}\cdot b}\lim_{\hbar\rightarrow 0}\mathcal{M}_4(\phi_1,\phi_2\rightarrow \phi_1,\phi_2)\,.
    \end{split}
\end{align}
It has been demonstrated in \cite{Bern:2020buy,Bjerrum-Bohr:2018xdl} that the Eikonal phase is related to the impulse upto 2PM (in the centre of mass frame) order takes the following form,
\begin{align}
    \begin{split}
        \Delta p_{\perp}= \frac{\partial \chi}{\partial b}\,.
    \end{split}
\end{align}
Later, it has been shown that the result can be extended to higher PM order \cite{Maybee:2019jus,Bern:2020buy}. In WQFT, one can identify the classical part of $\chi$ to be the free energy of the WQFT at tree level and hence given by,
\begin{align}
    \begin{split}
    e^{i\chi(\hat b_i,\hat v_i)}= Z_{\textrm{WQFT}}:=\mathcal{N}\rmint \mathcal{D}h_{\mu\nu}\mathcal{D}\varphi\,\rmint\prod_{k=1}^2\mathcal{D}z_{k}\exp\Big(iS_{g}+iS^k_{pm}\Big)
    \end{split}
\end{align}
where, $\hat b$ and $\hat v$ can be related to the averaged incoming momenta $\hat p_i$ which satisfies, $\hat p_1\Delta p_1=0$. Therefore, the formula for impulse is given by,
\begin{align}
    \begin{split}
        \Delta p_{1\mu}=-\frac{\partial \chi}{\partial \hat b_1^\mu}.
    \end{split}
\end{align}
If one can compute the impulse using the Eikonal method, one may lose some extra terms in the impulse, which are proportional to $v_i^\mu$. If one blindly computes the impulse, for example, for the following diagram, 
\begin{align}
    \begin{split}
        \chi&\propto \begin{minipage}[h]{0.12\linewidth}
	\scalebox{1.5}{\includegraphics[width=\linewidth]{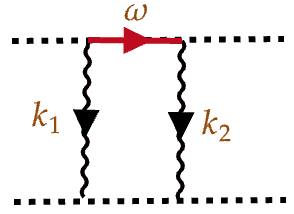}}\end{minipage} \\ &
= \rmint  \check{\mathcal{D}}[h_{\mu\nu}, \varphi,\{z_{i}\}]\rmint_{\{k_i,\textasciicaron{\omega}_j\}}\hat{\delta}(k_1\cdot v_2)\hat{\delta}(k_2\cdot v_2) \varphi(k_1) \varphi(k_2) e^{i(k_1+k_2)\cdot b_2}\\ &
        \hspace{3.7 cm}\hat\delta(k_3\cdot v_1+\omega_1)\hat\delta(k_4\cdot v_1+\omega_2)\varphi(k_3)\varphi(k_4)z^{\rho}(\omega_1)z^{\sigma}(\omega_2)\{2\omega_1 (v_1)_{\rho}+(k_{3})_{\rho}\}\\ &
        \hspace{3.7 cm}
        \{2\omega_2(v_1)_{\sigma}+(k_4)_{\sigma}\}\hat\delta^{(4)}(k_2+k_4)\hat\delta^{(4)}(k_1+k_3)\delta(\omega_1+\omega_2)\,,\\&
        =\rmint_{\{k_i,\omega_j\}}\hat{\delta}(k_1\cdot v_2)\hat{\delta}(k_2\cdot v_2)\hat\delta(k_3\cdot v_1+\omega_1)\hat\delta(k_4\cdot v_1+\omega_2)\hat\delta^{(4)}(k_1+k_3)\hat\delta^{(4)}(k_2+k_4)\delta(\omega_1+\omega_2)\\ &
       \hspace{2 cm} \frac{1}{(k_1^2-m^2)(k_2^2-m^2)\omega_1^2}\{4\omega_1\omega_2+2\omega_1 v_1\cdot k_4+2 \omega_2 v_1\cdot k_3+k_3\cdot k_4\}e^{i(k_1+k_2)\cdot b_2}e^{i(k_3+k_4)\cdot b_1}\,,\\ &
=\rmint_{k,k_1,\omega_1}\frac{\hat\delta(k_1\cdot v_2)\hat\delta(k\cdot v_2)\hat\delta(k\cdot v_1)\hat\delta(\omega_1-k_1\cdot v_1)}{(k_1^2-m^2)[(k-k_1)^2-m^2]\omega_1^2}e^{ik \cdot b}\\&\hspace{4.2 cm}[-4\omega_1^2+2\omega_1v_1\cdot(k_1-k)+2\omega_1v_1\cdot k_1+k_1\cdot (k-k_1)]\,.
    \end{split}
\end{align}
Integrating over $\delta(\omega_1-k_1\cdot v_1)$ and using the fact that the integral has support at $k\cdot v_1=0$, we left with the following integral,
{
}
{
The integral can be done 
\begin{align}
    \begin{split}
\chi&\sim\rmint_{k,k_1}\frac{\hat\delta(k_1\cdot v_2)\hat\delta(k\cdot v_2)\hat\delta(k\cdot v_1)}{(k_1^2-m^2)[(k-k_1)^2-m^2](k_1\cdot v_1)^2}[k_1\cdot(k-k_1)]e^{ik\cdot b}\,,\\ &
        =\frac{1}{2}\rmint_{k} e^{ik\cdot b}{\hat\delta(k\cdot v_1)\hat\delta(k\cdot v_2)}(k^2-2m^2)\underbrace{\rmint_{k_1}\frac{\hat\delta (k_1\cdot v_2)}{(k_1^2-m^2)[(k-k_1)^2-m^2](k_1\cdot v_1)^2}}_{\hat\chi_{k}}\,.\label{B.8a}
    \end{split}
\end{align}
One can see that the delta function inside the $k_1$ integral reduces the four dimensional integral into a three dimensional integral which reads,
\begin{align}
    \begin{split}
        \hat\chi_{k}\sim 4\rmint_{\vec k_1} \frac{1}{(\vec k_1^2+m^2)[(\vec k-\vec k_1)^2+m^2](2\,\vec k_1\cdot \vec v_1)^2}\,.\label{B.9c}
    \end{split}
\end{align}
The integral can be done by introducing the alpha parametrization \cite{smirnov2}, 
\begin{align}
    \frac{1}{(-A)^\lambda}=\frac{i^\lambda}{\Gamma(\lambda)}\rmint_{0}^{\infty}d\alpha \, \alpha^{\lambda-1}e^{iA\alpha}
\end{align}
which gives,
\begin{align}
    \begin{split}
      (2\pi)^3  \hat\chi_{k}&=\frac{4}{\Gamma(1)^2\Gamma(2)}\rmint d^{3-2\epsilon}k_1\rmint_0^\infty  \prod_{i=1}^3 d\alpha_i\,\alpha_3 \exp\Big[i( \vec k_1^2+m^2)\alpha_1+i[(\vec k_1-\vec k)^2+m^2]\alpha_2 +i\,2(\vec k_1\cdot \vec v_1)\alpha_3\Big]\,,\\ &
        =\frac{4}{\Gamma(1)^2\Gamma(2)}\rmint_0^\infty \prod_{i=1}^3 d\alpha_i\,\alpha_3\,e^{i(\alpha_1+\alpha_2)m^2}e^{i\alpha_2\vec k^2}\rmint d^{3-2\epsilon}k_1\exp\Big[i\vec k_1^2(\alpha_1+\alpha_2)-2\vec k_1\cdot (\alpha_2 \vec k-\alpha_3 \vec v_1)\Big]\,,\\ &
        =\frac{4}{\Gamma(1)^2\Gamma(2)}e^{i\frac{\pi}{2}(\epsilon-\frac{1}{2})}\pi^{\frac{3}{2}-\epsilon}\rmint_0^\infty  \prod_{i=1}^3 d\alpha_i\,\alpha_3\,e^{i(\alpha_1+\alpha_2)m^2}(\alpha_1+\alpha_2)^{\epsilon-\frac{3}{2}}\exp\Big(i\frac{\alpha_1\alpha_2\vec k^2-\alpha_3^2\vec v_1^2}{\alpha_1+\alpha_2}\Big)\,.
    \end{split}
\end{align}
Now doing the integral over $\alpha_3$ we get,
\begin{align}
    \begin{split}
      (2\pi)^3  \hat\chi_{k}&= \frac{4}{\Gamma(2)\Gamma(1)^2}e^{i\frac{\pi}{2}(\epsilon-\frac{1}{2})}\pi^{\frac{3}{2}-\epsilon}\rmint_0^\infty  \prod_{i=1}^2 d\alpha_i \,(\alpha_1+\alpha_2)^{\epsilon-\frac{3}{2}}\,e^{i(\alpha_1+\alpha_2)m^2}\exp\Big(i\frac{\vec{k}^2\alpha_1\alpha_2-\alpha_3^2 \gamma^2\beta^2}{\alpha_1+\alpha_2}\Big)\,,\\ &
        =-\frac{4i}{\gamma^2\beta^2}e^{i\frac{\pi}{2}(\epsilon-\frac{1}{2})}\pi^{\frac{3}{2}-\epsilon}\rmint_0^\infty  \prod_{i=1}^2 d\alpha_i (\alpha_1+\alpha_2)^{\epsilon-\frac{1}{2}}\underbrace{e^{i(\alpha_1+\alpha_2)m^2}}_{\textrm{extra term coming due to non zero mass}}\exp\Big(i\frac{\vec{k}^2\alpha_1\alpha_2}{\alpha_1+\alpha_2}\Big)\,.
    \end{split}
\end{align}
Now doing the following variable change,
$$\alpha_1+\alpha_2\to \eta\,\,\,\textrm{and}\,\,\, \frac{\alpha_1}{\alpha_1+\alpha_2}\to \xi$$ we get,
\begin{align}
    \begin{split}
     (2\pi)^3   \hat\chi_{k}=&-\frac{4i}{(\gamma^2-1)} e^{i\frac{\pi}{2}(\epsilon-\frac{1}{2})}\pi^{3/2-\epsilon}\rmint_0^1 d\xi \rmint_0^\infty d\eta \,\eta^{\epsilon+\frac{1}{2}}\exp\Big[i\eta\{m^2+\vec{k}^2\xi(1-\xi)\}\Big]\,,\\ &
        =\frac{4}{(\gamma^2-1)}\lim_{\epsilon\to 0}e^{i\epsilon\pi}\pi^{5/2-\epsilon}\frac{    \sec (\pi  \epsilon ) \left(B_{z_1}\left(-\epsilon -\frac{1}{2},-\epsilon -\frac{1}{2}\right)-B_{z_2}\left(-\epsilon -\frac{1}{2},-\epsilon -\frac{1}{2}\right)\right)}{k\,\left(k^2+4 m^2\right)^{\epsilon +1}  \Gamma \left(-\epsilon -\frac{1}{2}\right)}\label{B.12}
    \end{split}
\end{align}
where,
\begin{align}
   z_1= \frac{1}{2}-\frac{|\vec{k}|}{2 \sqrt{\vec{k}^2+4 m^2}} \,\textrm{and,}\,\, z_2=\frac{1}{2}+\frac{|\vec{k}|}{2 \sqrt{\vec{k}^2+4 m^2}}\,.
\end{align}
The beta functions can be written in a simplified manner to obtain,
\begin{align}
\begin{split}
    & B_{z_1}(-\epsilon-\frac{1}{2} ,-\epsilon-\frac{1}{2} )-B_{z_2}(-\epsilon-\frac{1}{2} ,-\epsilon-\frac{1}{2} )\\ &=-\rmint_{z_1}^{z_2}dt \,t^{-\epsilon-\frac{3}{2}}(1-t)^{-\epsilon-\frac{3}{2}}\,,\\ &
      =\frac{1}{\epsilon +\frac{1}{2}}\left(\frac{2m^2}{\vec k^2+4m^2}\right)^{-\epsilon -\frac{1}{2}} \Big[\left(\frac{|\vec k|}{\sqrt{\vec k^2+4 m^2}}+1\right)^{\epsilon +\frac{1}{2}} \,\, _2F_1\left(-\epsilon -\frac{1}{2},\epsilon +\frac{3}{2};\frac{1}{2}-\epsilon ;\frac{1}{2}-\frac{|\vec k|}{2 \sqrt{ \vec k^2+4 m^2}}\right)\\ &-\left(1-\frac{|\vec k|}{\sqrt{\vec k^2+4 m^2}}\right)^{\epsilon +\frac{1}{2}} \, _2F_1\left(-\epsilon -\frac{1}{2},\epsilon +\frac{3}{2};\frac{1}{2}-\epsilon ;\frac{|\vec k|}{2 \sqrt{\vec k^2+4 m^2}}+\frac{1}{2}\right)\Big]\,.
     \end{split}
     \end{align}
     Expanding around $\epsilon=0$ we get,
     \begin{align}
         \begin{split}
             (2\pi)^3\hat\chi_{k}&=\frac{8\,\pi^{5/2}}{(\gamma^2-1)}\frac{1}{m\,(\vec k^2+4m^2)}.\label{B.16ab}
         \end{split}
     \end{align}
     The integral $\hat\chi_k$ defined in \eqref{B.8a} can be done using Integral-by-Parts (IBP) reduction in a more sophisticated way. We have,
     \begin{align}
      \chi_k=   \int_{k_1}\frac{\hat\delta (k_1\cdot v_2)}{(k_1^2-m^2)[(k-k_1)^2-m^2](k_1\cdot v_1)^2}
     \end{align}
By unitarity cut the integral can be re written as,
\begin{align}
    \begin{split}
        \chi_k\sim \int_{k_1} \frac{1}{(k_1^2-m^2)[(k-k_1)^2-m^2](k_1\cdot v_1)^2}\Bigg[\frac{1}{k_1\cdot v_2+i\sigma}-\frac{1}{k_1\cdot v_2-i\sigma}\Bigg]\label{B.16a}
    \end{split}
    \end{align}
  The integral in \eqref{B.16a} belongs to the following family of integral,
  \begin{align}
      \begin{split}
          \mathcal{G}^{(\pm)}_{n_1,n_2,n_3,n_4}=\int_{k_1}\frac{1}{D_1^{n_1}D_2^{(\pm)n_2}D_3^{n_3}D_4^{n_4}}
      \end{split}
  \end{align}
  where,
  \begin{align}
      D_1=k_1\cdot v_1,\,D_2^{(\pm)}=k_1\cdot v_2\pm i\sigma,\,D_3=k_1^2-m^2, \,\textrm{and},\,D_4=(k_1-k)^2-m^2
  \end{align}
  Hence it is evident that  the integral $\chi_k= \mathcal{G}_{2,1,1,1}$ can be evaluated  in terms of master integrals using \textbf{LiteRed} \cite{Lee:2012cn} as follows,
  \begin{align}
  \begin{split}
    \chi_k&\sim  \mathcal{G}^{(+)}_{2,1,1,1}- \mathcal{G}^{(-)}_{2,1,1,1}\\ &
    =-\frac{2}{m^2(\gamma^2-1)(4m^2-k^2)}\Big[\mathcal{G}_{0,1,0,1}^{(+),\textrm{master}}-\mathcal{G}_{0,1,0,1}^{(-),\textrm{master}}\Big]+\mathcal{O}(\sigma)\\ &
    =-\frac{2}{m^2(\gamma^2-1)(4m^2-k^2)}\int_{k_1}\frac{\hat\delta(k_1\cdot v_2)}{(k_1-k)^2-m^2},\,(\textrm{via reverse unitarity.})\\ &
    \propto \frac{2}{m(\gamma^2-1)(4m^2+\vec k^2)}
    \end{split}
  \end{align}
  which exactly matches with the result given in \eqref{B.16ab}. In the last line use use the fact that $\hat\delta(k\cdot v_i)=0\,,\forall i=1,2$. Finally, the Eikonal phase is given by,
     \begin{align}
         \begin{split}
             \chi&\sim\frac{1}{2} \rmint_0^\infty dk\,k\, J_{0}(k|b|)(k^2+2m^2)\chi_k(m)\,,\\ &
             =\frac{1}{m\sqrt{\pi}(\gamma^2-1)}\rmint_0^\infty dk\, k\,J_{0}(k|b|)\frac{(k^2+2m^2)}{k^2+4m^2}\,,\\ &
             =\frac{1}{m\sqrt{\pi}(\gamma^2-1)}\rmint_0^\infty dk\, k\,J_{0}(k|b|)\Big[1-\frac{2m^2}{k^2+4m^2}\Big]\,.\label{B.15}
         \end{split}
     \end{align}
     The first term in \eqref{B.15} does not have any non-zero finite value (purely UV divergent) and hence can be ignored in the classical limit. Now the second term gives,
     \begin{align}
         \begin{split}
             \chi\sim -\frac{2 m}{\sqrt{\pi}(\gamma^2-1)}K_{0}(2 m|b|)\,.
         \end{split}
     \end{align}
We can see that the massive expression in \eqref{B.15} is non-zero in general.
 We will now show that there will be a non-vanishing. contribution to the $a_1$ 
 start with:
\begin{align}
    \begin{split}
        \mathcal{K}^\mu= a_1k^\mu+a_2 v_1^\mu+a_3 v_2^\mu.\label{B.18b}
    \end{split}
\end{align}
$a_2$ and $a_3$ s are already derived. Now contracting both side of \eqref{B.18b} with $k^\mu$ we get,
\begin{align}
    \begin{split}
        k\cdot \mathcal{K}= a_1 k^2.
    \end{split}
\end{align}
Therefore,
\begin{align}
    \begin{split}
        a_1&=\frac{1}{k^2}\rmint_{k_1}\hat\delta(k_1\cdot v_2)\frac{k_1\cdot (k-k_1)\,k\cdot (k-k_1)}{(k_1^2-m^2)[(k_1-k)^2-m^2](k_1\cdot v_1+i\epsilon)^2}\,,\\ &
        =\frac{1}{2k^2}\rmint_{k_1}\hat\delta(k_1\cdot v_2)k_1\cdot (k-k_1)\Big[\frac{k^2}{(k_1^2-m^2)[(k_1-k)^2-m^2](k_1\cdot v_1+i\epsilon)^2}+\frac{1}{(k_1^2-m^2)(k_1\cdot v_1+i\epsilon)^2}\\ &
        \hspace{6 cm }-\frac{1}{[(k_1-k)^2-m^2](k_1\cdot v_1+i\epsilon)^2}\Big]\,.\label{B.20a}
    \end{split}
\end{align}
The second and third terms in \eqref{B.20a} will cancel each other by  virtue of relabelling the third term by $k_1-k\to -k_1$ and taking into account $\hat\delta(k\cdot v_i)$\footnote{Note that, implicitly we take $\epsilon\to 0$ beforehand. But if we take this limit at the end, it will result in the same. For  specified $i\epsilon$ prescription the last two terms for \eqref{B.20a} takes the form,\\
\begin{align}
\begin{split}\int_{k_1}\hat\delta(k_1\cdot v_2)\frac{k_1\cdot (k-k_1)}{k_1^2-m^2}\hat\delta'(k_1\cdot v_1)&\to -\int_{\vec k_1}\frac{d}{dk_1^{(1)}}\hat\delta(k_1^{(1)})+\int_{ k_1^{(2,3)}}(-m^2+\vec k_1\cdot \vec k)\int_{k_1^{(1)}}\frac{1}{k_1^{(1)2}+k_1^{(2)2}+k_1^{(3)2}+m^2}\frac{d}{dk_1^{(1)}}\hat\delta(k_1^{(1)})\\ &
\to 0
\end{split}
\end{align}}. Therefore, we are left with,
\begin{align}
    \begin{split}
        a_1&=\frac{1}{2}\rmint_{k_1}\hat{\delta}(k_1\cdot v_2)\frac{k_1\cdot(k-k_1)}{(k_1^2-m^2)[(k_1-k)^2-m^2](k_1\cdot v_1+ i \epsilon)^2}\,,\\ &
        =\frac{1}{4}(k^2-2m^2)\hat\chi(k,m)\,.
    \end{split}
\end{align}
Hence, the correction to the impulse takes the following form,
\begin{align}
    \begin{split}
        \Delta p_1^\mu\Big|_{\textrm{corr.}}&=im_1\Big(\frac{s_1m_2 s_2}{8m_p^2}\Big)^2\rmint_{k}e^{ik\cdot b}\hat\delta(k\cdot v_1)\hat\delta(k\cdot v_2)k^\mu(k^2-2m^2)\,\hat\chi_{k}(k,b)\,,\\ &
        =-\frac{m_1}{2\pi}\Big(\frac{s_1m_2 s_2}{8m_p^2}\Big)^2\frac{1}{\sqrt{\gamma^2-1}}\frac{b^\mu}{|b|}\partial_{|b|}\rmint_{0}^{\infty} dk\,k\,J_{0}(k|b|)(k^2+2m^2)\,\hat\chi_{k}(k,m)\,,\\ &
        =-m_1\Big(\frac{s_1m_2 s_2}{8m_p^2}\Big)^2\frac{1}{\pi^{3/2}(\gamma^2-1)^{3/2}} \frac{b^\mu}{|b|}\Big(m K_1(2 |b| m)\Big)\,.\label{B.22aa}
    \end{split}
\end{align}
}
\textcolor{black}{
\eqref{B.22aa} explicitly indicates that if one computes the impulse from the Eikonal phase, one may get only a part of the impulse which is proportional to $b^\mu$, and other parts (proportional to $v_i^\mu$) will be lost. Along with that \eqref{B.22aa} has a smooth massless limit which reads, $$ \Delta p_1^\mu|_{\textrm{corr.}}\sim \frac{b^\mu}{b^2}\,.$$ }
However, the result in the massless limit direct contradicts with the result in \cite{Kalin:2020mvi}, where they showed $a_1=0$ by ignoring the radiation poles. But it seems there is no such radiation region due to the presence of delta function.
\section{Two master integrals used for the computation of  waveform}\label{app1}
We analyze the following two integrals,
\begin{align}
    \begin{split}
     &   I^{\mu_1\cdots\mu_n}_n:=\rmint_{q}\hat\delta(q\cdot v_1-k\cdot v_1)\hat\delta(q\cdot v_2)\frac{e^{-iq\cdot b}}{q^2}q^{\mu_1}\cdots q^{\mu_n}\,,\\ &
     J^{\mu_1\cdots\mu_n}_n:=\rmint_{q}\hat\delta(q\cdot v_1-k\cdot v_1)\hat\delta(q\cdot v_2)\frac{e^{-iq\cdot b}}{q^2(k-q)^2}q^{\mu_1}\cdots q^{\mu_n}\,.
    \end{split}
\end{align}
The two particular cases that we used in the main text are: $J^{\mu}_{1}$ and $J_{0}$. We first start with $J_{0}\,.$

\begin{align}
    \begin{split}
        J_{(0)}(k)=\rmint_{q}\hat\delta(q\cdot v_1-k\cdot v_1)\hat\delta(q\cdot v_2)\frac{e^{-iq\cdot b}}{q^2(k-q)^2}\,.\label{A.2}
    \end{split}
\end{align}
We will do this integral by implementing a certain choice of frame,
\begin{align}
    v_2^\mu=\delta^\mu_0,\,v_1^\mu=(\gamma,\gamma\beta \,\hat{e}_v)\,b=(0,|b|\,\hat{e}_b)\,,
\end{align}
such that, the unit vectors $\hat{e}_v$ and $\hat{e}_b$ are mutually orthogonal i.e. $\hat{e}_b \cdot \hat{e}_v=0$.
We further proceed with this integral by using the Feynman parametrization method along with the on-shell condition: $k^2=0$. Therefore, $J_0$ can be written as,
\begin{align}
    \begin{split}
        J_{(0)}(k)&=\rmint_0^1dy\rmint_q \hat\delta(q\cdot v_1-k\cdot v_1)\hat\delta(q\cdot v_2)\frac{e^{-iq\cdot b}}{(q-yk)^4},\,k^2=0\,,\\ &
        \xrightarrow[]{\bar q\rightarrow q-yk}\rmint_0^1 dy e^{-iy k\cdot b} \rmint_{\bar q}\hat\delta(\bar q\cdot v_1-(1-y)k\cdot v_1)\hat\delta(\bar q\cdot v_2+y k\cdot v_2)\frac{e^{-i\bar q\cdot b}}{\bar q^4}\,,\\ &
        =\frac{1}{\gamma}\rmint_{0}^1dy\,e^{-iyk\cdot b}\rmint_0^{\infty}dt\,t\rmint_{\tilde q}\exp\Big[i\tilde q\cdot b-t\tilde q^2-t\Delta(y)^2\Big]\,,\\ &
        =\frac{1}{\gamma}\rmint_{0}^1dy\,e^{-iyk\cdot b}\rmint_0^{\infty}\frac{dt}{4\pi}\, \exp(-\frac{|b|^2}{4t}-t\Delta(k,y)^2)\,,\\ &
       = \frac{b}{\gamma}\rmint_{0}^1dy\,e^{-iyk\cdot b}\frac{ K_1\left(|b| {\Delta(k,y) }\right)}{4 \pi  {\Delta(k,y) }}\label{C.4 mm}
    \end{split}
\end{align}
where, 
\begin{align}
    \begin{split}
       \Delta(k,y)&=\sqrt{-y^2(k\cdot v_2)^2+\frac{y^2}{\beta^2}(k\cdot v_2)^2+\frac{(1-y)^2}{\gamma^2\beta^2}(k\cdot v_1)^2+2\frac{y(1-y)}{\gamma\beta^2}(k\cdot v_2)(k\cdot v_1)}\,,\\ &
        =\frac{1}{\sqrt{\gamma^2-1}}\sqrt{y^2(k\cdot v_2)^2+(1-y)^2(k\cdot v_1)^2+2y(1-y)\gamma (k\cdot v_2)(k\cdot v_1)}\,.\label{A.4}
    \end{split}
\end{align}
In the waveform calculation, we have to further  perform a integral over $k$, which in general has the following form,
\begin{align}
    \begin{split}
        f(x):=\rmint_{k}e^{-ik\cdot x}T_{\mu_1\cdots \mu_n}J^{\mu_1\cdots \mu_n}(k)\,.
    \end{split}
\end{align}
In the case of $\varphi^3$ interaction we have,
\begin{align}
    \begin{split}
        f(x)&=\rmint_{k}e^{-ik\cdot x}J_{0}(k), \,k=\Omega n\,,\\ &
        =\frac{b}{4\pi\gamma}\rmint_{\Omega}e^{-i\Omega n\cdot x}\rmint_{0}^1 dy\,e^{-i\,y\,\Omega(n\cdot b)}\frac{K_1[|\Omega|\,|b|\bar\Delta(y)]}{|\Omega|\bar\Delta(y)}\,\,,\bar\Delta(y)\equiv\frac{\Delta(k,y)}{|\Omega|}\,.\label{A.6}
    \end{split}
\end{align}
Now to do the integral over $\Omega$, use the following representation of the modified Bessel function:
\begin{align}
    \begin{split}
        K_{\nu}(x)=\frac{1}{2}(\frac{x}{2})^{\nu}\rmint_{0}^{\infty}dt\,\exp\Big(-t-\frac{x^2}{4t}\Big)\frac{1}{t^{\nu+1}}\,.
    \end{split}
\end{align}
Therefore,
\begin{align}
    \begin{split}
        f(x)&=\frac{|b|}{4\pi\gamma}\rmint_{0}^{1}dy\, \frac{1}{\Bar\Delta(y)}\rmint_{\Omega}e^{-i\Omega n\cdot(x+yb)}\frac{|b|}{4}\Bar{\Delta(y)}\rmint \frac{dt}{t^2}\,\exp\Big(-t-\frac{\Omega^2|b|^2\bar\Delta(y)^2}{4t}\Big)\,,\\ &
        =\frac{|b|^2}{16\pi\gamma}\rmint_{0}^{1}dy\, \rmint_{0}^{\infty} \frac{dt}{t^2}e^{-t}\rmint d\Omega\,e^{-i\,\Omega\, l}\exp\Big(-\frac{\Omega^2|b|^2\bar\Delta(y)^2}{4t}\Big)\,,\\ &
         =\frac{|b|^2}{16\pi\gamma}\rmint_{0}^{1}dy\, \rmint_{0}^{\infty} \frac{dt}{t^2}e^{-t}\frac{2 \sqrt{\pi } \sqrt{t} e^{-\frac{l^2 t}{|b|^2 \bar\Delta(y) ^2}}}{|b| \bar\Delta(y) }\,,\\ &
          =-\frac{|b|}{4\gamma}\rmint_{0}^{1}dy\,\frac{1}{\Bar\Delta(y)}\sqrt{\frac{l^2}{|b|^2\bar\Delta^2}+1},\,l:=n\cdot(x+yb)\,.\label{A.8}
    \end{split}
\end{align}
Now come discuss the other important vector integral:
\begin{align}
    \begin{split}
J_{(1)}^{\mu}&=\rmint_q\hat\delta(q\cdot v_1-k\cdot v_1)\hat\delta(q\cdot v_2)e^{-iq\cdot b}\frac{q^{\mu}}{q^2(k-q)^2}\,,\\ &
=\rmint_{0}^1dy\, e^{-iyk\cdot b}\rmint_{q}\hat\delta(q\cdot v_1-(1-y))\hat\delta(q\cdot v_2+yk\cdot v_2)\frac{e^{-iq\cdot b}}{q^4}(q^\mu+yk^\mu)\,,\\ &
:=\mathcal{J}^{\mu}+\frac{|b|\,k^\mu}{\gamma}\rmint_{0}^1dy\,y\,\,e^{-iyk\cdot b}\frac{K_1(|b|\Delta(k,y))}{4\pi\Delta(k,y)}\,.\label{A.9 m}
    \end{split}
\end{align}
$\mathcal{J}^{\mu}$ can be solved by reducing the vector integral to a scalar integral as follows,
\begin{align}
    \begin{split}
\mathcal{J}^{\mu}=\rmint_{0}^1dy\,e^{-iyk\cdot b}\,[\lambda_b b^{\mu}+\lambda_1 v_1^{\mu}+\lambda_2v_2^\mu]\,,\label{C.11a}
    \end{split}
\end{align}
Now we need to solve the following equations:\par
    \textbullet $\,\,$ Contracting with $ b_{\mu}$ in the both side of \eqref{C.11a} we will get,
\begin{align}
    \begin{split}
 \rmint_{0}^1dy\, e^{-iyk\cdot b}\rmint_{q}\hat\delta(q\cdot v_1-(1-y))\hat\delta(q\cdot v_2+yk\cdot v_2)\frac{e^{-iq\cdot b}}{q^4}q\cdot b=\mathcal{J}\cdot b=-\rmint_0^1 dy\,e^{-iyk\cdot b}b^2\lambda_b\,.
    \end{split}
\end{align}
To do the the $q$ integral of the LHS, we introduce a new parameter $\kappa$ as,
\begin{align}
    \begin{split}
        -|b|^2\lambda_b&=\lim_{\kappa \rightarrow 1}\rmint_{q}\hat\delta(q\cdot v_1-(1-y))\hat\delta(q\cdot v_2+yk\cdot v_2)\frac{e^{-i\kappa q\cdot b}}{q^4}q\cdot b\,,\\ &
        =i\lim_{\kappa \rightarrow 1}\frac{\partial }{\partial \kappa}\rmint_{q}\hat\delta(q\cdot v_1-(1-y))\hat\delta(q\cdot v_2+yk\cdot v_2)\frac{e^{-i\kappa q\cdot b}}{q^4}\,,\\ &
=i\lim_{\kappa\rightarrow 1}\frac{\partial}{\partial \kappa}\hat{J}_0(\kappa |b|),\,\hat J_{0}(|b|)\equiv \frac{|b|}{\gamma}\frac{K_{1}(|b|\Delta(k,y))}{4\pi \Delta(k,y)}\,.
    \end{split}
\end{align}
Hence, 
\begin{align}
    \begin{split}
        \lambda_b=\frac{ i\, K_0(|b| \Delta(k,y) )}{4\pi\gamma }\,.
    \end{split}
\end{align}
 \textbullet $\,\,$ Contracting with $ v_{1\mu}$ in the both side of \eqref{C.11a} we will get,
\begin{align}
    \begin{split}
  \rmint_0^1 dy\,e^{-iyk\cdot b}\rmint_{q}\hat\delta(q\cdot v_1-(1-y)k\cdot v_1)\hat\delta(q\cdot v_2+yk\cdot v_2)\frac{e^{-iq\cdot b}}{q^4}q\cdot v_1=\mathcal{J}\cdot v_1=  \rmint_0^1 dy\,e^{-iyk\cdot b}[\lambda_1+\gamma \lambda_2]\,.
    \end{split}
\end{align}
Again, the $q$ integral in the LHS can be done again by introducing an auxiliary parameter $\kappa$ as,
\begin{align}
    \begin{split} \label{B.16}
        \lambda_1+\gamma \lambda_2&=\rmint_{q}\hat\delta(q\cdot v_1-(1-y)k\cdot v_1)\hat\delta(q\cdot v_2+yk\cdot v_2)\frac{e^{-iq\cdot b-i \kappa q\cdot v_1}}{q^4}q\cdot v_1\,,\\ &
      =  \rmint_{q}\hat\delta(q\cdot v_1-(1-y)k\cdot v_1)\hat\delta(q\cdot v_2+yk\cdot v_2)\frac{e^{-iq\cdot b}}{q^4}(1-y)k\cdot v_1\,,\\ &
      =(1-y)k\cdot v_1 \hat J_0\,.
    \end{split}
\end{align}
 \textbullet $\,\,$ Contracting with $ v_{2\mu}$ in the both side of \eqref{C.11a} we will get,
\begin{align}
    \begin{split}
    \rmint_0^1 dy\,e^{-iyk\cdot b}\rmint_{q}\hat\delta(q\cdot v_1-(1-y)k\cdot v_1)\hat\delta(q\cdot v_2+yk\cdot v_2)\frac{e^{-iq\cdot b}}{q^4}q\cdot v_2=\mathcal{J}\cdot v_2=  \rmint_0^1 dy\,e^{-iyk\cdot b}[\gamma\lambda_1+\lambda_2]\,.
    \end{split}
\end{align}
Hence,
\begin{align}
    \begin{split} \label{B.18}
        \gamma \lambda_1+\lambda_2&=-\rmint_{q}\hat\delta(q\cdot v_1-(1-y)k\cdot v_1)\hat\delta(q\cdot v_2+yk\cdot v_2)\frac{e^{-iq\cdot b}}{q^4}y(k\cdot v_2)\,,\\ &
        =-y k\cdot v_2\hat J_0 \,.
    \end{split}
\end{align}
Now solving (\ref{B.16}) and (\ref{B.18}) we get,
\begin{align}
    \begin{split}
      &  \lambda_2=\frac{y(k\cdot v_2)+\gamma (1-y)k\cdot v_1}{\gamma^2-1}\hat{J}_0\,,\\ &
      \lambda_1=\frac{-\gamma y k\cdot v_2-(1-y)k\cdot v_1}{\gamma^2-1}\hat{J}_0\,.
    \end{split}
\end{align}
Therefore, the contribution to the waveform takes the following form,
\begin{align}
   \begin{split}
       f(x)&=T_{\mu}\rmint_{\Omega}e^{-ik\cdot x}\mathcal{J}^{\mu}\,,\\ &
=T_{\mu}\rmint_{\Omega}e^{-ik\cdot x}\rmint_0^1dy\, e^{-iyk\cdot b}[\lambda_b b^\mu+\lambda_1 v_1^\mu+\lambda_2 v_2^\mu]\,,\\ &
=f_1+f_2+f_3\,,\label{A.19 m}
   \end{split}
\end{align}
where, 
\begin{align}
    \begin{split}
        f_1(x|l)&=\frac{-iT\cdot b}{4\pi \gamma}\rmint_{\Omega}e^{-i\Omega\,(n\cdot x)}\rmint_{0}^1dy\,K_0(|b||\Omega|\bar\Delta(y))\,,\\ &
        =\frac{-iT\cdot b}{4\pi \gamma}\rmint_{0}^1 dy\rmint_{0}^\infty \frac{dt}{2t}e^{-t}\rmint_{\Omega}d\Omega\,e^{-i\Omega l}\exp(-\frac{|b|^2\Omega^2\bar\Delta^2}{4t})\,,\\ &
       =\frac{-iT\cdot b}{4 \gamma |b|}\rmint_{0}^1 dy\frac{1}{\bar\Delta(y)}\sqrt{\frac{|b|^2\bar\Delta^2}{|b|^2\bar\Delta^2+l^2}}\,,
    \end{split}
\end{align}
\begin{align}
    \begin{split}
        f_2(x|l)&=T\cdot v_1\, \bar\lambda_1\rmint_{\Omega}\Omega\, e^{-i\Omega (n\cdot x)}\rmint_0^1 dy\,e^{-i\Omega y(n\cdot b)}\frac{b}{\gamma}\frac{K_{1}(b|\Omega|\Delta(k,y))}{4\pi |\Omega|\bar\Delta(y)}\,,\\ &
        =\frac{|b|^2\, T\cdot v_1\, \bar\lambda_1}{16\pi\gamma}\rmint_0^1 dy\rmint_{0}^\infty dt\frac{e^{-t}}{t^2}\rmint_{\Omega} e^{-i\Omega l}\Omega \,\exp\Big(-\frac{\Omega^2 |b|^2 \bar\Delta(y)^2}{4t}\Big)\,,\\ &
        =-\frac{i |b|^2\, T\cdot v_1\, \bar\lambda_1}{4\gamma}\rmint_0^1 dy\, \frac{l}{|b|^3 \bar\Delta^3}\frac{1}{\sqrt{\frac{l^2}{|b|^2 \Delta ^2}+1}}\,.
    \end{split}
\end{align}
and,
\begin{align}
    \begin{split}
        f_3(x|l)= -\frac{i \, T\cdot v_2\, \bar\lambda_2}{4\gamma |b|} \rmint_0^1 dy\, \frac{l}{ \bar\Delta^3}\frac{1}{\sqrt{\frac{l^2}{|b|^2 \Delta ^2}+1}}\,,
    \end{split}
\end{align}
where, $\bar\lambda_i=\frac{\lambda_i}{\Omega}$.
\section{Analysis through method of regions for $\lambda_4\varphi^4$ vertex contribution in the  waveform}\label{App2}
We analyze the integral mentioned in (\ref{5.18k}) using \textit{method of regions}. The method of regions \cite{Beneke:1997zp} is a universal method for expanding Feynman integrals in various limits of momenta and masses.
We split the integration into two regions, one where $q\sim b^{-1}\gg k_{3}$ and $k_{3}\gg q\sim b^{-1}$. The integral \eqref{4PHI} for the region $q\gg k_{3}$ reduces to 
\begin{align}
    \begin{split}
    \label{FstRe1}f_{\varphi}(x)&=\rmint_{\Omega}e^{-i k\cdot x+i k\cdot b_{1}}\rmint_{k_3,q} \frac{\hat\delta(k\cdot v_1-q\cdot v_1)\hat\delta(k_3\cdot v_2)\hat\delta(k_3\cdot v_2-q\cdot v_2)}{k_3^2 \, q^2(q-k)^2}e^{i q\cdot b}\,.
    \end{split}
\end{align}
The above integration can be done easily in the rest frame of the second particle. Hence, taking $v^{\mu}_{2}=(1,0,0,0)$, the above integral simplifies to 
\begin{align}
    \begin{split}
  \label{FstRe}   f_{\varphi}(x)&=\rmint_{\Omega}e^{-i k\cdot x+i k\cdot b_{1}}\rmint_{\vec{k}_3,\vec{q}} \frac{\hat\delta(k\cdot v_1-\vec{q}\cdot \vec{v}_1)}{|\vec{k}_3^2 |\, \vec{q}^2(q-k)^2|_{q^{0}=0}}e^{i \vec{q}\cdot \vec{b}}\,.
    \end{split}
\end{align}
The integral \eqref{FstRe} is a pure divergent for the $\vec{k}_3$ integral. Similarly, for the limit $k_3\gg q$, the integral \eqref{4PHI} reduces to 
\begin{align}
    \begin{split}
    \label{2NdRe}f_{\varphi}(x)&=\rmint_{\Omega}e^{-i k\cdot x+i k\cdot b_{1}}\rmint_{k_3,q} \frac{\hat\delta(k\cdot v_1-q\cdot v_1)\hat\delta(k_3\cdot v_2)\hat\delta(k_3\cdot v_2-q\cdot v_2)}{k_3^4 \,(q-k)^2}e^{i q\cdot b}
    \end{split}
\end{align}
and solving it in the rest frame of the second particle, the $\vec{k}_{3}$ integral and the $\vec{q}$ integral factorizes and the $\vec{k}_{3}$ integral equates to zero. Thus, 
\begin{align}
    \begin{split} \label{nocont}
    f_{\varphi}(x) =&
\rmint_{\Omega}e^{-i k\cdot x}\rmint_{k_3\gg q} \frac{\hat\delta(k\cdot v_1-q\cdot v_1)\hat\delta(k_3\cdot v_2)\hat\delta(k_3\cdot v_2-q\cdot v_2)}{k_3^2 (q-k)^2(q-k_3)^2}e^{i (k-q)\cdot b_1}e^{i(k_3-q)\cdot b_2}e^{ik_3\cdot b_2}\, + \\
&\rmint_{\Omega}e^{-i k\cdot x}\rmint_{q\gg 
k_3} \frac{\hat\delta(k\cdot v_1-q\cdot v_1)\hat\delta(k_3\cdot v_2)\hat\delta(k_3\cdot v_2-q\cdot v_2)}{k_3^2 (q-k)^2(q-k_3)^2}e^{i (k-q)\cdot b_1}e^{i(k_3-q)\cdot b_2}e^{ik_3\cdot b_2}
 \end{split}
\end{align}
is purely divergent. 

\section{Massive integral coming from derivative interaction} \label{appnew1}
The integral of interest is the following,
\begin{align}
    \begin{split}
        J_{(2)}^{\mu\nu}&= \rmint_{q}\hat\delta(q\cdot v_1-k\cdot v_1)\hat\delta(q\cdot v_2)\frac{q^\mu q^\nu\,e^{-i q\cdot b}}{(q-k)^2 (q^2-m^2)}\,,\\ &
        =\rmint_0^1 dy\,e^{-iy\,k\cdot b}\rmint_q \hat\delta(q\cdot v_1-(1-y)k\cdot v_1)\hat\delta(q\cdot v_2+yk\cdot v_2)\frac{e^{-i q\cdot b}}{[q^2-(1-2y+y^2)m^2]^2}(q^{\mu}+yk^\mu)(q^{\nu}+yk^\nu)\,.\label{A.22 m}
    \end{split}
\end{align}
In \eqref{A.22 m} the term proportional to $k^\mu k^\nu$ and $q^{\mu} k^\nu$ can be done using the same procedure as discussed in Appendix~(\ref{app1}). We will concentrate on the term which is proportional to $q^\mu q^\nu$ which can be expanded in term of basis vectors.
\begin{align}
    \begin{split}
        \mathcal{J}_{(2)}^{\mu\nu}(\sim q^\mu q^\nu)=\rmint_0^1 dy\, e^{-iy\,k\cdot b}[\lambda_{\eta}\eta^{\mu\nu}+\lambda_{bb}b^\mu b^\nu+\lambda_{1b}b^{(\mu}v_1^{\nu)}+\lambda_{2b}b^{(\mu}v_2^{\nu)}+\lambda_{11}v_1^\mu v_1^\nu+\lambda_{22}v_2^\mu v_2^\nu+\lambda_{12}v_1^{(\mu}v_2^{\nu)}]\,.\label{A.23 m}
    \end{split}
\end{align}
Now to extract the coefficients in \eqref{A.23 m} one needs to contract the LHS with the tensor structure of the RHS. The equations we need to solve are the following,\\
    \textbullet $\,\,$ Contracting with $b_\mu b_\nu$ gives:
    \begin{align}
       \rmint_q \hat\delta(q\cdot v_1-(1-y)k\cdot v_1)\hat\delta(q\cdot v_2+yk\cdot v_2)\frac{e^{-i q\cdot b}}{[q^2-(1-2y+y^2)m^2]^2} (q\cdot b)^2=\mathcal{X}_1=-\lambda_{\eta}|b|^2+\lambda_{bb}|b|^4\,.\label{A.24 m}
    \end{align}
    \textbullet $\,\,$ Contracting with $b_{(\mu}v_{1\nu)}$ gives:
    \begin{align}
    \begin{split}
        \rmint_q \hat\delta(q\cdot v_1-(1-y)k\cdot v_1)\hat\delta(q\cdot v_2+yk\cdot v_2)\frac{e^{-i q\cdot b}}{[q^2-(1-2y+y^2)m^2]^2} (q\cdot b)(q\cdot v_1)&=\mathcal{X}_2\\ &\hspace{-2 cm}=-\lambda_{1b}\frac{|b|^2}{2}-\lambda_{2b}\frac{\gamma |b|^2}{2}\,.
        \end{split}
    \end{align}
    \textbullet $\,\,$ Contracting with $b_{(\mu}v_{2\nu)}$ gives:
    \begin{align}
        \begin{split}
        \rmint_q \hat\delta(q\cdot v_1-(1-y)k\cdot v_1)\hat\delta(q\cdot v_2+yk\cdot v_2)\frac{e^{-i q\cdot b}}{[q^2-(1-2y+y^2)m^2]^2} (q\cdot b)(q\cdot v_2)&=\mathcal{X}_3\\ &=-\frac{|b|^2}{2}(\lambda_{2b}+\gamma \lambda_{1b})\,.
        \end{split}
    \end{align}
     \textbullet $\,\,$ Contracting with $v_{1\mu} v_{1\nu}$ gives:
    \begin{align}
        \begin{split}
             \rmint_q \hat\delta(q\cdot v_1-(1-y)k\cdot v_1)\hat\delta(q\cdot v_2+yk\cdot v_2)\frac{e^{-i q\cdot b}}{[q^2-(1-2y+y^2)m^2]^2}(q\cdot v_1)^2&=\mathcal{X}_4\\ &
           \hspace{-2 cm}  =\lambda_{\eta}+\lambda_{11}+\lambda_{22}\gamma^2+\lambda_{12}\gamma\,.
        \end{split}
    \end{align}
   \textbullet $\,\,$ Contracting with $v_{2\mu} v_{2\nu}$ gives:
    \begin{align}
        \begin{split}
              \rmint_q \hat\delta(q\cdot v_1-(1-y)k\cdot v_1)\hat\delta(q\cdot v_2+yk\cdot v_2)\frac{e^{-i q\cdot b}}{[q^2-(1-2y+y^2)m^2]^2}(q\cdot v_2)^2&=\mathcal{X}_5\\ &
              \hspace{-2 cm}=\lambda_\eta+\lambda_{22}+\lambda_{11} \gamma^2+\lambda_{12}\gamma\,.
        \end{split}
    \end{align}
    \textbullet $\,\,$ Contracting with $v_{1(\mu}v_{2\nu)}$ gives:
    \begin{align}
        \begin{split}
             \rmint_q \hat\delta(q\cdot v_1-(1-y)k\cdot v_1)\hat\delta(q\cdot v_2+yk\cdot v_2)\frac{e^{-i q\cdot b}}{[q^2-(1-2y+y^2)m^2]^2}(q\cdot v_1)(q\cdot v_2)&=\mathcal{X}_6\\ &
              \hspace{-2.8 cm}=\lambda_{\eta}\gamma+\gamma(\lambda_{11}+\lambda_{22})+\frac{\lambda_{12}}{2}(\gamma^2+1)\,.
        \end{split}
    \end{align}
      \textbullet $\,\,$ Contracting with $\eta_{\mu\nu}$ gives:
      \begin{align}
          \begin{split}
                \rmint_q \hat\delta(q\cdot v_1-(1-y)k\cdot v_1)\hat\delta(q\cdot v_2+yk\cdot v_2)\frac{e^{-i q\cdot b}}{[q^2-(1-2y+y^2)m^2]^2}q^2&=\mathcal{X}_7\\ &
                \hspace{-2 cm}=4\lambda_\eta-\lambda_{bb}|b|^2+\lambda_{11}+\lambda_{22}+\lambda_{12}\gamma\,.\label{A.30 m}
          \end{split}
      \end{align}
Solving \eqref{A.24 m} to \eqref{A.30 m} we will get,
\begin{align}
    \begin{split}
     &  \lambda _{\eta }\to \frac{\gamma ^2 \mathcal{X}_{7}-2 \gamma  \mathcal{X}_{6}+\mathcal{X}_{4}+\mathcal{X}_{5}-\mathcal{X}_{7}}{\gamma ^2-1}+\frac{\mathcal{X}_{1}}{|b|^2},\lambda _{\text{bb}}\to \frac{2 \mathcal{X}_{1}}{|b|^4}+\frac{\gamma ^2 \mathcal{X}_{7}-2 \gamma  \mathcal{X}_{6}+\mathcal{X}_{4}+\mathcal{X}_{5}+\mathcal{X}_{7}}{|b|^2 \left(\gamma ^2-1\right)},\\ & \lambda _b\to -\frac{2 \left(\gamma  \mathcal{X}_{3}-\mathcal{X}_{2}\right)}{|b|^2 \left(\gamma ^2-1\right)},\lambda _{2 b}\to -\frac{2 \left(\gamma  \mathcal{X}_{2}-\mathcal{X}_{3}\right)}{|b|^2 \left(\gamma ^2-1\right)},\\ &\lambda _{12}\to \frac{2 \left(-|b|^2 \left(\gamma ^3 \left(-\mathcal{X}_{7}\right)+3 \gamma ^2 \mathcal{X}_{6}-2 \gamma  \mathcal{X}_{4}-2 \gamma  \mathcal{X}_{5}+\gamma  \mathcal{X}_{7}+\mathcal{X}_{6}\right)+\gamma  \left(\gamma ^2-1\right) \mathcal{X}_{1}\right)}{|b|^2 \left(\gamma ^2-1\right)^2},\\ & \lambda _{11}\to \frac{\frac{\left(\gamma ^2-1\right) \mathcal{X}_{1}}{|b|^2}+\gamma ^2 \mathcal{X}_{7}+\left(\gamma ^2+1\right) \mathcal{X}_{5}-4 \gamma  \mathcal{X}_{6}+2 \mathcal{X}_{4}-\mathcal{X}_{7}}{\left(\gamma ^2-1\right)^2},\\ &\lambda _{22}\to \frac{\frac{\left(\gamma ^2-1\right) \mathcal{X}_{1}}{|b|^2}+\gamma ^2 \mathcal{X}_{7}+\left(\gamma ^2+1\right) \mathcal{X}_{4}-4 \gamma  \mathcal{X}_{6}+2 \mathcal{X}_{5}-\mathcal{X}_{7}}{\left(\gamma ^2-1\right)^2}\,.
    \end{split}
\end{align}
Next, we list down the values of the integrals $\mathcal{X}_i's\,.$
\begin{align}
    \begin{split}
        \mathcal{X}_1&= -\lim_{\kappa\to 1}\frac{\partial^2}{\partial \kappa^2}   \rmint_q \hat\delta(q\cdot v_1-(1-y)k\cdot v_1)\hat\delta(q\cdot v_2+yk\cdot v_2)\frac{e^{-i\kappa q\cdot b}}{[q^2-(1-2y+y^2)m^2]^2}\,,\\ &
        =-\frac{\kappa |b|}{4\pi \gamma}\lim_{\kappa\to 1}\frac{\partial^2}{\partial \kappa^2}\frac{K_1(\kappa |b|\sqrt{\Delta^2+(1-2y+y^2)m^2})}{\sqrt{\Delta^2+(1-2y+y^2)m^2}}\,,\\ &
        =-\frac{|b|}{4\pi \gamma}\Big[\frac{[|b|^2 \left(\Delta ^2+m^2 (y-1)^2\right)+2]K_1\left(|b| \sqrt{m^2 (y-1)^2+\Delta ^2}\right)}{\sqrt{\Delta ^2+m^2 (y-1)^2}}-|b| K_2\left(|b| \sqrt{m^2 (y-1)^2+\Delta ^2}\right)\Big]\,.
    \end{split}
\end{align}
\begin{align}
    \begin{split}
        \mathcal{X}_2&=i(1-y)k\cdot v_1\lim_{\kappa\to 1}\frac{\partial}{\partial \kappa}\rmint_q \hat\delta(q\cdot v_1-(1-y)k\cdot v_1)\hat\delta(q\cdot v_2+yk\cdot v_2)\frac{e^{-i\kappa q\cdot b}}{[q^2-(1-2y+y^2)m^2]^2}\,,\\ &
        =-i(1-y)k\cdot v_1\frac{|b|^2 K_0\left(|b| \sqrt{m^2 (y-1)^2+\Delta ^2}\right)}{4 \pi  \gamma }\,.
    \end{split}
\end{align}
\begin{align}
    \begin{split}
        \mathcal{X}_3&=-iy\,k\cdot v_2\lim_{\kappa\to 1}\frac{\partial}{\partial \kappa}\rmint_q \hat\delta(q\cdot v_1-(1-y)k\cdot v_1)\hat\delta(q\cdot v_2+yk\cdot v_2)\frac{e^{-i \kappa\,q\cdot b}}{[q^2-(1-2y+y^2)m^2]^2}\,,\\ &
        =i y \,k\cdot v_2 \frac{|b|^2 K_0\left(|b| \sqrt{m^2 (y-1)^2+\Delta ^2}\right)}{4 \pi  \gamma }\,.
    \end{split}
\end{align}
\begin{align}
    \begin{split}
        \mathcal{X}_4&=(1-y)^2 (k\cdot v_1)^2\rmint_q \hat\delta(q\cdot v_1-(1-y)k\cdot v_1)\hat\delta(q\cdot v_2+yk\cdot v_2)\frac{e^{-i q\cdot b}}{[q^2-(1-2y+y^2)m^2]^2}\,,\\ &
        =(1-y)^2 (k\cdot v_1)^2\frac{|b|}{4\pi \gamma}\frac{K_1( |b|\sqrt{\Delta^2+(1-2y+y^2)m^2})}{\sqrt{\Delta^2+(1-2y+y^2)m^2}}\,.
    \end{split}
\end{align}
\begin{align}
    \begin{split}
         \mathcal{X}_5&=y^2 (k\cdot v_2)^2\rmint_q \hat\delta(q\cdot v_1-(1-y)k\cdot v_1)\hat\delta(q\cdot v_2+yk\cdot v_2)\frac{e^{-i q\cdot b}}{[q^2-(1-2y+y^2)m^2]^2}\,,\\ &
        =y^2 (k\cdot v_2)^2\frac{|b|}{4\pi \gamma}\frac{K_1( |b|\sqrt{\Delta^2+(1-2y+y^2)m^2})}{\sqrt{\Delta^2+(1-2y+y^2)m^2}}\,.
    \end{split}
\end{align}
\begin{align}
    \begin{split}
        \mathcal{X}_{6}&=-y(1-y)(k\cdot v_1)(k\cdot v_2)\rmint_q \hat\delta(q\cdot v_1-(1-y)k\cdot v_1)\hat\delta(q\cdot v_2+yk\cdot v_2)\frac{e^{-i q\cdot b}}{[q^2-(1-2y+y^2)m^2]^2}\,,\\ &
        =-y(1-y)(k\cdot v_1)(k\cdot v_2)\frac{|b|}{4\pi \gamma}\frac{K_1( |b|\sqrt{\Delta^2+(1-2y+y^2)m^2})}{\sqrt{\Delta^2+(1-2y+y^2)m^2}}\,.
    \end{split}
\end{align}
The integral of $\mathcal{X}_7$ is little bit involved. One can do this integral in the following way,
\begin{align}
    \begin{split}
        \mathcal{X}_7&=\rmint_q \hat\delta(q\cdot v_1-(1-y)k\cdot v_1)\hat\delta(q\cdot v_2+yk\cdot v_2)\frac{e^{-i q\cdot b}}{[q^2-(1-2y+y^2)m^2]^2}\Big[q^2-(1-y)^2m^2+(1-y)^2 m^2\Big]\,.\label{A.38}
    \end{split}
\end{align}
The integral in \eqref{A.38} has two parts and can be evaluated separately,
\begin{align}
    \begin{split}
        \mathcal{X}_7^{(1)}&\equiv  \rmint_q \hat\delta(q\cdot v_1-(1-y)k\cdot v_1)\hat\delta(q\cdot v_2+yk\cdot v_2)\frac{e^{-i q\cdot b}}{[q^2-(1-2y+y^2)m^2]}\,,\\ &
        =\frac{1}{\gamma}\rmint_0^\infty \frac{dt}{4\pi}\,\frac{1}{t}\exp\Big(-\frac{|b|^2}{4t}-t\Delta^2-t(1-y)^2m^2\Big)\,,\\ &
        =\frac{1}{2\pi \gamma}K_{0}(|b|\sqrt{\Delta^2+(1-y)^2m^2})\,.
    \end{split}
\end{align}
and the second term takes the form,
\begin{align}
    \begin{split}
        \mathcal{X}_7^{(2)}&=(1-y)^2 m^2\,\rmint_q \hat\delta(q\cdot v_1-(1-y)k\cdot v_1)\hat\delta(q\cdot v_2+yk\cdot v_2)\frac{e^{-i q\cdot b}}{[q^2-(1-2y+y^2)m^2]^2}\,,\\ &
        =(1-y)^2 m^2\frac{|b|}{4\pi\gamma} \frac{K_1( |b|\sqrt{\Delta^2+(1-2y+y^2)m^2})}{\sqrt{\Delta^2+(1-2y+y^2)m^2}}\,.
    \end{split}
\end{align}

\section{Computation of worldline radiation diagram using large velocity approximation}\label{app22}
\begin{figure}
    \centering
    \includegraphics[scale=0.15]{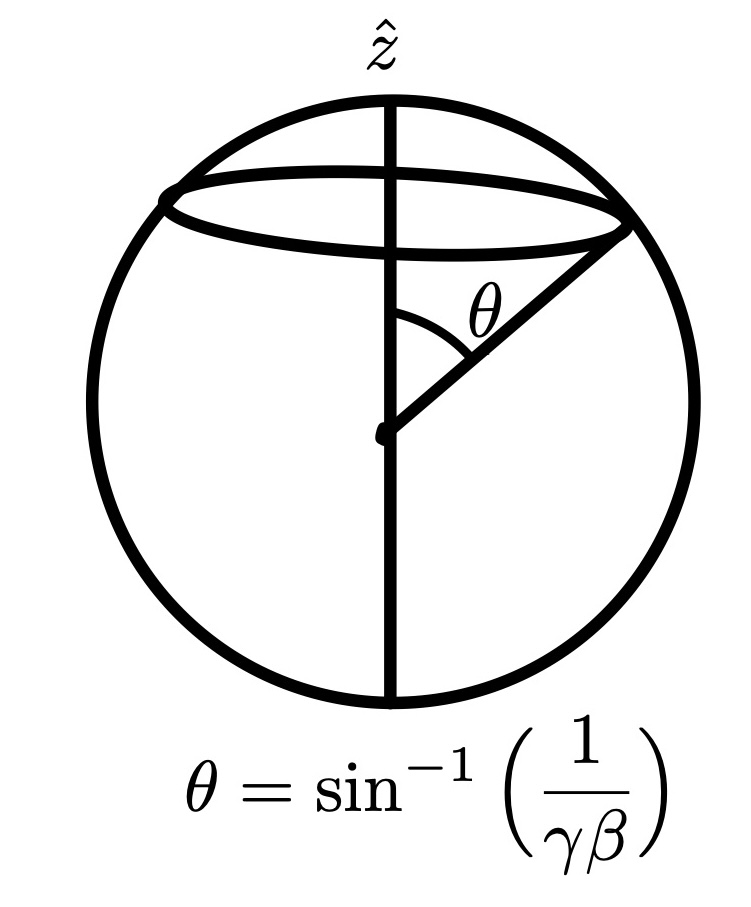}
    \caption{Sketch of the celestial sphere under consideration.}
    \label{celestial}
\end{figure}

We start with \eqref{6.4 mm} (omitting the overall constant which is irrelevant to explain our point).
\begin{align}
    \begin{split}
        f_{\varphi}(x)&\propto\rmint_{\Omega}e^{-ik\cdot (x+b_1)}\rmint_{\omega,k_1}e^{-i k_1\cdot b}\hat{\delta}(k_1\cdot v_2)\hat{\delta}[\Omega\, n(m,\Omega)\cdot v_1-\omega]\hat\delta(k_1\cdot v_1-\omega)\frac{\Omega(n\cdot k_1)}{\omega^2 (k_1^2-m^2)}\,.\label{7.7}
    \end{split}
\end{align}
To do the integration over $\Omega$ in (\ref{7.7}) we first find the roots of the equation $f(\Omega):=\Omega\, n(m,\Omega)\cdot v_1-\omega=0$, which gives,
\begin{align}
    \begin{split}
       & \hat{\delta}[\Omega\, n(m,\Omega)\cdot v_1-\omega]=\frac{\hat\delta(\Omega-\Omega_1)}{|f'(\Omega_1)|}+\frac{\hat\delta(\Omega-\Omega_2)}{|f'(\Omega_2)|},\,\, \\ &
        \text{where,}\,\, \Omega_{1}(\omega)=\fbox{$\frac{\tilde\beta  \sqrt{\left(\tilde\beta ^2-1\right) \gamma ^2 m^2+\omega ^2}+\omega }{\left(1-\tilde\beta ^2\right) \gamma }\text{ if }\gamma  m<\omega \lor \frac{\omega }{\gamma }<m<\frac{\omega }{\sqrt{1-\tilde\beta ^2} \gamma }$}\,\,\,\,\text{and,}\\ & \vspace{0.5cm}
    \hspace{1.2 cm}\Omega_2(\omega)=\fbox{$\frac{\tilde\beta  \sqrt{\left(\tilde\beta ^2-1\right) \gamma ^2 m^2+\omega ^2}-\omega }{\left(\tilde\beta ^2-1\right) \gamma }\text{ if }\frac{\omega }{\gamma }<m<\frac{\omega }{\sqrt{1-\tilde\beta ^2} \gamma }\,.$}
    \end{split}
\end{align}
As the observed frequency can not be negative, only the $\Omega_1$ solution is relevant to us. Therefore, the integral in (\ref{7.7}) can be written as (with $\tilde{x}=x-b_1$),
\begin{align}
    \begin{split}
        f_{\varphi}(x)&\propto  \sum_{a=1}^2\rmint_{k_1}e^{-i \Omega_a\, n(m,\Omega_a)\cdot (x-b_1)}e^{-ik_1\cdot b}\hat\delta(k_1\cdot v_2)\frac{\Omega_a\,n(m,\Omega_a)\cdot k_1}{|f'(\Omega_a)|(k_1\cdot v_1+i\epsilon)^2(k_1^2-m^2)}\,.
    \end{split}
\end{align}
\textcolor{black}{The integral could be significantly simplified by putting a suitable IR cutoff in the $\omega$ and $\Omega$ integral with the limit $\frac{m^2}{\Omega^2}\ll 1$.\textcolor{black}{The argument in favour of this statement is the following: $\frac{1}{m}$ is order of cosmological scales and $\Omega$ for a compact binary of roughly an hour of orbital period is of the order of $10^{-3}$ Hz(1/m $\sim$ 1 Mpc which makes m $\sim$ $10^{-14}$Hz)\,. } One point has to be noted is that, when we are putting the cutoff on the $\omega$ and $\Omega$ integral, it automatically implies that there should be a cutoff on the integral over $k_1$ also (coming from the condition $k_1\cdot v_1=\omega$). However, one could, in principle, argue that, instead of giving the cutoff on the $k_1$ integral, one should put a suitable condition on $|v_1|$ such that $k_1\cdot v_1\gg m$, even if $|k_1|\rightarrow 0$. Now the part $\sqrt{(k_1\cdot v_1)^2-m^2\gamma^2(1-\tilde\beta^2)}$ coming from  the solutions of $\Omega$ takes the following form in terms the parametrization that we have used here: $\sqrt{\gamma^2\beta^2 k_1^{(x)2}-m^2\gamma^2(1-\tilde\beta^2)}=\gamma\beta \sqrt{k_1^{(x)2}-\frac{m^2}{\gamma^2\beta^2}+m^2\sin^2\theta}$. Now, we assume that we can place our at at particular direction on the celestial sphere as shown in the Fig.~(\ref{celestial}), such that, $\sin^2\theta=\frac{1}{\gamma^2\beta^2}\,.$ For further simplification, we choose $\gamma\beta\rightarrow \infty$ so that $\sin \theta\rightarrow 0$.}
Hence, the integration that we have to perform is of the form as shown below,
\begin{align}
    \begin{split}
        f_{\varphi}(x)\sim \rmint_{\Omega \in| \Omega|_{\textrm{IR}}\gg m}d\Omega\rmint_{\omega \in |\omega|_{\textrm{IR}}\gg m}d\omega\rmint_{k_1,|k_1\cdot v_1|\gg m}\cdots\,.
    \end{split}
\end{align}

Finally we get,
\begin{align}
    \begin{split}
        f_{\varphi}(x)\propto\frac{n^{\mu}}{(n\cdot v_1)^2}\underbrace{\rmint_{k_1}e^{-i k_1\cdot w_1}\hat{\delta}(k_1\cdot v_2)\frac{k_{1\mu}}{(k_1\cdot v_1+i\epsilon)(k_1^2-m^2)}}_{\mathcal{K}_{\mu}^{(2)}},\,\, w_1\equiv\frac{n\cdot (x-b_1)}{n\cdot v_1}v_1+b
    \end{split}
\end{align}  
where,
\begin{align}
    \begin{split}
        \mathcal{K}_{\mu}^{(2)}\to-\rmint_{-\infty}^{\infty}d\tau\,\theta(\tau)\frac{(\vec w_1-\tau \vec v_1)_{i}}{|\vec w_1-\tau \vec v_1|^3}\,[1+m |\vec w_1-\tau \vec v_1|]\,\exp(-m |\vec w_1-\tau \vec v_1|)\,.
    \end{split}
\end{align}
Thenn we proceed as follows:
\begin{align}
    \begin{split}
        n\cdot \mathcal{K}^{(2)}&\rightarrow -\rmint d\tau\,\theta(\tau) \frac{\vec n\cdot (\vec w_1-\tau \vec v_1)}{|\vec w_1-\tau v_1|^3}[1+m |\vec w_1-\tau v_1|]\,\exp(-m |\vec w_1-\tau \vec v_1|)\,,\\ &
        =\rmint_{-\infty}^{\infty} d\tau\, \theta(\tau)\frac{(u_1-\tau)\gamma\tilde\beta+\chi}{(\gamma^2-1)^{3/2}\Big[\tau^2-u_1^2-2\tau u_1+\frac{b^2}{\gamma^2-1}\Big]^{3/2}}\Bigg[1+m(\gamma^2-1)^{1/2}\sqrt{\tau^2-u_1^2-2\tau u_1+\frac{|b|^2}{\gamma^2-1}}\Bigg]\\ &
        \hspace{2 cm}\exp\Big[-m(\gamma^2-1)^{1/2}\sqrt{\tau^2-u_1^2-2\tau u_1+\frac{|b|^2}{\gamma^2-1}}\Big]\,.
    \end{split}
\end{align}
In the limit of $\sin\theta\rightarrow 0$, the integral has a closed-form and smooth massless limit.\\\\
\hfsetfillcolor{gray!10}
\hfsetbordercolor{black}
\begin{equation}\label{e}\begin{split}
\tikzmarkin[disable rounded corners=false]{x}(0.3,-0.9)(-0.5,1.1)  f_{\varphi}(x)\propto \frac{\gamma}{(n\cdot v_1)^2(\gamma^2-1)} \frac{ \exp\Big({-m\sqrt{\gamma ^2-1}  \sqrt{\frac{|b|^2}{\gamma ^2-1}-u_1^2}}\,\Big)}{\sqrt{\frac{|b|^2}{\gamma ^2-1}-u_1^2}}\,.
\tikzmarkend{x}\\
        \end{split}
\end{equation}
\bibliography{ref}
\bibliographystyle{utphysmodb}

    \end{document}